\documentclass[lettersize,journal]{IEEEtran}
 \usepackage{cite}
\usepackage{booktabs}
\usepackage{tabularx}
\usepackage{multirow}
\usepackage{xcolor, colortbl}
\usepackage{enumitem}
\usepackage{amsfonts}
\usepackage[acronym]{glossaries}
\usepackage{enumitem,kantlipsum}
\usepackage{textcomp}
\usepackage{pifont}

\usepackage{stfloats}
\usepackage{url}
\usepackage{caption}
\usepackage{graphics} 
\usepackage{verbatim}
\usepackage{mathtools}
\usepackage{pifont}
\usepackage{float}
\usepackage{amsmath}
\usepackage{cancel}
\usepackage{amsmath}
\usepackage{graphicx}
\usepackage{wrapfig}
\usepackage{float}
\usepackage{lscape}
\usepackage{ltxtable}
\usepackage{hhline}
\usepackage{caption}
\usepackage{wasysym}
\usepackage{xurl}
\usepackage{subcaption}
\usepackage{algorithm} 
\usepackage{color} 
\usepackage[export]{adjustbox}
\usepackage{colortbl}
\usepackage{algpseudocode}
\usepackage{verbatim}
\usepackage{lipsum}
\usepackage{tikz}
\usetikzlibrary{shapes,arrows}
\usetikzlibrary{intersections}
\usetikzlibrary{automata, arrows.meta, positioning}
\usetikzlibrary{decorations.pathreplacing}
\usepackage{amsthm}
\usepackage{booktabs}
\usepackage{amsmath}
\usepackage{algorithm}
\usepackage[T1]{fontenc} 
\usepackage[utf8]{inputenc} 
\usepackage{pmboxdraw} 
\usepackage{tikz}
\usepackage{graphicx}
\usetikzlibrary{arrows.meta, positioning, shapes.misc, fit, decorations.pathreplacing, shadows}
\usepackage{caption}
\usepackage{inputenc}
\usepackage{adjustbox}
\usepackage{url} 
\usepackage[hidelinks]{hyperref}
\hypersetup{breaklinks=true}
\usepackage{multirow}
\usepackage{color,soul}
\usepackage{lipsum}
\usepackage{mathtools}
\usepackage{subcaption}
\usepackage{cuted}
\usepackage{wasysym}
\usepackage{pgfplotstable}
\usepackage{mdframed}
\usetikzlibrary{shapes, arrows.meta, positioning, decorations.pathreplacing}
\usepackage[normalem]{ulem}
\usetikzlibrary{arrows.meta}
\usepackage{smartdiagram}
\definecolor{lemon}{rgb}{1.0, 1.0, 0.13}
\usetikzlibrary{arrows.meta}
\usepackage{smartdiagram}
\usetikzlibrary{positioning, fit, calc, shapes, arrows,trees}
\usetikzlibrary{shapes.geometric, arrows.meta, positioning, shadows} 
\usepackage{fontawesome5} 

\usetikzlibrary{3d,calc}
\usetikzlibrary{positioning, shapes.geometric, arrows.meta, decorations.pathmorphing, shadows}

\definecolor{columbiablue}{rgb}{0.61, 0.87, 1.0}

\usepackage{algpseudocode}
\usetikzlibrary{mindmap,backgrounds,shadows.blur} 
\setlist[itemize]{leftmargin=*, nosep, topsep=1pt, align=parleft}
\newcommand{\checkmarkpq}{\checkmark\textsubscript{PQC}}
\newcommand{\checkmarkqc}{\checkmark\textsubscript{QC}}
\newcommand{\checkmarkc}{\checkmark\textsubscript{C}}
\definecolor{BulletsColor}{rgb}{0, 0, 0.9}
\newlist{myBullets}{itemize}{1}

\setlist[myBullets]{
 label={\textbullet},
 leftmargin=*,
 topsep=1ex,
 partopsep=0ex,
 parsep=0ex,
 itemsep=0ex,
}
\newcounter{module}


\usetikzlibrary{mindmap,shapes,backgrounds}
\usepackage{amsmath}

\usepackage[cmintegrals]{newtxmath}
\usepackage{tikz}
\usepackage{adjustbox}
\usetikzlibrary{mindmap,trees,backgrounds}

\usepackage{tikz}
\usepackage{graphicx}
\usetikzlibrary{arrows.meta, positioning, shapes.misc, fit, decorations.pathreplacing, shadows}

\definecolor{compositecolor}{RGB}{102,51,153} 
\definecolor{accentcolor}{RGB}{255,87,51} 
\definecolor{consensusgrey}{RGB}{52,73,94} 
\definecolor{blockcolor}{RGB}{236,240,241} 
\definecolor{systemcolor}{RGB}{46,134,193} 
\definecolor{signaturecolor}{RGB}{26,188,156} 
\definecolor{reviewer2}{rgb}{0.0, 0.0, 0.8}   
\definecolor{reviewer3}{rgb}{0, 0.8, 0}  
\usetikzlibrary{backgrounds, shapes.geometric}

\usepackage{hhline}
\definecolor{columbiablue}{rgb}{0.61, 0.87, 1.0}

\tikzset{%
 thick arrow/.style={
 -{Triangle[angle=120:1pt 1]},
 line width=1cm, 
 draw=gray!40
 },
 arrow label/.style={
 text=black,
 align=center
 },
 set mark/.style={
 insert path={
 node [midway, arrow label, node contents=#1]
 }
 }
}
\newcommand\deleted{\bgroup\markoverwith{\textcolor{red}{\rule[0.5ex]{2pt}{0.4pt}}}\ULon}

\usetikzlibrary{pgfplots.groupplots}

\newcommand\doublecheck{\textcolor{black}{\checkmark\kern-0em\checkmark}}
\newcommand\semidoublecheck{\textcolor{black}{\checkmark\kern-0em\bcancel{\checkmark}}}

\definecolor{lemon}{rgb}{1.0, 1.0, 0.13}

\newcommand\low{\cellcolor{green!60}L}
\newcommand\med{\cellcolor{lemon!80}M}
\newcommand\high{\cellcolor{red!80}H}

\newacronym{AI}{AI}{Artificial Intelligence}
\newacronym{AML}{AML}{Anti-Money Laundering}
\newacronym{BC}{BC}{Blockchain}
\newacronym{BFT}{BFT}{Byzantine Fault Tolerance}
\newacronym{DAG}{DAG}{Directed Acyclic Graph}
\newacronym{DAO}{DAO}{Decentralized Autonomous Organization}
\newacronym{DApp}{DApp}{Decentralized Application}
\newacronym{DDoS}{DDoS}{Distributed Denial of Service}
\newacronym{DeFi}{DeFi}{Decentralized Finance}
\newacronym{DI-QRNG}{DI-QRNG}{Device-Independent Quantum Random Number Generator}
\newacronym{DoS}{DoS}{Denial of Service}
\newacronym{DSA}{DSA}{Digital Signature Algorithm}
\newacronym{ECDSA}{ECDSA}{Elliptic Curve Digital Signature Algorithm}
\newacronym{ECIES}{ECIES}{Elliptic Curve Integrated Encryption Scheme}
\newacronym{FIFO}{FIFO}{First In, First Out}
\newacronym{FL}{FL}{Federated Learning}
\newacronym{HNDL}{HNDL}{Harvest-Now, Decrypt-Later}
\newacronym{HTLC}{HTLC}{Hash Time-Locked Contract}
\newacronym{HTLCs}{HTLCs}{Hash Time-Locked Contracts}
\newacronym{I}{I}{Impact}
\newacronym{IDS/IPS}{IDS/IPS}{Intrusion Detection and Prevention Systems}
\newacronym{IoT}{IoT}{Internet of Things}
\newacronym{KDF}{KDF}{Key Derivation Function}
\newacronym{KEM/ENC}{KEM/ENC}{Key Encapsulation Mechanism/Encryption}
\newacronym{KYC}{KYC}{Know Your Customer}
\newacronym{L}{L}{Likelihood}
\newacronym{MAC}{MAC}{Message Authentication Code}
\newacronym{ML}{ML}{Machine Learning}
\newacronym{MPC}{MPC}{Multi-Party Computation}
\newacronym{NFT}{NFT}{Non-Fungible Token}
\newacronym{NIST}{NIST}{National Institute of Standards and Technology}
\newacronym{PoA}{PoA}{Proof of Authority}
\newacronym{PoR}{PoR}{Proof of Randomness}
\newacronym{PoS}{PoS}{Proof of Stake}
\newacronym{POSet}{POSet}{Partially Ordered Set}
\newacronym{PoW}{PoW}{Proof of Work}
\newacronym{PQC}{PQC}{Post-Quantum Cryptography}
\newacronym{PRF}{PRF}{Pseudorandom Function}
\newacronym{QC}{QC}{Quantum Computing}
\newacronym{QML}{QML}{Quantum Machine Learning} 
\newacronym{R}{R}{Risk}
\newacronym{RNG}{RNG}{Random Number Generator}
\newacronym{SegWit}{SegWit}{Segregated Witness}
\newacronym{STRIDE}{STRIDE}{Spoofing, Tampering, Repudiation, Information Disclosure, Denial of Service, Elevation of Privilege}
\newacronym{TLS}{TLS}{Transport Layer Security}
\newacronym{VDF}{VDF}{Verifiable Delay Function}
\newacronym{VRF}{VRF}{Verifiable Random Function}
\newacronym{ZKP}{ZKP}{Zero-Knowledge Proof}
\makeglossaries
\begin{document}
\title{Blockchain Security Risk Assessment in Quantum Era, Migration Strategies and Proactive Defense}

\author{Yaser Baseri, Abdelhakim Hafid, Yahya Shahsavari, Dimitrios Makrakis and Hassan Khodaiemehr
 
\IEEEcompsocitemizethanks{\IEEEcompsocthanksitem Yaser Baseri, Abdelhakim Hafid, and Yahya Shahsavari are with the Montreal Blockchain Lab, Université de Montréal, Canada. Emails: \{yaser.baseri, ahafid, yahya.shahsavari\}@umontreal.ca. 
Dimitrios Makrakis is with the School of Electrical Engineering and Computer Science, University of Ottawa, Canada. Email: dmakraki@uottawa.ca. 
Hassan Khodaiemehr is with the School of Engineering, The University of British Columbia, Canada. Email: hassan.khodaiemehr@ubc.ca.\protect
}}
\markboth{}%
{Baseri \MakeLowercase{\textit{et al.}}: A Sample Article Using IEEEtran.cls for IEEE Journals}


\maketitle
\begin{abstract}

The advent of \gls{QC} poses significant threats to the cryptographic foundations of \gls{BC} systems, as quantum algorithms like Shor's and Grover's undermine the security of public-key cryptography and hash functions. This research conducts a comprehensive risk assessment of quantum vulnerabilities across critical \gls{BC} components, including consensus mechanisms, smart contracts, and digital wallets. Leveraging the STRIDE threat modeling framework, we analyze threat vectors specific to \gls{QC}, identifying key areas most susceptible to quantum-enabled attacks, such as private key compromise, consensus disruptions, and smart contract integrity risks.
Our contributions provide actionable mitigation strategies, including a detailed security blueprint for quantum resilience, encompassing the integration of \gls{PQC} and the adoption of quantum-resistant hash functions. We offer implementation best practices, focusing on key management, secure coding, and network security to strengthen \gls{BC} components against quantum threats.
To mitigate the risk of \gls{QC} during transition from classical to quantum-resistant \glspl{BC}, we present two hybrid \gls{BC} architectures. As part of a comprehensive quantum resilience strategy, these architectures facilitate a secure and scalable migration by integrating platform-specific adaptations that balance security, adaptability, and operational efficiency. Our analysis extends to major \gls{BC} platforms, including Bitcoin, Ethereum, Ripple, Litecoin, and Zcash, providing platform-specific vulnerability assessments and highlighting unique weaknesses in the quantum era.
By identifying vulnerabilities, developing proactive defense strategies, and adopting a structured hybrid migration approach, this research equips \gls{BC} stakeholders with a robust framework to achieve long-term quantum resilience. Finally, we explore challenges and research directions for integrating emerging technologies, including quantum machine learning, \gls{AI}, and Web3, with \gls{BC} systems, and discuss new threats that may arise from this convergence in the \gls{QC} era.
\end{abstract}


\section{Introduction}\label{sec:intro}
\IEEEPARstart{T}{he} advent of \gls{QC} represents a paradigm shift that fundamentally challenges the cryptographic foundations of \gls{BC} security. Quantum algorithms exploit the computational hardness assumptions underlying contemporary cryptographic schemes—integer factorization, discrete logarithms, unstructured search, and collision resistance—thereby compromising the digital signatures, encryption protocols, and hash functions that guarantee transaction authenticity, data confidentiality, and \gls{BC} immutability~\cite{kearney2021vulnerability,allende2023quantum}. Shor's   algorithm~\cite{shor1999polynomial} efficiently solves both integer factorization and discrete logarithm problems, rendering public-key cryptosystems such as RSA, \gls{ECDSA}, and Elliptic-curve Diffie–Hellman (ECDH) vulnerable and jeopardizing transaction verification mechanisms in platforms including Bitcoin and Ethereum. While symmetric cryptography demonstrates greater resilience, Grover's quantum search algorithm~\cite{grover1996fast} reduces effective key strength from $2^n$ to $2^{n/2}$ operations, necessitating doubled key lengths to preserve security equivalence. Hash function security faces similar degradation through the Brassard et al. quantum collision search algorithm~\cite{brassard1997quantum}, which reduces collision resistance from $\mathcal{O}(2^n)$ to $\mathcal{O}(2^{n/3})$, amplifying risks of block manipulation and consensus disruption. These quantum advances collectively create a systemic vulnerability across existing \gls{BC} infrastructures, making the transition to post-quantum cryptographic standards not merely advisable but imperative for preserving the integrity and trustworthiness of distributed ledger ecosystems.

This study investigates risks associated with transitioning from non-quantum-safe to quantum-resistant cryptographic methods across \gls{BC} components. It assesses vulnerabilities posed by quantum attacks on key components—networks, mining pools, transaction verification, smart contracts, and user wallets. Such vulnerabilities can compromise \gls{BC} integrity and security, posing risks of unauthorized access, data manipulation, and financial loss. To counter these evolving threats during transition, organizations must adopt a holistic security approach integrating quantum-resistant cryptography with system design and continuous monitoring~\cite{joseph2022transitioning,cisa2023quantum,nist2023migration}. A hybrid migration strategy, gradually transitioning from classical to quantum-resistant cryptography, is essential to reduce transition risks~\cite{kwon2024compact, ghinea2022hybrid}. This approach is critical for ensuring robust security, as organizations may encounter challenges such as increased key sizes requiring network traffic handling adjustments, and implementation complexities throughout the transition~\cite{nist2024transition,nather2024migrating}.

While transitioning to quantum-resistant cryptography is essential, it does not fully address all potential threats, as \gls{BC} systems remain vulnerable to sophisticated attack vectors beyond encryption weaknesses~\cite{CISAPQC, leng2020blockchain}. In the post-transition stage, organizations face multifaceted challenges across system architecture, network infrastructure, and operational processes~\cite{lim2024managing, chawla2023roadmap}. {Quantum-resistant algorithms introduce challenges such as larger key sizes, increased payloads, and higher computational overhead, resulting in latency, bandwidth strain, and fragmented network traffic. These performance bottlenecks can delay block propagation, thereby increasing fork probability and destabilizing consensus. Adversaries may exploit such instability to reorder transactions, trigger chain reorganizations, or manipulate protocol execution. Moreover, integrating quantum-resistant cryptography with existing \gls{BC} infrastructure creates novel attack vectors that can compromise smart contracts and transaction workflows.} Additionally, the coexistence of legacy and quantum-resistant protocols in hybrid architectures may create security gaps, while side-channel attacks could exploit hardware-level details in cryptographic operations. These risks 
underscore the need for continuous refinement and vigilant monitoring of \gls{BC} ecosystems well beyond cryptographic upgrades.

This paper addresses these challenges by leveraging the STRIDE (Spoofing, Tampering, Repudiation, Information Disclosure, Denial of Service, and Elevation of Privilege) framework~\cite{van2021descriptive,baseri2024evaluation} to identify and prioritize threats. We analyze vulnerabilities before, during, and after transitioning to quantum-safe algorithms, providing insights for effective countermeasures. As quantum threats escalate, especially during transition to quantum-resistant \glspl{BC}, proactive risk assessment and defense strategies become imperative. By advocating early vulnerability identification and integrating quantum-resistant solutions into system design, this study aims to safeguard \gls{BC} network integrity and ensure long-term resilience in the quantum era.

\begin{table}[!ht]
\centering 
\caption{List of Abbreviations}
\label{tab:abbreviation}
\scriptsize
\resizebox{\linewidth}{!}{%
\begin{tabular}{|c|p{0.755\linewidth}|}
\hline
\textbf{Abbreviation} & \textbf{Description} \\ \hline
{AI}&{Artificial Intelligence}\\ \hline
AML & Anti-Money Laundering \\ \hline
\gls{BC} & Blockchain \\ \hline
BFT & Byzantine Fault Tolerance \\ \hline
DAG & Directed Acyclic Graph \\ \hline
DAO & Decentralized Autonomous Organization \\ \hline
DApp & Decentralized Application \\ \hline
DDoS & Distributed Denial of Service \\ \hline
DeFi&Decentralized Finance\\ \hline
DH & Diffie–Hellman\\ \hline
DI-QRNG & Device-Independent Quantum Random Number Generator \\ \hline
DoS & Denial of Service \\ \hline
DSA & Digital Signature Algorithm \\ \hline
ECC & Elliptic-Curve Cryptography \\ \hline
ECDSA & Elliptic Curve Digital Signature Algorithm \\ \hline
ECIES & Elliptic Curve Integrated Encryption Scheme \\ \hline
FL & Federated Learning \\ \hline
HNDL&Harvest-Now, Decrypt-Later\\ \hline
HTLC&Hash Time-Locked Contract\\ \hline
I & Impact \\ \hline
{IDS/IPS}&{Intrusion Detection and Prevention Systems}\\ \hline
IoT & Internet of Things \\ \hline
KEM/ENC & Key Encapsulation Mechanism/Encryption \\ \hline
KYC & Know Your Customer \\ \hline
L & Likelihood \\ \hline
MAC & Message Authentication Code \\ \hline
{ML}&{Machine Learning} \\ \hline
MPC & Multi-Party Computation \\ \hline
NFT & Non-Fungible Token \\ \hline
NIST & National Institute of Standards and Technology \\ \hline
PoA & Proof of Authority \\ \hline
PoR & Proof of Randomness \\ \hline
PoS & Proof of Stake \\ \hline
POSet&Partially Ordered Set\\ \hline
PoW & Proof of Work \\ \hline
PQC & Post-Quantum Cryptography \\ \hline
PRF & Pseudorandom Function \\ \hline
QC & Quantum Computing \\ \hline
QML & Quantum Machine Learning \\ \hline
R & Risk \\ \hline
RNG & Random Number Generator \\ \hline
{SegWit}&{Segregated Witness} \\ \hline
STRIDE & Spoofing, Tampering, Repudiation, Information Disclosure, Denial of Service, Elevation of Privilege \\ \hline
TLS & Transport Layer Security \\ \hline
VDF&Verifiable Delay Function\\ \hline
VRF & Verifiable Random Function \\ \hline
ZKP & Zero-Knowledge Proof \\ \hline
\end{tabular}%
}
\vspace{-0.5cm}
\end{table}
\subsection{Contributions}\label{Contribution}
{This survey offers substantial advancements in the field of} \gls{BC} security within the context of emerging \gls{QC} technologies. The key contributions are as follows:
\begin{enumerate}[label=\textbf{\arabic*.}]
\item \textbf{Comprehensive Risk Assessment:} We evaluate \gls{BC} vulnerabilities and potential hazards posed by \gls{QC}, identify critical attack vectors, and assess risk severity through a structured,  {NIST}-aligned methodology, enabling risk prioritization and supporting the design of proactive defense strategies.

\item \textbf{Platform-Specific Vulnerability Analysis:} We perform a dual analysis of vulnerabilities in leading legacy platforms (Bitcoin, Ethereum, Ripple, Litecoin, Zcash) and examine the architectural and security properties of emerging post-quantum \gls{BC} platforms, highlighting migration challenges and resilience factors not comprehensively addressed in prior surveys.

\item \textbf{Hybrid \gls{BC} Approach for Secure Migration:} We introduce innovative hybrid \gls{BC} architectures designed to facilitate a seamless transition from legacy systems to quantum-resistant cryptography, balancing security, adaptability, and migration efficiency to ensure long-term \gls{BC} resilience.

\item \textbf{Actionable Mitigation Strategies:} We provide practical guidance for secure migration planning and proactive measures to strengthen \gls{BC} components against emerging quantum-era risks, emphasizing the adoption of quantum-resistant solutions and robust defense mechanisms.
\end{enumerate}
These contributions collectively advance the understanding of quantum threats to \gls{BC} security, offering actionable guidance
for stakeholders to fortify their systems against emerging quantum-induced cyber threats.

\subsection{Related Works}\label{Related}

{Recent literature on \gls{BC} security in the quantum era generally falls into six main categories: (1) \textit{threat modeling and vulnerability assessment}~\cite{wicaksana2025survey, khodaiemehr2023navigating, kearney2021vulnerability, kaushik2023demystifying}; (2) \textit{cryptographic migration strategies}, which involve either NIST-standardized \gls{PQC}~\cite{gharavi2024post, karakaya2024survey, allende2023quantum, kumar2021survey, khodaiemehr2023navigating} or quantum cryptography like Quantum Key Distribution (QKD)~\cite{yang2024survey, liu2023secure, yang2022decentralization, edwards2020review}; (3) \textit{platform-specific security analyses} focusing on classical \glspl{BC}~\cite{kearney2021vulnerability} or post-quantum Distributed Ledger Technologies (DLTs)~\cite{gharavi2024post}; (4) \textit{hybrid \gls{BC} architectures} that integrate classical systems with quantum or PQC-based layers~\cite{yang2024survey, liu2023secure, edwards2020review}; (5) \textit{post-migration security analysis}~\cite{gharavi2024post, yang2024survey, allende2023quantum}; and (6) \textit{actionable migration guidance and transition planning}~\cite{allende2023quantum, kaushik2023demystifying, yang2024survey}. While these studies offer valuable insights into specific areas, they often examine individual aspects in isolation. There's a notable absence of an integrated framework that unifies risk modeling, hybrid system design, cryptographic migration, and post-transition readiness. Our work aims to fill this gap by providing a comprehensive and systematic framework.}

{Our research distinguishes itself by offering a holistic approach to \gls{BC} quantum-security. First, we provide a comprehensive risk assessment framework that systematically evaluates threats across all fundamental \gls{BC} components, including consensus mechanisms, cryptographic algorithms, smart contracts, and network protocols. This approach goes beyond previous studies that focus primarily on specific vulnerabilities~\cite{wicaksana2025survey,khodaiemehr2023navigating, kearney2021vulnerability}.
Second, we conduct a detailed vulnerability analysis of the major \gls{BC} platforms, including Bitcoin, Ethereum, Ripple, Litecoin, and Zcash. Unlike previous works that provide general security discussions~\cite{kearney2021vulnerability}, our analysis offers in-depth platform-specific insights, including vulnerability analysis, threat modeling, and risk assessment, enabling a deeper understanding of quantum threats and their potential impact on specific cryptographic algorithms, consensus mechanisms, and smart contract vulnerabilities.
Third, we introduce novel hybrid \gls{BC} architectures that leverage \gls{PQC} to ensure a secure and efficient quantum-resistant future. Unlike other hybrid approaches that rely on quantum cryptography~\cite{edwards2020review, liu2023secure, yang2024survey}, our focus on \gls{PQC}, standardized by \gls{NIST}, provides a practical and sustainable solution. Our hybrid architectures combine the strengths of classical and quantum-resistant cryptographic primitives, offering a robust and flexible framework for securing \gls{BC} systems~\cite{joseph2022transitioning,giron2023post}.
Finally, we provide actionable guidance for stakeholders, empowering them to navigate the quantum era effectively. Unlike existing works that may focus on specific environments or pre-transition stages~\cite{allende2023quantum, khodaiemehr2023navigating}, our research offers a comprehensive roadmap that covers all phases of the quantum transition, including risk assessment, vulnerability analysis, migration strategies, and post-migration security considerations.}

\begin{table}[!h]
\scriptsize
{
\caption{{Comparison of Our Survey With Prior Work on \gls{BC} Security in the Quantum Era}}
\label{tab:survey}
\resizebox{\linewidth}{!}{%
\begin{tabular}{|c|
>{\centering\arraybackslash}b{0.45cm}|
>{\centering\arraybackslash}b{0.45cm}|
>{\centering\arraybackslash}b{0.45cm}|
>{\centering\arraybackslash}b{0.45cm}|
>{\centering\arraybackslash}b{0.45cm}|
>{\centering\arraybackslash}b{0.45cm}|
>{\centering\arraybackslash}b{0.45cm}|
>{\centering\arraybackslash}b{0.45cm}|
}
\hline
\rotatebox{90}{\textbf{Study}} &
\rotatebox{90}{\textbf{Risk Assessment}} &
\rotatebox{90}{\textbf{STRIDE-Based Threat Modeling}} &
\rotatebox{90}{\textbf{Vulnerability Analysis}} &
\rotatebox{90}{\textbf{Platform-Specific Analysis}} &
\rotatebox{90}{\textbf{NIST-Standard PQC Adoption}} &
\rotatebox{90}{\textbf{Post-Migration Security Analysis}} &
\rotatebox{90}{\textbf{Hybrid \gls{BC} Architecture}} &
\rotatebox{90}{\textbf{Actionable Migration Guidance $\ $}} \\ \hline
\cite{wicaksana2025survey} & 
$\times$ & 
$\times$ & 
\checkmark & 
$\times$ & 
\checkmark & 
$\times$ & 
$\times$ & 
\checkmark \\ \hline
\cite{yang2024survey} & 
$\times$ & 
$\times$ & 
\checkmark & 
$\times$ & 
$\times$ & 
$\times$ & 
\checkmarkqc & 
\checkmark \\ \hline
\cite{gharavi2024post} & 
$\times$ & 
$\times$ & 
\checkmark & 
\checkmarkpq & 
\checkmark & 
\checkmark & 
$\times$ & 
\checkmark \\ \hline
\cite{karakaya2024survey} & 
$\times$ & 
$\times$ & 
\checkmark & 
$\times$ & 
\checkmark & 
$\times$ & 
$\times$ & 
$\times$ \\ \hline
\cite{allende2023quantum} & 
$\times$ & 
$\times$ & 
\checkmark & 
$\times$ & 
\checkmark & 
$\times$ & 
$\times$ & 
$\times$ \\ \hline
\cite{kaushik2023demystifying} & 
$\times$ & 
$\times$ & 
\checkmark & 
$\times$ & 
$\times$ & 
$\times$ & 
$\times$ & 
$\times$ \\ \hline
\cite{liu2023secure} & 
$\times$ & 
$\times$ & 
\checkmark & 
$\times$ & 
$\times$ & 
$\times$ & 
\checkmarkqc & 
$\times$ \\ \hline
\cite{khodaiemehr2023navigating} & 
$\times$ & 
$\times$ & 
\checkmark & 
$\times$ & 
\checkmark & 
$\times$ & 
$\times$ & 
$\times$ \\ \hline
\cite{yang2022decentralization} & 
$\times$ & 
$\times$ & 
\checkmark & 
$\times$ & 
$\times$ & 
$\times$ & 
$\times$ & 
$\times$ \\ \hline
\cite{kumar2021survey} & 
$\times$ & 
$\times$ & 
\checkmark & 
$\times$ & 
\checkmark & 
$\times$ & 
$\times$ & 
$\times$ \\ \hline
\cite{kearney2021vulnerability} & 
$\times$ & 
$\times$ & 
\checkmark & 
\checkmarkc & 
$\times$ & 
$\times$ & 
$\times$ & 
$\times$ \\ \hline
\cite{edwards2020review} & 
$\times$ & 
$\times$ & 
$\times$ & 
$\times$ & 
$\times$ & 
$\times$ & 
\checkmarkqc & 
$\times$ \\ \hline
\textbf{Our Work} & 
\checkmark & 
\checkmark & 
\checkmark & 
\checkmark & 
\checkmark & 
\checkmark & 
\checkmark & 
\checkmark \\ \hline
\end{tabular}}}

\vspace{0.8em}
\footnotesize{
{\checkmark = addressed; $\times$ = not addressed or not a primary focus;
\checkmarkpq = addressed post-quantum platform only; 
\checkmarkc = addressed classic platform only; 
\checkmarkqc = addressed using quantum cryptography (e.g., QKD, QPoS)}}

\end{table}

{As shown in Table~\ref{tab:survey}, our study is distinct in offering an integrated strategy for BC security in the quantum era. It combines \gls{BC} vulnerability analysis, STRIDE-based threat modeling, platform-specific evaluations, and NIST-standardized PQC integration. It further addresses post-migration security, proposes practical hybrid \gls{BC} architectures, and offers actionable migration guidance for stakeholders, thereby providing a comprehensive and forward-looking strategy for quantum-resilient \gls{BC} infrastructures.}

\subsection{Survey Methodology and Literature Sourcing}\label{methodology}
{We adopted a systematic methodology to ensure comprehensive and unbiased coverage of quantum-resilient \gls{BC} technologies, combining structured database searches with targeted reviews of leading cryptography and AI research venues. Searches were conducted across IEEE Xplore, ACM Digital Library, SpringerLink, Scopus, and arXiv, covering publications from 2014--2025 to include both early conceptual proposals and the latest developments following NIST's post-quantum standardization. Search queries combined the term ``blockchain'' with terms such as ``post-quantum cryptography,'' ``quantum-resistant,'' and domain-specific keywords for consensus mechanisms, digital signatures, key exchange, and cryptographic protocols.
Targeted reviews of PQCrypto, CRYPTO, EUROCRYPT, and the \textit{Journal of Cryptology} captured state-of-the-art post-quantum primitives and their \gls{BC} integration. AI-related research from AAAI, IJCAI, ICLR, ICML, NeurIPS, and KDD was examined for quantum-resilient machine learning models, \gls{BC}-specific risk assessment frameworks, and adaptive security mechanisms.
Inclusion criteria required direct relevance to quantum threats in \gls{BC} systems, post-quantum ledger implementations, hybrid architectures, or AI-enhanced quantum-resistant protocols. Quality assessment considered venue prestige, citation impact, and methodological rigor. Screening was conducted independently by multiple researchers, with disagreements resolved by consensus.
This process yielded 187 papers from cryptography venues (38 shortlisted, 13 retained) and 51 from AI venues (17 retained). Table~\ref{tab:search-phrases-results} summarizes the search queries and results.}
\begin{table}[h!]
\centering
{
\caption{Summary of Search Phrases and Results}
\scriptsize
\label{tab:search-phrases-results}
\begin{tabular}{|p{0.72\linewidth}|c|}
\hline
\textbf{Search Phrase} & \textbf{Results Found} \\
\hline
blockchain AND (quantum OR post-quantum) AND (security OR risk assessment OR vulnerability OR threat OR attack) & 6,218 \\
\hline
blockchain AND (quantum OR post-quantum) AND (migration OR transition OR upgrade OR adaptation OR protocol migration OR defense OR mitigation OR countermeasure OR resilience OR proactive) & 5,199 \\ 
\hline
blockchain AND (quantum OR post-quantum) AND (digital signature OR consensus OR cryptography OR key exchange OR hash-based OR lattice-based OR zero knowledge) & 2,869 \\ 
\hline
blockchain AND (quantum OR post-quantum) AND (survey OR review OR taxonomy) & 2,003 \\ 
\hline
(``PQCrypto'' OR ``CRYPTO'' OR ``EUROCRYPT'' OR ``Journal of Cryptology'' OR ``IEEE S\&P'') AND blockchain AND (quantum OR post-quantum) & 145 \\ 
\hline
\end{tabular}}
\vspace{-0.2cm}
\end{table}

\subsection{Organization}\label{Organization}
Figure~\ref{fig:survey_structure} presents the structure of this survey, which addresses the quantum-safe transition of \gls{BC} systems across eleven sections. Section~\ref{risk-analysis-approach} presents a comprehensive methodology for assessing quantum-specific threats and vulnerabilities during \gls{BC} system migration.  
Section~\ref{sec:risk} reviews cryptographic standards and evaluates the cyber impact of \gls{QC} on classical and post-quantum algorithms.  
Section~\ref{sec:impact} examines quantum threats to key \gls{BC} components, including the network layer, mining pools, transaction verification, smart contracts, and user wallets.  
Section~\ref{sec:roles} analyzes evolving security responsibilities across the \gls{BC} ecosystem and proposes strategies for quantum-safe migration.  
Section~\ref{sec:navigate} discusses challenges in mitigating \gls{QC} threats and transitioning to quantum-secure \gls{BC} infrastructures.  
Section~\ref{sec:hybrid} introduces hybrid \gls{BC} architectures for seamless migration to quantum-resistant cryptographic primitives.  
Section~\ref{sec:platform} evaluates the quantum-era security posture of major \gls{BC} platforms, including Bitcoin, Ethereum, Ripple, Litecoin, and Zcash.  
Section~\ref{sec:PQ-BC} surveys emerging post-quantum \glspl{BC}, analyzing cryptographic foundations, security considerations, and interoperability challenges.  
Section~\ref{sec:future} presents future research directions on \gls{QC}-driven challenges, migration strategies, and systemic preparedness for \gls{BC} ecosystems. 
Finally, Section~\ref{sec:conclude} summarizes the key findings, highlights proactive defense strategies, and underscores the need for sustained collaboration to secure \gls{BC} systems in the quantum era.

\begin{figure}[!htbp]
\centering
\begin{adjustbox}{width=\linewidth}
\begin{tikzpicture}[node distance=0.7cm]

\definecolor{titlecolor}{RGB}{45, 62, 80}   
\definecolor{color0}{RGB}{220, 53, 69}  
\definecolor{color1}{RGB}{255, 87, 34}  
\definecolor{color2}{RGB}{255, 152, 0}  
\definecolor{color3}{RGB}{255, 193, 7}  
\definecolor{color4}{RGB}{205, 220, 57} 
\definecolor{color5}{RGB}{76, 175, 80}  
\definecolor{color6}{RGB}{0, 150, 136}  
\definecolor{color7}{RGB}{33, 150, 243} 
\definecolor{color8}{RGB}{63, 81, 181}  
\definecolor{color9}{RGB}{103, 58, 183} 
\definecolor{color10}{RGB}{156, 39, 176} 
\large

\tikzstyle{title} = [rectangle, draw=black, thick, fill=titlecolor!30, 
           text width=4.6cm, minimum height=2.2cm, 
           rounded corners=3pt, font=\large,align=justify]

\tikzstyle{section} = [rectangle, draw=black, thick, text width=6.1cm, text centered, minimum height=1.2cm, rounded corners=2pt,font=\large]

\tikzstyle{subsection} = [rectangle, draw=black, thick, text width=10.3cm, 
             text centered, minimum height=0.8cm, rounded corners=2pt]

\tikzstyle{arrow} = [thick, ->, >=stealth, color=black!70]

\node (title) [title] {{Blockchain Security Risk Assessment in the Quantum Era: Migration Strategies and Proactive Defense}};

\node (intro) [section, right of=title, xshift=5.5cm, yshift=12.6cm, fill=color0!50] 
{Section~\ref{sec:intro}. Introduction};

\node (risk) [section, below of=intro, yshift=-2.7cm, fill=color1!50] 
{Section~\ref{risk-analysis-approach}. Quantum-Safe Risk Assessment};

\node (standards) [section, below of=risk, yshift=-1.7cm, fill=color2!50] 
{Section~\ref{sec:risk}. Cryptographic Standards \& Quantum Risk};

\node (impact) [section, below of=standards, yshift=-3.2cm, fill=color3!50] 
{Section~\ref{sec:impact}. Quantum Impact on BC Components};

\node (roles) [section, below of=impact, yshift=-2.1cm, fill=color4!50,align=justify]
{Section~\ref{sec:roles}. Roles \& Responsibilities in Mitigating QC Impacts on BC};

\node (navigate) [section, below of=roles, yshift=-2.1cm, fill=color5!50] 
{Section~\ref{sec:navigate}. Post-Transition Challenges \& Strategies};

\node (hybrid) [section, below of=navigate, yshift=-4.3cm, fill=color6!50] 
{Section~\ref{sec:hybrid}. Hybrid Blockchain Approaches};

\node (platform) [section, below of=hybrid, yshift=-2.4cm, fill=color7!50] 
{Section~\ref{sec:platform}. Blockchain Platform Evaluation};

\node (pqbc) [section, below of=platform, yshift=-0.8cm, fill=color8!50] 
{Section~\ref{sec:PQ-BC}. Post-Quantum Blockchain Landscape};

\node (future) [section, below of=pqbc, yshift=-2.8cm, fill=color9!50] 
{Section~\ref{sec:future}. Future Research Directions};

\node (conclude) [section, below of=future, yshift=-2.6cm, fill=color10!50] 
{Section~\ref{sec:conclude}. Conclusion};{Section~\ref{sec:conclude}. Conclusion};

\node (intro1) [subsection, right of=intro, xshift=8.5cm, yshift=1.5cm, fill=color0!30] 
{\ref{Contribution}. Contributions};
\node (intro2) [subsection, below of=intro1, yshift=-0.3cm, fill=color0!30] 
{\ref{Related}. Related Works};
\node (intro3) [subsection, below of=intro2, yshift=-0.3cm, fill=color0!30] 
{\ref{methodology}. Methodology and Literature Sourcing};
\node (intro4) [subsection, below of=intro3, yshift=-0.3cm, fill=color0!30] 
{\ref{Organization}. Organization};

\node (risk1) [subsection, right of=risk, xshift=8.5cm, yshift=0.5cm, fill=color1!30] 
{\ref{Risk-Scoping}. Risk Scoping and Preparation};
\node (risk2) [subsection, below of=risk1, yshift=-0.3cm, fill=color1!30] 
{\ref{Conducting-Risk}. Conducting Risk Assessment};

\node (standards1) [subsection, right of=standards, xshift=8.5cm, yshift=0.5cm, fill=color2!30] 
{\ref{Classic-Cryptographic-Standards}. Classical Standards \& Quantum Risk Assessment};
\node (standards2) [subsection, below of=standards1, yshift=-0.3cm, fill=color2!30] 
{\ref{Selected-Quantum-Resistant-Cryptographic-Standards}. PQC Standards \& Cyber Risk Assessment};

\node (impact1) [subsection, right of=impact, xshift=8.5cm, yshift=2cm, fill=color3!30] 
{\ref{Classic-BC-Network}. BC Network and QC Threats};
\node (impact2) [subsection, below of=impact1, yshift=-0.3cm, fill=color3!30] 
{\ref{Classic-Mining-Pool}. Mining Pool and QC Threats};
\node (impact3) [subsection, below of=impact2, yshift=-0.3cm, fill=color3!30] 
{\ref{Classic-Transaction-Verification}. Transaction Verification Mechanism and QC Threats};
\node (impact4) [subsection, below of=impact3, yshift=-0.3cm, fill=color3!30] 
{\ref{Classic-Smart-Contract}. Smart Contract and QC Threats};
\node (impact5) [subsection, below of=impact4, yshift=-0.3cm, fill=color3!30] 
{\ref{Classic-User-Wallet}. User Wallet and QC Threats};

\node (nav1) [subsection, right of=navigate, xshift=8.5cm, yshift=2cm, fill=color5!30] 
{\ref{Post-BC-Network}. BC Network: Challenges \& Strategies};
\node (nav2) [subsection, below of=nav1, yshift=-0.3cm, fill=color5!30] 
{\ref{Post-Mining-Pool}. Mining Pool: Challenges \& Strategies};
\node (nav3) [subsection, below of=nav2, yshift=-0.3cm, fill=color5!30] 
{\ref{Post-Transaction-Verification}. Verification Mechanism: Challenges \& Strategies};
\node (nav4) [subsection, below of=nav3, yshift=-0.3cm, fill=color5!30] 
{\ref{Post-Smart-Contract}. Smart Contract: Challenges \& Strategies};
\node (nav5) [subsection, below of=nav4, yshift=-0.3cm, fill=color5!30] 
{\ref{Post-User-Wallet}. User Wallet: Challenges \& Strategies};

\node (hybrid1) [subsection, right of=hybrid, xshift=8.5cm, yshift=1.5cm, fill=color6!30] 
{\ref{Hybrid-Strategies}. Hybrid Strategies for Cryptographic Primitives};
\node (hybrid2) [subsection, below of=hybrid1, yshift=-0.3cm, fill=color6!30] 
{\ref{Architectural-Design}. Architectural Design Approaches};
\node (hybrid3) [subsection, below of=hybrid2, yshift=-0.3cm, fill=color6!30] 
{\ref{hybrid-Risk}. Risk Assessment and Security Posture};
\node (hybrid4) [subsection, below of=hybrid3, yshift=-0.3cm, fill=color6!30] 
{\ref{hybrid-comparative}. Comparative Analysis and Strategic Insights};

\node (platform1) [subsection, right of=platform, xshift=8.5cm,  fill=color7!30] 
{\ref{sec:patform-lesson}. Lessons
Learned and Strategic Roadmap};

\node (pqbc1) [subsection, right of=pqbc, xshift=8.5cm,  fill=color8!30] 
{\ref{sec:PQ-BC-lesson}. Lessons
Learned and Strategic Roadmap};
\node (future1) [subsection, right of=future, xshift=8.5cm, yshift=2cm, fill=color9!30] 
{\ref{quantum-ai-integration}. Quantum-Enhanced \gls{AI} and BC Integration};
\node (future2) [subsection, below of=future1, yshift=-0.3cm, fill=color9!30] 
{\ref{web3-quantum-security}.  Quantum Security and Interoperability for Web3};
\node (future3) [subsection, below of=future2, yshift=-0.3cm, fill=color9!30] 
{\ref{privacy-preserving-systems}. Advanced Privacy-Preserving Cryptographic Systems};
\node (future4) [subsection, below of=future3, yshift=-0.3cm, fill=color9!30] 
{\ref{federated-edge-iot}. FL, Edge Computing, and IoT Integration};
\node (future6) [subsection, below of=future4, yshift=-0.3cm, fill=color9!30] 
{\ref{cross-cutting-challenges}. Cross-Cutting Technical and Standardization};

\foreach \x in {intro,risk,standards,impact,navigate,hybrid,platform,pqbc,future,conclude}
 \draw [arrow] (title) |- (\x.west);

 \draw [arrow] (title.east) |- (roles.west);
\draw [arrow] (intro) |- (intro1.west); 
\draw [arrow] (intro.east) |- (intro2.west); 
\draw [arrow] (intro.east) |- (intro3.west);
\draw [arrow] (intro) |- (intro4.west);

\draw [arrow] (risk.east) |- (risk1.west);
\draw [arrow] (risk.east) |- (risk2.west);

\draw [arrow] (standards.east) |- (standards1.west);
\draw [arrow] (standards.east) |- (standards2.west);

\draw [arrow] (impact) |- (impact1.west); 
\draw [arrow] (impact) |- (impact2.west); 
\draw [arrow] (impact.east) |- (impact3.west);
\draw [arrow] (impact) |- (impact4.west); 
\draw [arrow] (impact) |- (impact5.west);

\draw [arrow] (navigate) |- (nav1.west); 
\draw [arrow] (navigate) |- (nav2.west);
\draw [arrow] (navigate.east) |- (nav3.west);
\draw [arrow] (navigate) |- (nav4.west); 
\draw [arrow] (navigate) |- (nav5.west);

\draw [arrow] (hybrid) |- (hybrid1.west); 
\draw [arrow] (hybrid.east) |- (hybrid2.west); 
\draw [arrow] (hybrid.east) |- (hybrid3.west);  
\draw [arrow] (hybrid) |- (hybrid4.west);
\draw [arrow] (platform.east) |- (platform1.west);
\draw [arrow] (pqbc.east) |- (pqbc1.west);
\draw [arrow] (future) |- (future1.west); 
\draw [arrow] (future) |- (future2.west);
\draw [arrow] (future.east) |- (future3.west);
\draw [arrow] (future) |- (future4.west); 
\draw [arrow] (future) |- (future6.west);

\end{tikzpicture}
\end{adjustbox}
\caption{{Organizational Structure of This Survey}}
\label{fig:survey_structure}
\end{figure}

\section{Quantum-Safe Migration Risk Assessment Approach}\label{risk-analysis-approach}
This study presents a systematic framework for evaluating the security risks of \gls{BC} migration to quantum-safe systems. Aligned with the four-step process defined in \gls{NIST} SP~800-30~\cite{NIST-SP800-30}—preparation, conduct, communication, and maintenance—the methodology incorporates STRIDE-based threat categorization to address quantum-specific vulnerabilities. Our analysis emphasizes the conduct stage, where threats, vulnerabilities, likelihood, and impact are examined to prioritize risks across \gls{BC} components. Figure~\ref{fig:risk-assesment} summarizes the overall approach.

\begin{figure*}[!ht]
 \begin{center} 
\resizebox{0.7\linewidth}{!}{
\begin{tikzpicture}[
 title/.style={minimum height=1cm,minimum width=4.1cm,font = {\large}},
 body/.style={draw,top color=white, bottom color=blue!20, rounded corners,minimum width=3.8cm,,minimum height=1cm,font = {\small}},
 typetag/.style={rectangle, draw=black!100, anchor=west}]
 \node (d0) [draw,top color=white, bottom color=blue!20, rounded corners,minimum height=1cm,minimum width=1.27\textwidth,font = {\large}] at (0,0) {\gls{NIST} Cybersecurity Framework

};
 
\node (d3) [title,below=of d0.center] {Detect
};
 \node (d31) [body,below=of d3.west, typetag, xshift=2mm,yshift=-0.9cm,minimum height=2.1cm] {\begin{minipage}[c]{3.4cm}\centering
Anomalies and Events 
\end{minipage}};
 \node (d32) [body,below=of d31.west, typetag,yshift=-1.4cm,minimum height=2.1cm] {\begin{minipage}[c]{3.4cm}\centering
Security Continuous Monitoring
\end{minipage}};
 \node (d33) [body,below=of d32.west, typetag,yshift=-1.35cm,minimum height=2.1cm] {\begin{minipage}[c]{3.4cm}\centering
Detection Processes
\end{minipage}};
\node [top color=blue!40, bottom color=blue!40, rounded corners,minimum height=1cm,draw=black!100,fill opacity=0.1, fit={ (d3) (d31) (d32) (d33)}] {};
 
\node (d2) [title, left of=d3,xshift=-0.2\textwidth] {Protect};
 \node (d21) [body,below=of d2.west, typetag, yshift=-0.35cm, xshift=2mm] {\begin{minipage}[c]{3.4cm}\centering
Identity Management and Access Control
\end{minipage}};
 \node (d22) [body,below=1.17 of d21.west, typetag] {\begin{minipage}[c]{3.4cm}\centering
Awareness and Training
\end{minipage}};
 \node (d23) [body,below=1.17 of d22.west, typetag] {\begin{minipage}[c]{3.4cm}\centering
Data Security 
\end{minipage}};
\node (d24) [body,below=1.17 of d23.west, typetag] {\begin{minipage}[c]{3.4cm}\centering
Info. Protection Processes and Procedures
\end{minipage}};
\node (d25) [body,below=1.17 of d24.west, typetag] {\begin{minipage}[c]{3.4cm}\centering
Maintenance
\end{minipage}};
\node (d26) [body,below=1.17 of d25.west, typetag]{\begin{minipage}[c]{3.4cm}\centering
Protective Technology
\end{minipage}};
\node [top color=blue!40, bottom color=blue!40, rounded corners,minimum height=1cm,draw=black!100,fill opacity=0.1, fit={ (d2) (d21) (d22) (d23) (d24) (d25) (d26)}] {};

\node (d1) [title,left of=d2,xshift=-0.2\textwidth] {Identify
};
 \node (d11) [body,below=of d1.west, typetag, yshift=-0.35cm, xshift=2mm] {\begin{minipage}[c]{3.4cm}\centering
Asset Management
\end{minipage}};
 \node (d12) [body,below=1.17 of d11.west, typetag] {\begin{minipage}[c]{3.4cm}\centering
Business Environment
\end{minipage}};
 \node (d13) [body,below=1.17 of d12.west, typetag] {\begin{minipage}[c]{3.4cm}\centering
Governance
\end{minipage}};
\node (d14) [body,below=1.17 of d13.west, typetag] {\begin{minipage}[c]{3.4cm}\centering
Risk Assessment
\end{minipage}};
\node (d15) [body,below=1.17 of d14.west, typetag] {\begin{minipage}[c]{3.4cm}\centering
Risk Management Strategy
\end{minipage}};
\node (d16) [body,below=1.17 of d15.west, typetag]{\begin{minipage}[c]{3.4cm}\centering Supply Chain Risk Management
\end{minipage}};
\node [top color=blue!40, bottom color=blue!40, rounded corners,minimum height=1cm,draw=black!100,fill opacity=0.1, fit={ (d1) (d11) (d12) (d13) (d14) (d15) (d16)}] {}; 
 
\node (d4) [title, right of=d3,xshift=0.2\textwidth] {Respond
};
 \node (d41) [body,below=of d4.west, typetag, yshift=-0.6cm, xshift=2mm,minimum height=1.55cm] {\begin{minipage}[c]{3.4cm}\centering
Response Planning
\end{minipage}};
 \node (d42) [body,below=of d41.west, typetag, yshift=-0.8cm,minimum height=1.55cm] {\begin{minipage}[c]{3.4cm}\centering
Communications 
\end{minipage}};
 \node (d43) [body,below=of d42.west, typetag, yshift=-0.75cm,minimum height=1.55cm] {\begin{minipage}[c]{3.4cm}\centering
Analysis
\end{minipage}};
\node (d44) [body,below=of d43.west, typetag, yshift=-0.8cm,minimum height=1.55cm] {\begin{minipage}[c]{3.4cm}\centering
Mitigation Improvements 
\end{minipage}};
\node [top color=blue!40, bottom color=blue!40, rounded corners,minimum height=1cm,draw=black!100,fill opacity=0.1, fit={ (d4) (d41) (d42) (d43) (d44)}] {};

\node (d5) [title, right of=d4,xshift=0.2\textwidth] {Recover};
 \node (d51) [body,below=of d5.west, typetag, xshift=2mm,yshift=-0.9cm,minimum height=2.1cm] {\begin{minipage}[c]{3.4cm}\centering
Recovery Planning
\end{minipage}};
 \node (d52) [body,below=of d51.west, typetag,yshift=-1.4cm,minimum height=2.1cm] {\begin{minipage}[c]{3.4cm}\centering
Improvements 
\end{minipage}};
 \node (d53) [body,below=of d52.west, typetag,yshift=-1.35cm,minimum height=2.1cm] {\begin{minipage}[c]{3.4cm}\centering
Communications
\end{minipage}};
\node (d1to5)[top color=blue!40, bottom color=blue!40, rounded corners,minimum height=1cm,draw=black!100,fill opacity=0.1, fit={ (d5) (d51) (d52) (d53)}] {};

 \node (steps) [below=of d1to5.south, yshift=- 0.3cm, xshift=-13.5cm]{
 \begin{tikzpicture}
 
 \fill[blue!30] (0,0) -- (7.5,0) -- (8,0.5) -- (7.5,1) -- (0,1) -- cycle;
 \node at (3.2,0.8) {\text{Identify Threat Sources and Events}};

 \fill[red!30] (0,-1.2) -- (8.5,-1.2) -- (9,-0.7) -- (8.5,-0.2) -- (0,-0.2) -- cycle;
 \node at (4.4,-0.4) {\text{Identify Vulnerabilities and Predisposing Conditions}};

 \fill[green!30] (0,-2.4) -- (9.5,-2.4) -- (10,-1.9) -- (9.5,-1.4) -- (0,-1.4) -- cycle;
 \node at (3.2,-1.6) {\text{Determine Likelihood of Occurrence}};

 \fill[orange!30] (0,-3.6) -- (10.5,-3.6) -- (11,-3.1) -- (10.5,-2.6) -- (0,-2.6) -- cycle;
 \node at (2.9,-2.8) {\text{Determine Magnitude of Impact}};

 \fill[blue!30] (0,-4.8) -- (11.5,-4.8) -- (12,-4.3) -- (11.5,-3.8) -- (0,-3.8) -- cycle;
 \node at (1.3,-4) {\text{Assess Risk}};

 \node [below=-1cm, align=center] at (6, -6) {\large{Risk Assessment Process Phases According to \gls{NIST}}};
 \end{tikzpicture}};

 \node (risk) [below=of d1to5.south, yshift=0.5cm, xshift=-1cm] {\renewcommand{\arraystretch}{1.8}{
\subfloat{}{\begin{math}
 \qquad \raisebox{-1\normalbaselineskip}{Likelihood$\ \left\{\rule{0pt}{3\normalbaselineskip}\right.$}
\hspace{-0.15cm}
\begin{tabular}{p{0.1cm}}\\
\multicolumn{1}{l}{Low} \\ 
\multicolumn{1}{l}{Medium} \\
\multicolumn{1}{l}{High}
\end{tabular}
\end{math}
}
\hspace{-0.3cm}
\begin{math}
 \overbrace{\
\begin{tabular}{p{1.5cm}p{1.5cm}p{1.5cm}}
Low & Medium & High \\ \hline
\multicolumn{1}{|l|}{\cellcolor{green!60}Low} & \multicolumn{1}{l|}{\cellcolor{green!60}Low} & \multicolumn{1}{l|}{\cellcolor{lemon!80}Medium} \\ \hline
\multicolumn{1}{|l|}{\cellcolor{green!60}Low} & \multicolumn{1}{l|}{\cellcolor{lemon!80}Medium} & \multicolumn{1}{l|}{\cellcolor{red!80}High} \\ \hline
\multicolumn{1}{|l|}{\cellcolor{lemon!80}Medium} & \multicolumn{1}{l|}{\cellcolor{red!80}High} & \multicolumn{1}{l|}{\cellcolor{red!80}High} \\ \hline
\end{tabular}
}^{\mbox{Impact}}
\end{math}
 }};

 \node (risk-evaluation) [below=of risk,xshift=2.2cm,yshift=1cm]{\large{Risk = Likelihood $\times$ Impact}};

\draw [thick arrow] (d14.west) -- ++(-1,0) |- (steps);
\draw [thick arrow] ($ (steps.east) + (0,-2) $)-| (risk-evaluation.south);

 \end{tikzpicture}
}
\caption{{Quantum-Safe Migration Risk Assessment Approach}}
\label{fig:risk-assesment}
\vspace{-0.5cm}
\end{center}
\end{figure*} 
\subsection{Risk Scoping and Preparation}\label{Risk-Scoping}

Aligned with \gls{NIST} guidelines~\cite{NIST-SP800-30}, this risk assessment framework defines the purpose, scope, assumptions, and risk model for prioritizing quantum-related threats across \gls{BC} components, including network infrastructure, mining pools, transaction verification, smart contracts, and wallets. It assumes   the emergence of quantum threats within the migration timeframe and the availability of \gls{PQC} solutions, while recognizing current efficiency and interoperability limitations.  A qualitative STRIDE-based model~\cite{van2021descriptive,baseri2024evaluation} classifies threats from literature and industry reports, applying High/Medium/Low likelihood and impact ratings to enable systematic prioritization and targeted mitigation strategies.

\subsection{Conducting Risk Assessment}\label{Conducting-Risk}
The risk assessment process involves five key tasks: (1) identifying potential threat sources and events; (2) identifying vulnerabilities within the system; (3) determining the likelihood of each identified vulnerability; (4) determining the impact of each vulnerability; and (5) assessing the risk level based on the evaluated likelihood and impact.
Figure~\ref{fig:risk-assesment} summarizes the assessment process.

\begin{enumerate}[wide, font=\itshape, labelwidth=!, labelindent=0pt, label*=\textit{Task} \arabic*.]
\item \textit{Identify Threat Sources and Events:} To identify potential threats, we consider quantum-specific risks such as quantum algorithm attacks (e.g., Shor's and Grover's algorithms), quantum side-channel attacks (exploiting timing or power differences), and quantum hardware attacks (malicious quantum devices). These threats can compromise \gls{BC} systems at various levels, including cryptographic algorithms, consensus mechanisms, and smart contracts. By understanding these risks, we can develop effective mitigation strategies to protect \gls{BC} systems in the quantum era.

\item \textit{Identify Vulnerabilities and Predisposing Conditions:} We conduct a comprehensive review of \gls{BC} components to identify vulnerabilities that could be exploited by quantum attacks. Key focus areas include quantum-resistant cryptographic algorithms (lattice-based, code-based, and hash-based), \gls{BC} protocols (consensus mechanisms, smart contracts, and network protocols), and quantum-safe key management (key generation, distribution, and storage).
\item \textit{Determine Likelihood of Occurrence:} 
Using a qualitative likelihood assessment approach, we evaluate the likelihood of vulnerabilities being exploited by a quantum attacker. 
We apply qualitative criteria—derived from NIST-SP~800-30 Appendix~G~\cite{NIST-SP800-30}—to categorize likelihood into three levels: 
\textit{Low (L)}, \textit{Medium (M)}, and \textit{High (H)}, as detailed in Table~\ref{table:likelihood}. 
The assessment considers factors such as exploitability, attack complexity, attacker motivation, availability of countermeasures, and their effectiveness.
  \begin{table}[!htbp]\vspace{-0.5cm}
\centering
\caption{Evaluation Criteria for Likelihood Levels}
\small
\resizebox{\linewidth}{!}{%
\begin{tabular}{|p{0.15\linewidth}|l|p{0.8\linewidth}|}
\hline
\multirow{14}{*}{\text{Likelihood}} & \cellcolor{red!80}\multirow{1}{*}{High} &
\begin{minipage}[c]{\linewidth}\vspace{3pt}
\begin{itemize}
 \item High likelihood of exploitation due to critical vulnerabilities or weak cryptographic protections.
 \item Broad network exposure, making exploitation accessible to adversaries with quantum capabilities.
 \item High probability of intent and capability from quantum threat actors.
\end{itemize}\vspace{3pt}
\end{minipage}
\\ \cline{2-3} 
&\cellcolor{lemon!80}\multirow{1}{*}{Medium} & 
\begin{minipage}[c]{\linewidth}\vspace{3pt}
\begin{itemize}
 \item Known vulnerabilities, but partial mitigations limit ease of exploitation.
 \item Moderate network exposure; some access restrictions, but feasible for a skilled quantum adversary.
 \item Exploitation possible with moderate quantum resources and expertise.
\end{itemize}\vspace{3pt}
\end{minipage} \\ \cline{2-3} 
&\cellcolor{green!60}\multirow{1}{*}{Low} & 
\begin{minipage}[c]{\linewidth}\vspace{3pt}
\begin{itemize}
 \item No significant vulnerabilities for quantum exploitation; strong quantum-resistant cryptography in place.
 \item Minimal network exposure; attacks require highly specialized access and resources.
 \item Unlikely exploitation due to high barriers and limited feasibility.
\end{itemize} \vspace{3pt}
\end{minipage}\\ \hline
\end{tabular}%
}
\label{table:likelihood}
\vspace{-0.5cm}
\end{table}

\begin{table}[!htbp]
\centering
\caption{Evaluation Criteria for Impact Levels}
\small
\resizebox{\linewidth}{!}{%
\begin{tabular}{|p{0.12\linewidth}|l|p{0.8\linewidth}|}
\hline
\multirow{14}{*}{\text{Impact}} & \cellcolor{red!80}\multirow{1}{*}{High} & 
\begin{minipage}[c]{\linewidth}\vspace{3pt}\begin{itemize}
 \item Severe compromise, including exposure of critical data or loss of cryptographic integrity.
 \item Extensive operational downtime or significant disruptions to \gls{BC} consensus mechanisms.
 \item High financial, regulatory, and reputational risks with long-term effects.
\end{itemize} \vspace{3pt}
\end{minipage} \\ \cline{2-3} 
& \cellcolor{lemon!80}\multirow{1}{*}{Medium} & 
\begin{minipage}[c]{\linewidth}\vspace{3pt}\begin{itemize}
 \item Partial compromise of data integrity or limited operational impact.
 \item Moderate data or service disruption with manageable recovery time.
 \item Moderate financial or reputational consequences; potential regulatory concerns.
\end{itemize} \vspace{3pt}
\end{minipage} \\ \cline{2-3} 
& \cellcolor{green!60}\multirow{1}{*}{Low} & 
\begin{minipage}[c]{\linewidth}\vspace{3pt}\begin{itemize}
 \item Minimal operational disruption or data exposure; negligible security implications.
 \item Temporary disruptions without lasting effects on \gls{BC} integrity or user trust.
 \item Limited financial or reputational risk.
\end{itemize}  \vspace{3pt}
\end{minipage}\\ \hline
\end{tabular}%
}
\label{table:impact}
\vspace{-0.2cm}
\end{table}
 \item \textit{Determine Magnitude of Impact:} {The impact levels, defined as \textit{Low (L)}, \textit{Medium (M)}, and \textit{High (H)}, reflect the potential severity of a quantum threat event (see Table}~\ref{table:impact} {for evaluation criteria). These criteria, based on NIST-SP 800-30 Appendix H}~\cite{NIST-SP800-30}, consider the possible harm to \gls{BC} assets, system stability, user satisfaction, data confidentiality, and organizational reputation.
 \item \textit{Assess Risk:} Risk is determined by combining likelihood and impact levels, visualized in a risk matrix (see Figure~\ref{fig:risk-matrix}). Risk levels—High, Medium, and Low—help prioritize vulnerabilities and guide mitigation strategies.
   \begin{figure}[!h]

\renewcommand{\arraystretch}{1.5}
 \scriptsize
\centering{
\subfloat{}{\begin{math}
 \qquad \raisebox{-1\normalbaselineskip}{Likelihood$\ \left\{\rule{0pt}{3\normalbaselineskip}\right.$}
\hspace{-0.15cm}
\begin{tabular}{p{0.1cm}}\\
\multicolumn{1}{l}{Low} \\ 
\multicolumn{1}{l}{Medium} \\
\multicolumn{1}{l}{High}
\end{tabular}
\end{math}
}
\hspace{-0.3cm}
\begin{math}
 \overbrace{\
\begin{tabular}{p{1cm}p{1cm}p{1cm}}
Low & Medium & High \\ \hline
\multicolumn{1}{|l|}{\cellcolor{green!60}Low} & \multicolumn{1}{l|}{\cellcolor{green!60}Low} & \multicolumn{1}{l|}{\cellcolor{lemon!80}Medium} \\ \hline
\multicolumn{1}{|l|}{\cellcolor{green!60}Low} & \multicolumn{1}{l|}{\cellcolor{lemon!80}Medium} & \multicolumn{1}{l|}{\cellcolor{red!80}High} \\ \hline
\multicolumn{1}{|l|}{\cellcolor{lemon!80}Medium} & \multicolumn{1}{l|}{\cellcolor{red!80}High} & \multicolumn{1}{l|}{\cellcolor{red!80}High} \\ \hline
\end{tabular}
}^{\mbox{Impact}}
\end{math}
 }\caption{Qualitative Risk Assessment based on Likelihood and Impact}\vspace{-0.2cm}
 \label{fig:risk-matrix}
\end{figure}
\end{enumerate}
 \section{Cryptographic Standards and Quantum Computing: Cyber Impact and Risk Assessment}\label{sec:risk}

This paper presents a security framework to protect \gls{BC} technology from \gls{QC} threats by securing cryptographic components essential to \gls{BC} integrity. Mitigating these threats requires understanding the risks \gls{QC} poses to classical cryptographic algorithms—the foundation of \gls{BC} security—and evaluating the readiness and vulnerabilities of \gls{NIST}-standardized \gls{PQC} schemes. We prioritize threats by exploitability and potential impact, considering both current vulnerabilities and the defensive capabilities of \gls{NIST}-standardized \gls{PQC} schemes as quantum-resistant alternatives. 
\subsection{Classic Cryptographic Standards and Quantum Computing: Cyber Impact and Risk Assessment}\label{Classic-Cryptographic-Standards}
\begin{table*}[!h]
\caption{Classic Cryptographic Standards and Quantum Computing: Cyber Impact and Risk Assessment}
\small
\renewcommand{\arraystretch}{1.2}
\label{tab:Pre-Migration-Alg}
\resizebox{\textwidth}{!}{%
\begin{tabular}{|l|l|l|l|l|l|p{0.16\linewidth}|p{0.52\linewidth}|l|l|l|p{0.24\linewidth}|}
\hline
\multirow{2}{*}{\textbf{Crypto Type}} & \multirow{2}{*}{\textbf{Algorithms}} & \multirow{2}{*}{\textbf{Variants}} & \multirow{2}{*}{\textbf{Key Length (bits)}} & \multicolumn{2}{l|}{\textbf{$\ \ $ Strengths (bits)}} & \multirow{2}{*}{\textbf{Vulnerabilities}} &\multirow{2}{*}{\textbf{STRIDE threats}}& \multirow{2}{*}{\textbf{L}} & \multirow{2}{*}{\textbf{I}} & \multirow{2}{*}{\textbf{R}} & \multirow{2}{*}{\textbf{QC-Resistant Alternatives}} \\ 
\cline{5-6}
 & & & & \multicolumn{1}{l|}{\textbf{Classic}} & \textbf{Quantum} & & & & & &\\ \hline
\multirow{8}{*}{Asymetric} & \multirow{3}{*}{ECC~\cite{RFC8813, RFC7748}} & ECC 256 & 256 & \multicolumn{1}{l|}{128} & 0 & \multirow{8}{*}{{\begin{minipage}{\linewidth}
Broken by Shor's Algorithm~\cite{shor1999polynomial}.
\end{minipage}}} & &\med &\high&\high& \multirow{8}{*}{{\begin{minipage}{\linewidth}Algorithms presented in Table~\ref{tab:qkd_pqc_comparison}.\end{minipage}}}\\\cline{3-6}\cline{9-11}
 & & ECC 384 & 384 & \multicolumn{1}{l|}{256} & 0 & & \multirow{6}{*}{{\begin{minipage}{\linewidth}
 \vspace{-0.3cm}
{For digital signature:}
\begin{myBullets}
\item {Spoofing: Shor's Algorithm allows forging of digital signatures.}
\item {Tampering: Integrity checks can be bypassed due to signature forgery.}
\item {Repudiation: Valid signatures can be forged, denying the origin of the message.}
\end{myBullets}
{For \gls{KEM/ENC}:}
\begin{myBullets}
\item {Info. Disclosure: \gls{KEM/ENC} algorithms can be broken, revealing encrypted data.}
\end{myBullets}
\end{minipage}}}&\med& \high &\high& \\\cline{3-6}\cline{9-11}
 & & ECC 521 & 521 & \multicolumn{1}{l|}{256} & 0 & & &\med & \high&\high& \\ \cline{2-6}\cline{9-11}
 & \multirow{2}{*}{FFDHE~\cite{rfc7919}} & DHE 2048 & 2048 & \multicolumn{1}{l|}{112} & 0 & &&\med & \high&\high &\\\cline{3-6}\cline{9-11}
 & & DHE 3072 & 3072 & \multicolumn{1}{l|}{128} & 0 & & &\med& \high& \high&\\ \cline{2-6}\cline{9-11}
 & \multirow{3}{*}{RSA~\cite{moriarty2016pkcs}} & RSA 1024 & 1024 & \multicolumn{1}{l|}{80} & 0 & 
& & \med& \high& \high & \\\cline{3-6}\cline{9-11}
 & & RSA 2048 & 2048 & \multicolumn{1}{l|}{112} & 0 & & &\med& \high &\high& \\\cline{3-6}\cline{9-11}
 & & RSA 3072 & 3072 & \multicolumn{1}{l|}{128} & 0 & & &\med& \high &\high& \\ \hline
\multirow{9}{*}{Symmetric} & \multirow{3}{*}{AES~\cite{FIPS197}} & AES 128 & 128 & \multicolumn{1}{l|}{128} & 64 & \multirow{3}{*}{{\begin{minipage}{\linewidth}
Weakened by Grover's Algorithm~\cite{grover1996fast}.
\end{minipage}}}
& \multirow{3}{*}{{\begin{minipage}{\linewidth}
\begin{myBullets}
\vspace{0.2cm}
\item {Info. Disclosure: Grover's algorithm reduces the effective key length, making brute-force attacks feasible.}
\end{myBullets}
\end{minipage}}}&\med& \med & \med & \multirow{3}{*}{{\begin{minipage}{\linewidth}Larger key sizes are needed.\end{minipage}}} \\
\cline{3-6}\cline{9-11}
 & & AES 192 & 192 & \multicolumn{1}{l|}{192} & 96 & & &\med & \med &\med & \\\cline{3-6}\cline{9-11}
 & & AES 256 & 256 & \multicolumn{1}{l|}{256} & 128 & & &\med& \low&\low& \\ 
 \cline{2-12}
 & \multirow{3}{*}{SHA2~\cite{eastlake2011us}} & SHA 256 & - & \multicolumn{1}{l|}{128} & {85} & \multirow{6}{*}{{\begin{minipage}{\linewidth}
Weakened by Brassard et al.'s Algorithm~\cite{brassard1997quantum}.
\end{minipage}}} & \multirow{6}{*}{{\begin{minipage}{\linewidth}
\begin{myBullets}
\item {Spoofing: Fake hash values can be created.}
\item {Tampering: Data integrity can be compromised by finding collisions.}
\end{myBullets}
\end{minipage}}}&\med& \med &\med & \multirow{6}{*}{{\begin{minipage}{\linewidth}Larger hash values are needed.\end{minipage}}} \\\cline{3-6}\cline{9-11} 
 & & SHA 384 & - & \multicolumn{1}{l|}{192} & 128 & & &\med & \low&\low& \\\cline{3-6}\cline{9-11}
 & & SHA 512 & - & \multicolumn{1}{l|}{256} & 170 & & & \med& \low&\low& \\ \cline{2-6}\cline{9-11}
 & \multirow{3}{*}{SHA3~\cite{eastlake2011us}} & SHA3 256 & - & \multicolumn{1}{l|}{128} & {85} && &\med& \med &\med & \\\cline{3-6}\cline{9-11} 
 & & SHA3 384 & - & \multicolumn{1}{l|}{192} & 128 & & &\med & \low&\low & \\\cline{3-6}\cline{9-11}
 & & SHA3 512 & - & \multicolumn{1}{l|}{256} & 170 & & &\med& \low&\low & \\ \hline
\end{tabular}%
}  \vspace{0cm}
\end{table*}
Widely used classical cryptographic algorithms, both symmetric and asymmetric, are increasingly vulnerable to QC. Quantum algorithms such as Shor's, Grover's and Brassard et al.'s algorithm~\cite{brassard1997quantum} can compromise the hard problems underlying modern cryptography, threatening the security of diverse applications and communications. This section analyzes the vulnerabilities of existing cryptographic systems to motivate the need for quantum-resistant alternatives.
 \begin{enumerate}[topsep=1ex, itemsep=1ex, wide, font=\itshape, labelwidth=!, labelindent=0pt, label*=A.\arabic*.]
\item \textit{Risk Assessment:} 
Understanding the risks posed by \gls{QC} to classical cryptosystems requires anticipating the emergence of large-scale quantum computers and their cryptanalytic impact. This study examines the projected 5–30 year timeline for such systems and evaluates the cumulative likelihood of significant quantum threats to classical cryptosystems over this period. Figure~\ref{fig:chart1} synthesizes expert assessments of the ``quantum threat'' timeline~\cite{mosca2022quantum}, defined as the capability to break RSA-2048 encryption within 24 hours using a quantum computer. These assessments can be extended to evaluate the likelihood of breaking other cryptographic algorithms based on their quantum security level. A comprehensive risk assessment therefore requires evaluating each algorithm’s ``quantum strength'' relative to the RSA-2048 benchmark, considering susceptibility to Shor’s, Grover’s, and Brassard et al.’s algorithms.
\begin{figure}[!h]
 \centering
 \includegraphics[trim=0.2cm 0.1cm 0.2cm 0.1cm, clip=true, width=0.9\linewidth, height=0.54\linewidth,frame]{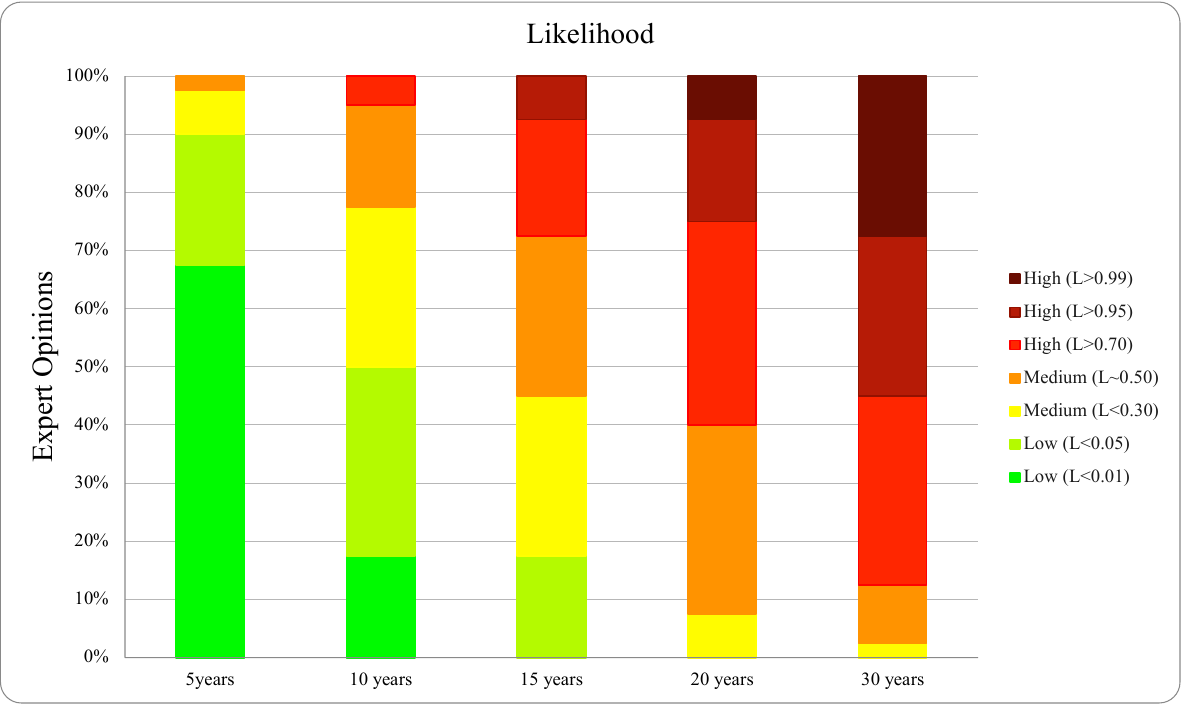}
 \captionof{figure}{Cumulative Expert Opinions Related to Quantum Threat to Classic Cryptography}
 \label{fig:chart1}
 \vspace{-0.3cm}
\end{figure}%

 \begin{enumerate}[topsep=1ex, itemsep=1ex, wide, font=\itshape, labelwidth=!, labelindent=0pt, label*=\arabic*.]
\item \textit{Expected Likelihood of Quantum Threat:} 
To assess the \textit{expected likelihood} of a quantum threat to classical cryptosystems over time horizons of 5, 10, 15, 20, and 30 years, we analyze aggregated expert predictions from Figure~\ref{fig:chart1}. For each period $j$, the expected likelihood is computed as:
\begin{equation*}
 E_{\text{period}_j}[\text{likelihood}] = \hspace{-0.2cm}\sum_{\omega_i\subseteq [0,1]} \hspace{-0.2cm}\text{likelihood}_{\text{period}_j} (\omega_i) \times \text{Pr}_{\text{period}_j} (\omega_i),
\end{equation*}
where $\omega_i$ denotes a subset of the likelihood range $[0,1]$, and $\text{Pr}_{\text{period}_j}(\omega_i)$ is the proportion of expert opinions assigning that likelihood to period $j$. For qualitative risk assessment, the computed likelihoods are classified into three categories: \textit{Low}, \textit{Medium}, and \textit{High}. As depicted in Figure~\ref{fig:Likelihood}, the results show a \textit{low} threat within 10 years, \textit{medium} within 15 years, and \textit{high} beyond 20 years.
 \begin{figure}[!h]
 \centering
\includegraphics[trim=0.2cm 0.1cm 0.1cm 0.6cm, clip=true, width=0.9\linewidth, height=0.54\linewidth,frame]{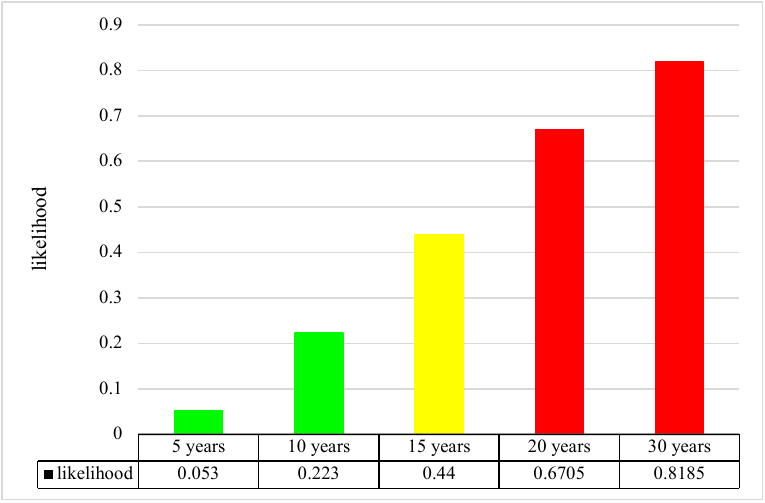}
 \captionof{figure}{Expected Likelihood of Quantum Threat for Classic Cryptography Within 30 Years}
 \label{fig:Likelihood}
 \vspace{-0.3cm}
\end{figure}
\item \textit{Quantum Impact Assessment:} 
To assess the impact of quantum threats on classical cryptographic algorithms, we evaluate their \textit{quantum security strength}, which measures resilience against quantum attacks~\cite{NISTpostquantum,NISTIR8545}. This metric is critical for identifying vulnerabilities and prioritizing mitigation efforts. The impact is classified as \textit{High} if quantum strength is $<64$ bits, \textit{Low} if $\geq 128$ bits, and \textit{Medium} for values in between, as shown in Figure~\ref{fig:classic-impact}.
\begin{figure}[!h]\vspace{-0.3cm}
 \centering
\includegraphics[trim=0.1cm 0.1cm 0.1cm 0.8cm, clip=true, width=0.9\linewidth, height=0.54\linewidth,frame]{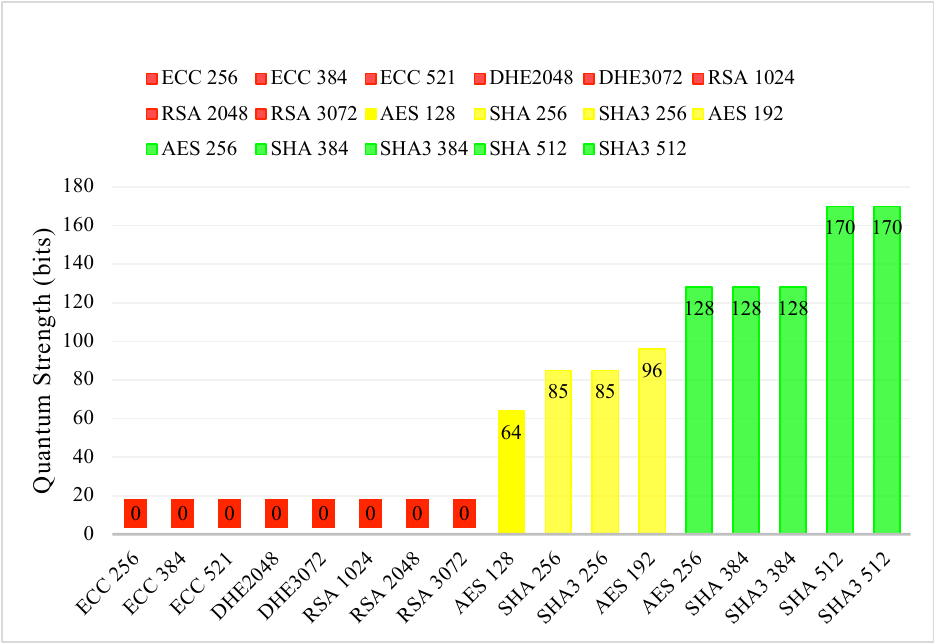}
\caption{Expected Impact of Quantum Threat for Classic Cryptography}
\label{fig:classic-impact}
\vspace{-0.3cm}
\end{figure}
\item \textit{Evaluating Risk:}
The overall \gls{R} of a cryptographic algorithm combines exploitation \gls{L} and potential \gls{I} as illustrated in the risk matrix (see Figure~\ref{fig:risk-matrix}), providing a systematic framework for evaluating current security posture and directing quantum-resistant migration strategies. Table~\ref{tab:Pre-Migration-Alg} provides detailed analysis of pre-migration classical algorithms, including their classical and quantum security strengths, vulnerabilities, quantum threats, associated risks, and potential quantum-resistant alternatives. Our analysis assumes a 15-year timeline with \textit{Medium} quantum threat likelihood (see Figure~\ref{fig:Likelihood}), though this parameter can be adjusted for other periods.
\end{enumerate}
\end{enumerate}
\subsection{Selected Quantum-Resistant Cryptographic Standards: Cyber Impact and Risk Assessment}\label{Selected-Quantum-Resistant-Cryptographic-Standards}
\begin{table*}[!htbp]
\scriptsize
\caption{NIST-Standardized Quantum-Resistant Cryptographic Algorithms: Cyber Impact and Risk Assessment}
\label{tab:qkd_pqc_comparison}
\resizebox{\textwidth}{!}{%
\begin{tabular}{|l|p{0.13\linewidth}|p{0.1\linewidth}|p{0.15\linewidth}|p{0.45\linewidth}|p{0.35\linewidth}|l|l|l|}
\hline
 \multirow{2}{*}{\textbf{Algorithms}} & \multirow{2}{*}{\textbf{Description}} & \multirow{2}{*}{\textbf{FIPS Compliance}} & \multirow{2}{*}{\textbf{Attacks}} & \multirow{2}{*}{\textbf{Possible Countermeasures}} & \multirow{2}{*}{\textbf{STRIDE Threats}} & \multirow{2}{*}{\textbf{L}} & \multirow{2}{*}{\textbf{I}} & \multirow{2}{*}{\textbf{R}} \\
 &&&&&&&& \\ \hline
 \multirow{20}{*}{Kyber~\cite{bos2018crystals}} & \multirow{20}{*}{\begin{minipage}{\linewidth}Key encapsulation mechanism 
 based on the Module Learning with Errors (M-LWE) problem, in conjunction with cyclotomic rings\end{minipage}}&\multirow{20}{*}{\begin{minipage}{\linewidth}Pending FIPS
certification (FIPS
203~\cite{FIPS203})\end{minipage}}& Fault Attacks~\cite{ravi2019number,oder2016practical,ravi2020drop} & 
{\begin{minipage}{\linewidth}
\begin{myBullets}
\vspace{0.1cm}
\item Masking decryption process by splitting secret key~\cite{oder2016practical,ravi2020drop},
 \item Checking the secret and error components of the LWE instances for known trivial weaknesses~\cite{ravi2019number}.
\vspace{0.1cm}\end{myBullets}
\end{minipage}}
 & 
{\begin{minipage}{\linewidth}
\begin{myBullets}
\vspace{0.1cm}
\item Info. Disclosure (by recovery of message and key~\cite{ravi2019number,oder2016practical,ravi2020drop}).
\vspace{0.1cm}\end{myBullets}
\end{minipage}}
 &\med &\med &\med \\ \cline{4-9} 
& & &Simple Power Analysis~\cite{hamburg2021chosen} & 
 {\begin{minipage}{\linewidth}
\begin{myBullets}
\vspace{0.1cm}
 \item Masking of input~\cite{hamburg2021chosen},
 \item Randomizing the order of executed operations within an NTT computation or by inserting random dummy operations inside the NTT~\cite{hamburg2021chosen}.
\vspace{0.1cm}\end{myBullets}
\end{minipage}}
 & 
 {\begin{minipage}{\linewidth}
\begin{myBullets}
\vspace{0.1cm}
 \item Info. Disclosure (by recovery of key~\cite{hamburg2021chosen}).
\vspace{0.1cm}\end{myBullets}
\end{minipage}}
 &\med & \med&\med \\ \cline{4-9} 
& & &Advanced Power Analysis~\cite{pessl2019more,kamucheka2021power, dubrova2022breaking} & {\begin{minipage}{\linewidth}
\begin{myBullets}
\vspace{0.1cm}
 \item Masking the Number Theoretic Transform (NTT), which is an integral part of efficient implementations of many lattice-based schemes~\cite{pessl2019more}.
 \item No countermeasures for the attack mentioned in~\cite{dubrova2022breaking}.
\vspace{0.1cm}\end{myBullets}
\end{minipage}}
 & 
 {\begin{minipage}{\linewidth}
\begin{myBullets}
\vspace{0.1cm}
 \item Info. Disclosure (recovery of the transmitted symmetric key~\cite{pessl2019more}).
\vspace{0.1cm}\end{myBullets}
\end{minipage}}
 &\high & \med&\high\\ \cline{4-9} 
 & & &Electromagnetic Attacks~\cite{ravi2020generic,xu2021magnifying,ravi2020drop} & 
 {\begin{minipage}{\linewidth}
 \begin{myBullets}
\vspace{0.1cm}
\item Masking the ECC procedures, including masking the decryption/decapsulation operations~\cite{ravi2020generic,oder2016practical}, masking to protect the FO transform operations in the CCA setting~\cite{ravi2020generic}, masking to protect the secret key~\cite{xu2021magnifying}.
 \item Discarding ciphertexts with a special structure or low entropy~\cite{xu2021magnifying}.
 \item Splitting the secret into random shares and thereafter randomizing the entire decryption or decapsulation~\cite{xu2021magnifying}.
\vspace{0.1cm}\end{myBullets}
\end{minipage}}
 & 
 {\begin{minipage}{\linewidth}
\begin{myBullets}
\vspace{0.1cm}
\item Info. Disclosure (by full key extraction~\cite{xu2021magnifying,ravi2020generic} or by disclosing bits of the secret message~\cite{ravi2020drop}).
\vspace{0.1cm}\end{myBullets}
\end{minipage}}

 &\low &\med & \low\\ \cline{4-9} 
& & &Template Attacks~\cite{ravi2021exploiting} & 
 {\begin{minipage}{\linewidth}
\begin{myBullets}
\vspace{0.1cm}
\item No countermeasures for the attack mentioned in~\cite{ravi2021exploiting}.
\vspace{0.1cm}\end{myBullets}
\end{minipage}}
 & 
 {\begin{minipage}{\linewidth}
\begin{myBullets}
\vspace{0.1cm}
\item Info. Disclosure (by message recovery~\cite{ravi2021exploiting}).
\vspace{0.1cm}\end{myBullets}
\end{minipage}}
&\med & \med&\med\\ \cline{4-9} 
& & &Cold-Boot Attacks~\cite{albrecht2018cold} & 
 {\begin{minipage}{\linewidth}
\begin{myBullets}
\vspace{0.1cm}
 \item Storing the secret in the time domain instead of the frequency domain~\cite{albrecht2018cold}.
\vspace{0.1cm}\end{myBullets}
\end{minipage}}
 & 
 {\begin{minipage}{\linewidth}
\begin{myBullets}
\vspace{0.1cm}
 \item Info. Disclosure (recovery of the secret key~\cite{albrecht2018cold}).
\vspace{0.1cm}\end{myBullets}
\end{minipage}}
 & \low & \med & \low \\ \hline
 \multirow{18}{*}{Dilithium~\cite{ducas2018crystals}} & \multirow{18}{*}{\begin{minipage}{\linewidth}
Lattice-based digital signature scheme that uses Lyubashevsky's Fiat-Shamir with Aborts technique and rejection sampling to ensure compactness and relies on the hardness of module lattice problems
 \end{minipage}} &\multirow{18}{*}{\begin{minipage}{\linewidth}Pending FIPS
certification (FIPS
204~\cite{FIPS204})\end{minipage}}& Fault Attacks~\cite{ravi2019number,bruinderink2018differential} & 
{\begin{minipage}{\linewidth}
\begin{myBullets}
\vspace{0.1cm}
 \item Checking the secret and error components of the LWE instances for known trivial weaknesses~\cite{ravi2019number}.

\item Generic countermeasures including Double computation, Verification-after-sign, and Additional randomness~\cite{bruinderink2018differential}.
\vspace{0.1cm}\end{myBullets}
\end{minipage}}
 & 
{\begin{minipage}{\linewidth}
\begin{myBullets}
\vspace{0.1cm}
 \item Spoofing and Tampering (by recovery of key~\cite{ravi2019number,bruinderink2018differential}),
 \item Tampering and Repudiation (by forging a signature on any given message~\cite{bruinderink2018differential}),
 \item Elevation of Privilege (by elevating the privileges and gaining access to restricted systems or data via forged signatures~\cite{bruinderink2018differential}).
\vspace{0.1cm}\end{myBullets}
\end{minipage}}
&\med & \med&\med \\ \cline{4-9} 
 & & &Advanced Power Analysis~\cite{migliore2019masking,marzougui2022profiling} & 
 {\begin{minipage}{\linewidth}
\begin{myBullets}
\vspace{0.1cm}
 \item Masking using linear secret sharing scheme~\cite{migliore2019masking},
 \item Boolean and arithmetic masking by leveraging splitting and sharing sensitive variable~\cite{marzougui2022profiling}.
\vspace{0.1cm}\end{myBullets}
\end{minipage}}
 & 
{\begin{minipage}{\linewidth}
\begin{myBullets}
\vspace{0.1cm}
 \item Spoofing, Tampering, and Repudiation (by forging signature~\cite{marzougui2022profiling}),
 \item Elevation of Privilege (by elevating the privileges and gaining access to restricted systems or data via forged signature~\cite{marzougui2022profiling}).
 \item Spoofing, Tampering, Repudiation and Elevation of Privilege (by disclosing secret variables~\cite{migliore2019masking}).\vspace{0.1cm}\end{myBullets}
\end{minipage}}
 &\med & \med&\med \\ \cline{4-9} 
& & &Electromagnetic Attacks~\cite{ravi2019exploiting,singh2024end} & 
{\begin{minipage}{\linewidth}
\begin{myBullets}
\vspace{0.1cm}
\item Re-ordering of operations within the signing procedure and embedding the vulnerable addition operation deep enough inside the signing procedure~\cite{ravi2019exploiting},
\item Bit-slicing design for NTT, the most critical sub-block, to provide a spatial intra-instruction redundancy~\cite{singh2024end}.
\vspace{0.1cm}\end{myBullets}
\end{minipage}}
 & 
{\begin{minipage}{\linewidth}
\begin{myBullets}
\vspace{0.1cm}
 \item Spoofing, and Tampering (by disclosing some info. about secret key~\cite{singh2024end}),
 \item Tampering, Repudiation and Elevation of Privilege (by forging a signature on any given message~\cite{ravi2019exploiting}).\vspace{0.1cm}\end{myBullets}
\end{minipage}}
 &\low & \med&\low\\ \cline{4-9}

 & & &Template Attacks~\cite{berzati2023exploiting,cryptoeprint:2023/050} & 
 {\begin{minipage}{\linewidth}
\begin{myBullets}
\vspace{0.1cm}
\item Shuffling and Secret sharing~\cite{cryptoeprint:2023/050,berzati2023exploiting}.
\vspace{0.1cm}\end{myBullets}
\end{minipage}}
 & 
 {\begin{minipage}{\linewidth}
\begin{myBullets}
\vspace{0.1cm}
\item Spoofing and Tampering (by revealing information
on the signer’s secret key~\cite{cryptoeprint:2023/050,berzati2023exploiting}).
\item Repudiation and Elevation of Privilege (by forging signature via revealed signer’s secret key~\cite{cryptoeprint:2023/050,berzati2023exploiting}).
\vspace{0.1cm}\end{myBullets}
\end{minipage}}
&\med & \med&\med\\ \hline 
 \multirow{5}{*}{SPHINCS+~\cite{bernstein2019sphincs+}} & \multirow{3}{*}{\begin{minipage}{\linewidth}
 Stateless hash-based
signature scheme relying on the hardness of finding collisions in hash functions
\end{minipage}} &\multirow{5}{*}{\begin{minipage}{\linewidth}Pending FIPS
certification (FIPS
205~\cite{FIPS205})\end{minipage}}& Fault Attacks~\cite{castelnovi2018grafting,genet2018practical} & 
{\begin{minipage}{\linewidth}
\begin{myBullets}
\vspace{0.1cm}
 \item Making the signature computation redundant~\cite{castelnovi2018grafting},
 \item Computing the index of the few-time signatures (FTS) from public values instead of secret ones~\cite{castelnovi2018grafting},
 \item Linking the different layers of the hyper-tree to detect faults in the computation of the tree, which results in a non-valid signature~\cite{castelnovi2018grafting},
 \item Detecting faults by recomputing of sub-trees with swapped nodes, as well as an enhanced hash function that inherently protects against faults~\cite{genet2018practical},
\item Computing and storing one-time signatures to reuse the results whenever needed~\cite{genet2018practical},
\item Recomputing the vulnerable instructions on different hardware modules to detect mismatches~\cite{genet2018practical}. 
\vspace{0.1cm}\end{myBullets}
\end{minipage}}
 & 
\begin{minipage}{\linewidth}
\begin{myBullets}
\vspace{0.1cm}
 \item Spoofing and Tampering (by recovering parts of the
secret key~\cite{castelnovi2018grafting} or universal signature forgery~\cite{genet2018practical}),
 \item Tampering and Repudiation (by forging any message signature~\cite{castelnovi2018grafting} or by creating a universal forgery with a voltage glitch injection
on the targeted platform and collecting faulty
signatures to create~\cite{genet2018practical}).
\vspace{0.1cm}\end{myBullets}
\end{minipage}
&\med & \med&\med \\ \cline{4-9} 
& & &Advanced Power Analysis~\cite{kannwischer2018differential} & 
{\begin{minipage}{\linewidth}
\begin{myBullets}
\vspace{0.1cm}
 \item Hiding the order of the Mix procedures~\cite{kannwischer2018differential}.
\vspace{0.1cm}\end{myBullets}
\end{minipage}}
 & 
{\begin{minipage}{\linewidth}
\begin{myBullets}
\vspace{0.1cm}
 \item Spoofing and Tampering (by recovering secret key~\cite{kannwischer2018differential}),
 \item Tampering, Repudiation and Elevation of Privilege (by generating signature on arbitrary messages~\cite{kannwischer2018differential}).
\vspace{0.1cm}\end{myBullets}
\end{minipage}}
 &\med & \med&\med \\ \hline
 \multirow{14}{*}{Falcon~\cite{fouque2018falcon}} & \multirow{14}{*}{{\begin{minipage}{\linewidth}
Lattice-based 
digital signature
algorithm
based on the hardness of the shortest vector problem in structured NTRU lattices
 \end{minipage}}} & \multirow{14}{*}{\begin{minipage}{\linewidth}Pending FIPS
certification \end{minipage}}&Fault Attacks~\cite{mccarthy2019bearz} & 
{\begin{minipage}{\linewidth}
\begin{myBullets}
\vspace{0.1cm}
 \item Double computation of signature~\cite{mccarthy2019bearz},
 \item Immediate verification after signing~\cite{mccarthy2019bearz}, 
 \item Zero checking of the sampled vector~\cite{mccarthy2019bearz}.\vspace{0.1cm}\end{myBullets}
\end{minipage}}
 & 
{\begin{minipage}{\linewidth}
\begin{myBullets}
\vspace{0.1cm}
 \item Spoofing and Tampering (by retrieving the private-key~\cite{mccarthy2019bearz}),
 \item Repudiation and Elevation of Privilege (by forging signature via retrieved private key~\cite{mccarthy2019bearz}).
\vspace{0.1cm}\end{myBullets}
\end{minipage}}
&\med & \med&\med \\ \cline{4-9} 
 & & &Timing Attacks~\cite{mccarthy2019bearz} & 
 {\begin{minipage}{\linewidth}
\begin{myBullets}
\vspace{0.1cm}
 \item Blind-Vector algorithm extended the use of the Fisher-Yates shuffling procedure to enhance random shuffles for side-channel protection~\cite{mccarthy2019bearz}, 
 \item Sample discard performing extra cache read from random addresses to distort statistics~\cite{mccarthy2019bearz}.
\vspace{0.1cm}\end{myBullets}
\end{minipage}}
 & 
 {\begin{minipage}{\linewidth}
\begin{myBullets}
\vspace{0.1cm}
 \item Spoofing and Tampering (by retrieving the private-key~\cite{mccarthy2019bearz}),
 \item Repudiation and Elevation of Privilege (by forging signature via retrieved private key~\cite{mccarthy2019bearz}).\vspace{0.1cm}\end{myBullets}
\end{minipage}}
&\med & \med&\med \\ \cline{4-9} 
 & & &Simple Power Analysis~\cite{guerreau2022hidden} & 
 {\begin{minipage}{\linewidth}
\begin{myBullets}
\vspace{0.1cm}
 \item Practically lower the Hamming weight gap~\cite{guerreau2022hidden}.
\vspace{0.1cm}\end{myBullets}
\end{minipage}}
 & 
 {\begin{minipage}{\linewidth}
\begin{myBullets}
\vspace{0.1cm}
 \item Spoofing, Tampering, Repudiation and Elevation of Privilege (by complete recovery of the secret keys~\cite{guerreau2022hidden}).\vspace{0.1cm}\end{myBullets}
\end{minipage}}
&\med & \med&\med\\ \cline{4-9} 
& & &Electromagnetic Attacks~\cite{karabulut2021falcon} & {\begin{minipage}{\linewidth}
\begin{myBullets}
\vspace{0.1cm}
 \item Hiding by making power consumption constant,~\cite{karabulut2021falcon}),
 \item Masking using randomizing the intermediates values~\cite{karabulut2021falcon}).
\vspace{0.1cm}\end{myBullets}
\end{minipage}}
 & 
 {\begin{minipage}{\linewidth}
\begin{myBullets}
\vspace{0.1cm}
 \item Spoofing and Tampering (by extracting the secret signing keys~\cite{karabulut2021falcon}),
 \item Tampering, Repudiation and Elevation of Privilege(by forging signatures on arbitrary messages~\cite{karabulut2021falcon}).
\vspace{0.1cm}\end{myBullets}
\end{minipage}}
 & \low & \med& \low \\ \hline
\multirow{10}{*}{{{HQC}~\cite{melchor2018hamming}} }&
\multirow{10}{*}{\begin{minipage}{\linewidth}
{Code-based \gls{KEM/ENC} relying on the hardness of decoding random linear codes in the Hamming metric.}
\end{minipage}} &
\multirow{10}{*}{\begin{minipage}{\linewidth}
{Pending FIPS certification (FIPS 207)~\cite{alagic2025status}}
\end{minipage}} & Fault Attacks~\cite{cayrel2020message,xagawa2021fault} &
\begin{minipage}{\linewidth}
\begin{myBullets} \vspace{0.1cm}
\item Constant-time error handling to avoid secret overwrites~\cite{xagawa2021fault},
\item Instruction duplication and random delays to disrupt fault injection~\cite{xagawa2021fault}.
\item No known countermeasures reported in~\cite{cayrel2020message}. 
\vspace{0.1cm} \end{myBullets}
\end{minipage} &
\begin{minipage}{\linewidth}
\begin{myBullets} \vspace{0.1cm}
\item \textbf{Info. Disclosure:} via message recovery and key leakage~\cite{cayrel2020message,xagawa2021fault}.
\vspace{0.1cm} \end{myBullets}
\end{minipage} & \med & \med & \med \\ \cline{4-9}

& & & Timing Attacks~\cite{guo2020key,guo2022don,wafo2020practicable} &
\begin{minipage}{\linewidth}
\begin{myBullets} \vspace{0.1cm}
\item Constant-time decoding, field operations, and RNG~\cite{wafo2020practicable,guo2022don}.
\item No mitigation suggested in~\cite{guo2020key}.
\vspace{0.1cm} \end{myBullets}
\end{minipage} &
\begin{minipage}{\linewidth}
\begin{myBullets} \vspace{0.1cm}
\item \textbf{Info. Disclosure:} via key recovery through timing side-channels~\cite{guo2020key,guo2022don}.
\vspace{0.1cm} \end{myBullets}
\end{minipage} & \med & \med & \med \\ \cline{4-9}

& & & Simple Power Analysis~\cite{schamberger2020power} &
\begin{minipage}{\linewidth}
\begin{myBullets} \vspace{0.1cm}
\item No effective mitigation reported~\cite{schamberger2020power}.
\vspace{0.1cm} \end{myBullets}
\end{minipage} &
\begin{minipage}{\linewidth}
\begin{myBullets} \vspace{0.1cm}
\item \textbf{Info. Disclosure:} via partial key recovery from Hamming weight leakage~\cite{schamberger2020power}.
\vspace{0.1cm} \end{myBullets}
\end{minipage} & \high & \med & \high \\ \cline{4-9}

& & & Electromagnetic Attacks~\cite{goy2022new,paiva2025tu} &
\begin{minipage}{\linewidth}
\begin{myBullets} \vspace{0.1cm}
\item Linear secret sharing to mask sensitive values~\cite{goy2022new}.
\item No effective mitigation for the attack mentioned in~\cite{paiva2025tu}.
\vspace{0.1cm} \end{myBullets}
\end{minipage} &
\begin{minipage}{\linewidth}
\begin{myBullets} \vspace{0.1cm}
\item \textbf{Info. Disclosure:} via full key recovery from EM leakage~\cite{goy2022new,paiva2025tu}.
\vspace{0.1cm} \end{myBullets}
\end{minipage} & \med & \med & \med \\ \hline
\end{tabular}%
}\vspace{3pt}
\scriptsize{$^*$ We perform risk evaluation with the presumption of considering the countermeasures mentioned in the table.}
  \vspace{-0.3cm}
\end{table*}
The threat of \gls{QC} necessitates a transition to \glsfirst{PQC}. \gls{NIST} has led a standardization effort and
\gls{PQC} algorithms are designed to safeguard public-key cryptography (key encapsulation/encryption and digital signatures) against quantum attacks. 
Several promising \gls{PQC} categories have emerged, including lattice-based~\cite{bos2018crystals, ducas2018crystals, fouque2018falcon}, code-based~\cite{melchor2018hamming}, and hash-based~\cite{bernstein2019sphincs+} approaches. However, ongoing research remains vital to ensure their long-term security. 
\gls{NIST} announced the first four \gls{PQC} candidates for standardization in 2022, along with candidates for a fourth round of analysis~\cite{nist-pqc-2022}. Additionally, \gls{NIST} has solicited comments on the initial public drafts of three Federal Information Processing Standards: FIPS 203~\cite{FIPS203}, FIPS 204~\cite{FIPS204}, and FIPS 205~\cite{FIPS205}. These drafts outline quantum-resistant key establishment and digital signature schemes to safeguard against future quantum attacks~\cite{nist2024fips}. {\gls{NIST} further solidified its commitment to \gls{PQC} by selecting HQC in 2025, a code-based \gls{KEM/ENC}, for standardization, with a draft FIPS expected in 2026~\cite{alagic2025status}.}
 \begin{enumerate}[topsep=1ex, itemsep=1ex, wide, font=\itshape, labelwidth=!, labelindent=0pt, label*=B.\arabic*.]
\item \textit{Security Evaluation and Vulnerability Analysis:}  
Recent discoveries of side-channel vulnerabilities in \gls{NIST}-standardized \gls{PQC} algorithms underscore the importance of sustained and systematic security evaluation (see Figure~\ref{fig:Side}). This subsection reviews identified attack vectors, corresponding mitigation strategies, and associated risk profiles. Quantum-capable adversaries may exploit information leakage from cryptographic implementations—arising from power consumption, electromagnetic emanations, or timing variations—to recover secret keys or forge digital signatures. Table~\ref{tab:qkd_pqc_comparison} systematically analyzes known vulnerabilities across \gls{PQC} algorithms, detailing attack vectors, countermeasures, STRIDE threat classifications, and corresponding risk assessments. This structured analysis contributes to ongoing \gls{NIST} standardization efforts and informs the design of resilient countermeasures for post-quantum deployment.

\begin{figure}[!h]
\centering
\large
\resizebox{0.9\linewidth}{!}{
\begin{tikzpicture}[line width=0.035cm]
\node (d2) [draw,top color=white, bottom color=blue!20, rounded corners,minimum height=1cm,minimum width=4cm] at (0,0) {Side-channel Attacks};
\node (d21) [draw,top color=white, bottom color=blue!20, rounded corners,minimum height=1cm,minimum width=4cm,below right of=d2,xshift=4cm,yshift=-1.5cm] {Active Attacks};
\node (d211) [draw,top color=white, bottom color=blue!20, rounded corners,minimum height=1cm,minimum width=6.5cm, right of=d21,xshift=5cm] {Fault Attacks};

\node (d22) [draw,top color=white, bottom color=blue!20, rounded corners,minimum height=1cm,minimum width=4cm, above right of=d2,
 ,xshift=4cm,yshift=1.7cm] {Passive Attacks};
 \node (d223) [draw,top color=white, bottom color=blue!20, rounded corners,minimum height=1cm,minimum width=6.5cm, right of=d22,xshift=5cm,,yshift=2.3cm] {Electromagnetic Attacks};
 \node (d222) [draw,top color=white, bottom color=blue!20, rounded corners,minimum height=1cm,minimum width=6.5cm, right of=d22,xshift=5cm,,yshift=1.15cm] {Template Attacks};
 \node (d221) [draw,top color=white, bottom color=blue!20, rounded corners,minimum height=1cm,minimum width=6.5cm, right of=d22,xshift=5cm,,yshift=0cm] {Cold-Boot Attacks};
\node (d224) [draw,top color=white, bottom color=blue!20, rounded corners,minimum height=1cm,minimum width=6.5cm, right of=d22,xshift=5cm,,yshift=-1.15cm] { Advanced Power Analysis Attacks};
\node (d225) [draw,top color=white, bottom color=blue!20, rounded corners,minimum height=1cm,minimum width=6.5cm, right of=d22,xshift=5cm,,yshift=-2.3cm] {Simple Power Analysis Attacks};
\node (d226) [draw,top color=white, bottom color=blue!20, rounded corners,minimum height=1cm,minimum width=6.5cm, right of=d22,xshift=5cm,,yshift=-3.45cm] {Timing Attacks};
 \draw[-latex] (d2) |- (d21);
 \draw[-latex] (d2) |- (d22);
 \draw[-latex] (d21) -- (d211);
 \draw[-latex] (d22) -- (d221); 
 \draw[-latex] (d22) |- (d222);
 \draw[-latex] (d22) |- (d223);
 \draw[-latex] (d22) |- (d224);
 \draw[-latex] (d22) |- (d225); 
 \draw[-latex] (d22) |- (d226); 
\end{tikzpicture}}
\caption{{Taxonomy of Attacks for \gls{NIST}-Standardized Quantum-Resistant Cryptographic Algorithms}}
\label{fig:Side}
\vspace{-0.4cm}
\end{figure}


\item \textit{Risk Assessment:} 
To assess security risks associated with \gls{PQC} algorithms, we evaluate likelihood using the criteria in Table~\ref{table:likelihood}, derived from NIST-SP~800-30 Appendix~G~\cite{NIST-SP800-30}. Key factors include exploitability, attack complexity, attacker motivation, and the effectiveness of countermeasures (see Table~\ref{tab:qkd_pqc_comparison}). Likelihood is categorized into three levels: (a) \textit{High}, referring to remotely exploitable vulnerabilities without effective countermeasures, or vulnerabilities requiring minimal attacker sophistication or resources; (b) \textit{Medium}, referring to vulnerabilities that require physical access or specific conditions, or that lack robust countermeasures when exploited remotely, often involving moderate attack complexity or skill level; and (c) \textit{Low}, referring to scenarios with no known vulnerabilities, or only highly complex attacks requiring administrative or physical privileges, mitigated by effective countermeasures and limited attacker motivation.

The impact assessment, grounded in NIST-SP~800-30 Appendix~H~\cite{NIST-SP800-30}, evaluates the consequences of quantum attacks on user satisfaction, data confidentiality, and organizational reputation.  According to the criteria in Table~\ref{table:impact},  \gls{PQC} attacks have a \textit{Medium} impact. This assessment indicates that such attacks could lead to partial data compromise, limited operational disruption, and manageable recovery times, with moderate financial and reputational consequences. Risk evaluation combines likelihood and impact using the risk matrix (see Figure~\ref{fig:risk-matrix}). Table~\ref{tab:qkd_pqc_comparison} provides algorithm-specific insights essential for developing robust \gls{PQC} transition strategies to mitigate quantum threats and strengthen cybersecurity resilience.

\end{enumerate}
 \section{Quantum Impact on \gls{BC} Technology Components}\label{sec:impact}
Quantum attacks pose a significant threat to various components of \gls{BC} technology, including (a) \gls{BC} network, (b) mining pools, (c) transaction verification mechanisms, (d) smart contracts, and (e) user wallets. These attacks can compromise the trust and immutability that \gls{BC} technology aims to provide. Understanding these vulnerabilities and implementing effective mitigation strategies is crucial for safeguarding the integrity and security of \gls{BC} ecosystems. This section will delve deeper into each of these components and explore potential solutions to safeguard their integrity and security in the quantum age.

\subsection{\gls{BC} Network and Quantum Computing Threats}\label{Classic-BC-Network}
\gls{QC} poses a significant threat to \gls{BC} networks by potentially exploiting weaknesses in core functionalities. Here is a breakdown of some key vulnerabilities along with mitigation strategies. 
 \begin{enumerate}[topsep=1ex, itemsep=1ex, wide, font=\itshape, labelwidth=!, labelindent=0pt, label*=A.\arabic*.]
\item \textit{Cryptographic Hashing Vulnerabilities:}  
 \gls{BC} networks rely heavily on cryptographic hashing functions to ensure data integrity and immutability. However, advancements in \gls{QC} present a potential challenge to the security of widely used hash functions like SHA-256. As highlighted in Table~\ref{tab:Pre-Migration-Alg}, these vulnerabilities can weaken the security strength of hash functions and introduce risks such as transaction forgery or alteration
of ledger history~\cite{kearney2021vulnerability, dasgupta2019survey}.

Effective mitigation strategies to address these vulnerabilities include:  
(i) promoting the standardization of quantum-resistant hash functions that meet \gls{BC}-specific security requirements, as emphasized in \gls{NIST}'s \gls{PQC} transition guidelines~\cite{nist2024transition};  
(ii) increasing hash output length, which raises the quantum resource threshold for collision and preimage attacks under Grover's algorithm; and  
(iii) implementing migration plans that facilitate a seamless transition to quantum-resistant alternatives with minimal network disruption~\cite{baseri2024evaluation,petrenko2019assessing}.

\item \textit{False Message Attacks:} Quantum algorithms can compromise the security of digital signatures and hash functions, enabling attackers to forge signatures or manipulate data within the network. This could allow the injection of false information or the spread of misinformation, undermining the integrity of the \gls{BC} system~\cite{baseri2024evaluation,10737027}. Additionally, quantum computational power, including \textit{quantum parallelism} and \textit{exponential speedup} (via algorithms like Shor's and Grover's), could facilitate \textit{Sybil attacks}, where attackers deploy numerous fake nodes to manipulate identities or votes. This threat is particularly concerning for \glspl{BC} utilizing \gls{PoW} consensus mechanisms, where quantum acceleration could make it easier to create and manage fake nodes, potentially disrupting network operations~\cite{kearney2021vulnerability}. However, the risks are not exclusive to \gls{PoW}-based systems, as \gls{QC} could also affect cryptographic vulnerabilities in \gls{PoS} or \gls{BFT} systems, potentially undermining consensus or enabling false message injection by compromising digital signatures or private keys.

Effective mitigation strategies for these quantum threats include:
(i)  deploying \gls{PQC} algorithms, such as the NIST-Standardized \gls{PQC} approaches outlined in Table~\ref{tab:qkd_pqc_comparison}, to secure digital signatures and other cryptographic primitives~\cite{nist2023postquantum,alagic2025status};
(ii) developing and enhancing reputation-based filtering systems that help nodes identify and filter messages from unreliable sources, with adaptations to address quantum-specific vulnerabilities and ensure post-quantum resilience~\cite{feraudo2024diva}; and
(iii) designing quantum-aware incentive mechanisms that reward nodes for verifying information validity and penalize those spreading false data, adapting current effective systems to account for \gls{QC} vulnerabilities and ensure future effectiveness~\cite{smahi2024vfl}.


\item \textit{51\% Attacks:}  
A significant concern for \gls{PoW}{-based} \glspl{BC} is the possibility of a 51\% attack. In this scenario, a malicious actor or group could control more than half of the network's mining hash rate, allowing them to manipulate transaction confirmations, potentially double-spend cryptocurrency, and ultimately compromise the network's integrity. While achieving a 51\% attack with traditional computers is already computationally expensive, quantum computers could significantly reduce the resources needed, making it a more realistic threat for some \glspl{BC}~\cite{sayeed2019assessing, fernandez2020towards}.

Mitigation strategies to proactively defend against 51\% attacks on \gls{BC} networks include:
(i) transitioning to alternative consensus mechanisms, such as \gls{PoS}, which rely on coin ownership for validation rather than raw computational power and are less susceptible to 51\% attacks~\cite{ferdous2021survey}; quantum-resistant consensus mechanisms beyond those discussed above include QRL's \gls{PoS} (QPoS)\cite{QPoS} and quantum \gls{PoW} (QPoW)\cite{singh2025proof} algorithms;
(ii) implementing enhanced difficulty adjustment algorithms that dynamically adjust mining difficulty based on network hash rate to make it more expensive for attackers to acquire a controlling stake, thereby enhancing the network's resilience against 51\% attacks~\cite{kraft2016difficulty}; and
(iii) employing merged mining, which may strengthen smaller \glspl{BC} by sharing computational resources with larger networks. However, this introduces additional complexities: if a malicious actor gains control over a significant portion of the hash rate in a merged mining setup, they could potentially launch 51\% attacks on multiple \glspl{BC} simultaneously. Careful consideration is therefore required to manage these risks and ensure that the security of all participating networks is maintained~\cite{judmayer2017merged}.


\begin{table*}[!h]
\caption{Analysis of \gls{BC} Network Component: Quantum Threats, Vulnerabilities, Mitigations, and Risk Assessment}
\small
\resizebox{\textwidth}{!}{%
\begin{tabular}{|m{0.09\linewidth}|m{0.20\linewidth}|m{0.2\linewidth}|m{0.16\linewidth}|m{0.28\linewidth}|m{0.28\linewidth}|m{0.14\linewidth}|c|c|c|}
\hline
\textbf{Layer} & \textbf{Exploited Vulnerabilities} & \textbf{Attack Vector} & \textbf{Potential Impacts} & \textbf{STRIDE Threats} & \textbf{Mitigation Strategies} & \textbf{Actionable Parties} & \textbf{L} & \textbf{I} & \textbf{R} \\ \hline
\multirow{24}{*}{\begin{minipage}[c]{\linewidth}
 \gls{BC} Network
\end{minipage}}& Cryptographic Hashing Vulnerabilities & Exploiting hashing algorithms & \begin{myBullets} 
\item Transaction Malleability
\item Ledger history alteration. 
\vspace{-2ex} \end{myBullets} & \begin{myBullets} 
\item Tampering: Potential modification of transaction details or blocks
\item Spoofing: Forging or bypassing digital signatures. 
\vspace{-2ex} \end{myBullets} & \begin{myBullets} 
\item Standardization of Quantum-Resistant Hashing
\item Migration Plans
\item Increasing Hash Output Length
\item \gls{PQC} Standards Development. 
\vspace{-2ex} \end{myBullets} & \begin{myBullets} 
\item Developers 
\item Miners 
\item Auditors 
\vspace{-2ex} \end{myBullets} & \low & \high & \med \\ \cline{2-10}
& False Message Attacks & Injecting false information through forging signatures and Compromising the security of hash functions & \begin{myBullets} 
\item Disrupted Consensus Process
\item Invalid Transactions
\vspace{-2ex} \end{myBullets} &\begin{myBullets} 
\item Tampering: Manipulation of transaction data or blocks
\item Repudiation: Forged signatures allow attackers to deny responsibility or implicate others
\item Info. Disclosure: Exposure of sensitive transaction data 
\vspace{-2ex} \end{myBullets} & \begin{myBullets} 
\item \gls{PQC}
\item Network Reputation Systems
\item Incentivize Honest Behavior 
\vspace{-2ex} \end{myBullets} & \begin{myBullets} 
\item Developers 
\item Node Operators 
\item Governance Participants 
\vspace{-2ex} \end{myBullets} & \high & \high & \high \\ \cline{2-10}
 & 51\% Attacks & Control of network hash rate & \begin{myBullets} 
\item Double Spending
\item Network Disruption 
\vspace{-2ex} \end{myBullets} & \begin{myBullets} 
\item Spoofing: False confirmations of transactions
\item \gls{DoS}: Potential for network downtime or congestion due to malicious control of the mining hash rate 
\vspace{-2ex} \end{myBullets} & \begin{myBullets} 
\item Transition to Alternative Consensus Mechanisms
\item Enhanced Difficulty Adjustment Algorithms
\item Merged Mining (with careful risk management) 
\vspace{-2ex} \end{myBullets} & \begin{myBullets}
\item Miners 
\item Developers 
\item Community Participants 
\vspace{-2ex} \end{myBullets} & \med & \high & \high \\ \cline{2-10}
& {\gls{DoS} Attacks} & Overwhelming network resources & \begin{myBullets} 
\item Network Congestion
\item Service Outages. 
\vspace{-2ex} \end{myBullets} &\begin{myBullets} 
\item Leveraging quantum power for large-scale resource exhaustion, overwhelming nodes, disrupting transactions, and destabilizing the network.
\vspace{-2ex} \end{myBullets} &\begin{myBullets} 
\item Resource Reservation and Rate Limiting
\item Distributed Network Architecture
\item Redundancy and Fault Tolerance. 
\vspace{-2ex} \end{myBullets} & \begin{myBullets} 
\item Node Operators 
\item Service Providers 
\vspace{-2ex} \end{myBullets} & \high & \high & \high \\ \cline{2-10}
 & Data Privacy and Obfuscation Vulnerabilities & Interception of transaction data, exploiting vulnerabilities in obfuscation protocols (e.g., Bulletproofs, zk-SNARKs) and ECC-based cryptography & \begin{myBullets} 
\item Exposure of sensitive transaction data (amounts, identities)
\item Loss of privacy and trust. 
\vspace{-2ex} \end{myBullets} & \begin{myBullets} 
\item Info. Disclosure: Exposure of sensitive data (amounts, identities)
\item Tampering: Data alteration, including transaction amounts and metadata. 
\vspace{-2ex} \end{myBullets} & \begin{myBullets} 
\item \gls{PQC}
\item \glspl{ZKP}
\item Off-Chain Data Storage
\item Transition to Quantum-Resistant Cryptographic Protocols
\item Layered Security with Quantum-Resistant Techniques 
\vspace{-2ex} \end{myBullets} & \begin{myBullets} 
\item Developers 
\item Service Providers 
\item Regulators 
\vspace{-2ex} \end{myBullets} & \high & \high & \high \\ \hline

\end{tabular}%
}
  \vspace{-0.3cm}
\label{tab:blockchain_analysis}
\end{table*}
\item \textit{\gls{DoS} Attacks:}  
\gls{QC} significantly amplifies the threat of \gls{DoS} attacks on \gls{BC} networks. By leveraging their immense computational power, quantum attackers could overwhelm nodes or critical network infrastructure with a deluge of traffic, disrupting transaction processing, delaying block confirmations, and potentially centralizing control. Quantum-accelerated attacks can exploit network vulnerabilities more efficiently, leading to more severe and prolonged disruptions.

Mitigation strategies for quantum-enhanced \gls{DoS} attacks on \gls{BC} networks include: (i) implementing resource reservation and rate-limiting mechanisms to prevent malicious actors from monopolizing bandwidth or computational resources~\cite{alladi2022comprehensive};
(ii) maintaining a distributed network architecture to reduce the impact of localized failures and prevent cascading disruptions~\cite{riskhan2023adaptive};
(iii) designing systems with redundancy and fault-tolerant components to ensure operational continuity during \gls{DoS} events~\cite{deng2020distributed,kumari2021survey};
(iv) deploying network optimization techniques (e.g., traffic filtering, data compression, congestion control, and sharding) to reduce processing bottlenecks and enhance resilience~\cite{sanka2021systematic}; and
(v) transitioning to quantum-resistant cryptographic protocols and consensus mechanisms to mitigate vulnerabilities that quantum attackers may exploit in classical systems~\cite{nist2023postquantum}.


\item \textit{Data Privacy and Obfuscation Vulnerabilities:}  
The transparency and anonymity inherent in \gls{BC} transactions can be compromised by quantum attacks. Quantum computers could potentially intercept and manipulate transaction data, exposing sensitive information such as transaction amounts and participant identities. Additionally, quantum attacks could compromise the pseudonymity of \gls{BC} addresses, linking them to real-world identities and undermining privacy~\cite{conklin2023legal}.

Privacy-centric \glspl{BC} like Monero and Zcash employ advanced cryptographic protocols such as Bulletproofs and zk-SNARKs to protect transaction details. However, both of these protocols rely on ECC, which is vulnerable to Shor's algorithm. Additionally, the cryptographic hash functions used in these \glspl{BC} (such as Keccak in Monero and BLAKE2b in Zcash) could be affected by Grover's algorithm, potentially reducing their security strength. These quantum threats could expose sensitive transaction metadata, including amounts and participant identities~\cite{kearney2021vulnerability, aditya2021survey, fuchsbauer2019aggregate}. 
To mitigate the risk of data privacy and obfuscation vulnerabilities, several strategies can be adopted: (i) implementing quantum-resistant cryptographic algorithms to secure transaction data~\cite{nist2023postquantum, alagic2025status}; (ii) adopting post-quantum zero-knowledge proof systems (e.g., lattice-based \glspl{ZKP}) to enable private verification without revealing transaction contents~\cite{zhou2024leveraging}; (iii) combining ring signatures (which obfuscate transaction initiators) with quantum-resistant \glspl{ZKP} such as zk-STARKs (which validate amounts without revealing details) to enhance sender anonymity~\cite{xu2020layered}; (iv) transitioning privacy-focused \glspl{BC} like Monero and Zcash to quantum-resistant obfuscation protocols based on lattice cryptography and Dilithium~\cite{aditya2021survey,  liang2024identity}; and (v) using off-chain decentralized storage for non-critical metadata while ensuring robust security evaluation to avoid introducing new vulnerabilities~\cite{liu2022extending}. 
\item \textit{{Risk Assessment for Quantum Threat to} \gls{BC} {Network Component}:} 
To assess the risks posed by \gls{QC} to \gls{BC} network components, we evaluate both the likelihood of exploitation and the potential impact using the qualitative criteria in Tables~\ref{table:likelihood} and~\ref{table:impact}. Likelihood reflects both the probability of occurrence and the feasibility of an attack, determined by \gls{QC} availability\footnote{Likelihood evaluations across all \gls{BC} components assume the availability of large-scale quantum computers capable of breaking current cryptographic primitives. Projected \gls{QC} timelines may shift these assessments (see Figure~\ref{fig:Likelihood}).}, 
the nature of the vulnerability, the existence of effective mitigation strategies, and the resources required for exploitation.
Based on these factors, likelihood is classified as \textit{High}, \textit{Medium}, or \textit{Low}:
\begin{enumerate}[itemsep=0ex]
\item \textit{High:} \gls{DoS} attacks are highly likely because they rely on computational power to overwhelm network resources rather than cryptographic weaknesses. These attacks could exploit quantum parallelism to amplify their scale, leading to network destabilization and significant disruption of operations. Similarly, false message attacks pose a high risk, as quantum systems can break digital signature schemes (e.g., ECDSA) using Shor’s algorithm. This enables attackers to forge signatures, inject false messages, and compromise consensus. Furthermore, data privacy and obfuscation vulnerabilities are also highly likely, given that quantum computers could exploit weaknesses in zk-SNARKs or elliptic curve cryptography to expose sensitive transaction data.

\item \textit{Medium:} 
51\% attacks present a medium likelihood due to the potential of quantum acceleration to reduce the computational resources required to control the majority of network hash power. While such an attack would still demand substantial quantum hardware, ongoing developments elevate the feasibility of this threat. Achieving quantum dominance sufficient to compromise \gls{PoW} consensus mechanisms remains a concern as quantum technologies progress.
\item \textit{Low:} 
Cryptographic hashing vulnerabilities remain unlikely in the near term. Breaking robust hash functions like SHA-256 using Grover’s algorithm provides only a quadratic speedup, which is insufficient to pose a significant threat to adequately designed systems with sufficiently long hash outputs. Additionally, the requirement for large-scale, fault-tolerant quantum computers to execute such attacks remains speculative, keeping this threat at a low likelihood.
\end{enumerate}
Impact captures the severity of a successful attack, including privacy loss, service disruption, data manipulation, financial damage, and erosion of user trust. Given the critical nature of the \gls{BC} network component, all vulnerabilities in this component are categorized as \textit{High} impact due to their potential to cause significant and potentially irreversible consequences to network operations, user trust, and financial integrity.  Overall risk combines likelihood and impact using the risk matrix in Figure~\ref{fig:risk-matrix}. Table~\ref{tab:blockchain_analysis} details vulnerabilities, likelihood, impact, and mitigation strategies for addressing quantum-era threats. Adopting these strategies enables the \textit{\gls{BC} network component} to enhance resilience against cryptographic weaknesses, network disruptions, and privacy compromises, safeguarding transaction security and integrity in the post-quantum era.

\end{enumerate}

\subsection{Mining Pool and Quantum Computing Threats}\label{Classic-Mining-Pool}
Mining pools are pivotal in upholding the security of \gls{PoW} \glspl{BC} by dedicating computational resources to solve cryptographic puzzles and validate transactions. However, \gls{QC} introduces unique threats to mining pools, potentially undermining the decentralized nature of these networks. Here is a breakdown of some key concerns:

 \begin{enumerate}[topsep=1ex, itemsep=1ex, wide, font=\itshape, labelwidth=!, labelindent=0pt, label*=B.\arabic*.]
\item \textit{Disruption of Consensus Mechanisms via Malicious Nodes:}    \gls{QC} introduces significant risks to mining pools by enhancing attackers' ability to infiltrate with malicious nodes masquerading as legitimate participants. These attacks disrupt the consensus process through methods such as \textit{Pool Flooding} and \textit{Selfish Mining}.
In pool flooding, attackers leverage \gls{QC}’s immense computational power and parallelism to overwhelm the pool's infrastructure. By rapidly generating and injecting a large number of malicious nodes (Sybil nodes), attackers hinder transaction processing and disrupt consensus procedures~\cite{aitzhan2016security}. 
In selfish mining, using quantum-enhanced computational efficiency, attackers exploit vulnerabilities in reward distribution algorithms. They strategically withhold computational power or selectively participate in block validation to maximize personal rewards, disrupting fair reward distribution and reducing pool security~\cite{conti2018survey, strivemindz2023quantumthreat}. 
These disruptions can result in delayed block confirmations, inconsistent transactions, or even forks in the \gls{BC}~\cite{kearney2021vulnerability,strivemindz2023quantumthreat}. 

\gls{QC} amplifies the scale and efficiency of these attacks, necessitating tailored mitigation strategies: (i) adopting voting-based consensus protocols (e.g., \gls{BFT}) that tolerate a threshold of Byzantine nodes without compromising network integrity, ensuring consensus even under large-scale quantum-facilitated attacks~\cite{malkhi2019flexible, zhong2023byzantine}; (ii) enforcing stake-based admission policies requiring participants to deposit cryptocurrency, thereby increasing the economic cost of Sybil attacks even for quantum-enabled adversaries~\cite{liu2019survey}; and (iii) deploying decentralized reputation systems that detect and exclude nodes exhibiting suspicious behavior, limiting the influence of malicious quantum-generated participants~\cite{noorian2010state}.


\item \textit{Disruption of Consensus Mechanisms via Reward Distribution Manipulation:}  By leveraging \gls{QC}'s computational power, attackers can manipulate reward distribution mechanisms to gain an unfair advantage in receiving block rewards. Quantum-enhanced optimization and parallel processing enable them to exploit vulnerabilities in deterministic reward algorithms more efficiently, centralizing mining power and compromising network security~\cite{parida2023post,kearney2021vulnerability}.

To protect mining pools from consensus disruptions caused by reward distribution manipulation, several mitigation strategies can be employed: 
(i) implementing transparent, verifiable, and publicly auditable reward distribution mechanisms to deter manipulation attempts~\cite{rodrigues2022trust,weingartner2021prototyping};
(ii) employing \glspl{VRF} to introduce cryptographic randomness in block selection and reward assignment, preventing outcome prediction and adversarial manipulation~\cite{gilad2017algorand}; and 
(iii) promoting mining pool diversification to reduce the risk of centralization and limit the impact of a single compromised pool~\cite{gandotra2020cryptocurrency}.


\item \textit{Disruption of Consensus Mechanisms via \gls{PoW}:}  
\gls{QC} poses significant threats to \gls{PoW} systems by exploiting vulnerabilities in public-key cryptography and hashing functions, which are critical to maintaining \gls{BC} integrity and trust. While classical \gls{PoW} systems use public-key cryptography to secure communications and validate transactions, quantum attackers, leveraging Shor’s algorithm,  could potentially forge digital signatures, manipulate transactions, perform double-spending, disrupt consensus, and launch Sybil attacks. Although current systems remain secure today, future quantum advancements threaten to compromise these foundational cryptographic protocols.
In addition to cryptographic signatures, \gls{PoW} relies on hashing functions like SHA 256 to secure mining operations. Quantum computers, using Grover’s algorithm, could significantly reduce the computational effort required to solve these cryptographic puzzles, enabling attackers to generate blocks faster than legitimate miners. While this does not directly affect digital signatures, it exposes mining pools that rely on outdated hashing algorithms to quantum-enabled disruptions, jeopardizing the fairness and security of the \gls{BC}~\cite{strivemindz2023quantumthreat, sayeed2019assessing}.

Mitigation strategies to proactively defend against disruption of consensus mechanisms via \gls{PoW} include:
(i) adopting essential \gls{PQC} algorithms to secure digital signatures and transaction data from quantum manipulation. These algorithms align with NIST's ongoing post-quantum standardization efforts~\cite{nist2023postquantum,alagic2025status};
(ii) exploring hybrid consensus models that combine \gls{PoW} with \gls{PoS}. Since \gls{PoS} relies on coin ownership for validation, it is less vulnerable to manipulation based on computational power. This makes it more resilient to quantum-enabled attacks targeting \gls{PoW} systems~\cite{pass2016hybrid}; and
(iii) deploying quantum-safe communication protocols throughout the broader \gls{BC} ecosystem. This includes securing wallets, exchanges, and supporting infrastructure. Although not part of the mining process, these components are also exposed to quantum risks such as eavesdropping and message tampering~\cite{garcia2024quantum,allende2023quantum}.

\begin{table*}[!hbpt]
\caption{Analysis of Mining Pool Component: Quantum Threats, Vulnerabilities, Mitigations, and Risk Assessment}
\small
\resizebox{\textwidth}{!}{%
\begin{tabular}{|m{0.07\linewidth}|m{0.20\linewidth}|m{0.18\linewidth}|m{0.22\linewidth}|m{0.31\linewidth}|m{0.24\linewidth}|m{0.14\linewidth}|c|c|c|}
\hline
\textbf{Layer} & \textbf{Exploited Vulnerabilities} & \textbf{Attack Vector} & \textbf{Potential Impacts} & \textbf{STRIDE Threats} & \textbf{Mitigation Strategies} & \textbf{Actionable Parties} & \textbf{L} & \textbf{I} & \textbf{R} \\ \hline
\multirow{12}{*}{\begin{minipage}[c]{\linewidth}
 Mining Pool
\end{minipage}}
& Disruption of Consensus Mechanisms via Malicious Nodes 
& Overwhelming Pool Infrastructure, Exploiting Reward Distribution Algorithms 
& \begin{myBullets} 
\item Loss of network integrity 
\item \gls{DoS} 
\item Double-spending attacks 
\vspace{-2ex} \end{myBullets} 
& \begin{myBullets} 
\item {Spoofing:} Malicious nodes impersonate legitimate participants.
\item {Tampering:} Manipulation of reward algorithms disrupts fairness.
\item {\gls{DoS}:} Pool flooding disrupts operations.
\item {Elevation of Privilege:} Attackers gain unauthorized control.
\vspace{-2ex} \end{myBullets} 
& \begin{myBullets} 
\item Byzantine Fault Tolerance Protocols 
\item Stake-Based Admission 
\item Decentralized Reputation Systems 
\vspace{-2ex} \end{myBullets} 
& \begin{myBullets} 
\item Miners, 
\item Developers,
\item Node Operators,
\item Auditors.
\vspace{-2ex} \end{myBullets} 
& \high & \high & \high \\ \cline{2-10}
& Disruption of Consensus Mechanisms via Reward Distribution Manipulation 
& Exploiting Reward Distribution Systems 
& \begin{myBullets} 
\item Reduced Miner Participation 
\item Centralization of mining power 
\item Double-spending attacks 
\vspace{-2ex} \end{myBullets} 
& \begin{myBullets} 
\item {Repudiation:} Attackers deny manipulating rewards.
\item {Tampering:} Unfair reward distribution due to manipulation.
\vspace{-2ex} \end{myBullets} 
& \begin{myBullets} 
\item Transparent Reward Distribution Mechanisms 
\item Verifiable Random Functions 
\item Pool Diversification 
\vspace{-2ex} \end{myBullets} 
& \begin{myBullets} 
\item Miners,
\item Developers,
\item Auditors.
\vspace{-2ex} \end{myBullets} 
& \high & \high & \high \\ \cline{2-10}
& Disruption of Consensus Mechanisms via \gls{PoW} 
& Exploiting vulnerabilities in digital signatures or hashing algorithms 
& \begin{myBullets} 
\item Forged transactions (digital signatures) 
\item Re-mining blocks/double-spending (hashing functions) 
\item Loss of trust in the network 
\vspace{-2ex} \end{myBullets} 
& \begin{myBullets} 
\item {Spoofing:} Forged digital signatures allow impersonation.
\item {Tampering:} Attackers modify \gls{BC} data or re-mine blocks.
\item {Info. Disclosure:} Weak cryptography exposes transaction details.
 
\vspace{-2ex} \end{myBullets} 
& \begin{myBullets} 
 
\item Transition to \gls{PQC} 
\item Hybrid Mining Models 
\item Quantum-Safe Communication Protocols 
 
\vspace{-2ex} \end{myBullets} 
& \begin{myBullets} 
 
\item Developers,
\item Node Operators, 
\item Regulators, 
\item Community Participants. 
 
\vspace{-2ex} \end{myBullets} 
& \med & \high & \high \\ 
\hline
\end{tabular}%
}  \vspace{-0.3cm}
\label{tab:mining_pool_analysis}
\end{table*}

\item \textit{{Risk Assessment for Quantum Threats to Mining Pool Component}:}  
\gls{QC} advancements pose significant risks to mining pools, potentially impacting both \gls{BC} users and overall network stability. These risks are evaluated based on their likelihood and impact, following the criteria outlined in Tables~\ref{table:likelihood} and~\ref{table:impact}. 
Here is the categorization of the likelihood of \gls{QC} threats to mining pools:

\begin{enumerate}[itemsep=0ex]
\item \textit{High:} Disruption of consensus mechanisms due to malicious nodes and reward distribution manipulation is highly probable with the emergence of \gls{QC}. Quantum computers, leveraging their immense computational power, can facilitate Sybil attacks by rapidly generating numerous malicious nodes or exploiting selfish mining tactics. These actions disrupt consensus, delay block confirmations, and threaten mining pool stability. Additionally, quantum algorithms enable efficient exploitation of deterministic reward systems, centralizing mining power and undermining fair reward distribution. These vulnerabilities underscore the critical need for mitigation strategies such as \gls{BFT} protocols, stake-based admission, \glspl{VRF}, and pool diversification to ensure mining pool security.

\item \textit{Medium:} The disruption of consensus mechanisms via \gls{PoW} is moderately likely. \gls{QC} using Grover’s algorithm can accelerate the solving of cryptographic puzzles, enabling attackers to potentially generate blocks faster than legitimate miners. However, the high computational resources currently required for such attacks limit their immediate feasibility. As quantum technology progresses, this risk could increase, emphasizing the need to adopt hybrid consensus models or transition to alternative mechanisms such as \gls{PoS} to reduce long-term vulnerability.
\end{enumerate}
As with the \gls{BC} network component, successful attacks on mining pools would have a \textit{High} impact, causing disruptions like loss of network integrity, \gls{DoS} attacks, double-spending vulnerabilities, reduced miner participation, and centralization of mining power. These issues can significantly harm the security and stability of the entire network. Overall risk combines likelihood and impact via the risk matrix presented in Figure~\ref{fig:risk-matrix}. 
Table~\ref{tab:mining_pool_analysis} details vulnerabilities, likelihood, impact, and mitigation strategies for addressing quantum-era threats. Implementing these strategies enables the \textit{mining pool component} to strengthen resilience against malicious node infiltration, reward manipulation, and \gls{PoW} weaknesses, safeguarding \gls{BC} network security and stability.


\end{enumerate}
\subsection{Transaction Verification Mechanism and Quantum Computing Threats}\label{Classic-Transaction-Verification}

The process of verifying transactions on a \gls{BC} ensures the integrity of the network and prevents fraudulent activities. However, \gls{QC} poses a significant threat to these verification mechanisms, which can lead to manipulated transactions and compromised network security. Here is a detailed breakdown of potential vulnerabilities: 
\begin{enumerate}[topsep=1ex, itemsep=1ex, wide, font=\itshape, labelwidth=!, labelindent=0pt, label*=C.\arabic*.]
\item \textit{Double-Spending Vulnerabilities:} 
\gls{QC} threatens \gls{BC} security by accelerating attacks such as double-spending. Grover's algorithm~\cite{grover1996fast} can expedite the search for valid transaction signatures, enabling attackers to manipulate the \gls{BC} and fraudulently spend the same digital currency multiple times~\cite{kearney2021vulnerability}. Consequently, existing double-spending attacks, such as race conditions (exploiting the confirmation time gap for multiple transactions~\cite{khan2021empirical}) and Finney attacks (pre-mining a conflicting transaction~\cite{nicolas2020blockchain}), could become more efficient in the future.

Mitigation strategies to proactively defend against  double-spending vulnerabilities include:
(i) transitioning to NIST-recommended quantum-resistant signature schemes to replace current digital signatures vulnerable to quantum attacks~\cite{NISTPQC};
(ii) adopting enhanced consensus mechanisms like \gls{PoS} or \gls{BFT}, which rely on economic incentives or fault tolerance rather than on computational power that is vulnerable to quantum-accelerated hash computations~\cite{BunzCampenGunther2018};
(iii) employing sharding and \glspl{DAG} to reduce transaction confirmation times, limiting the attack window for quantum-accelerated double-spending~\cite{refc1,dang2019towards};
(iv) integrating \glspl{VDF} to enforce provable delays in transaction processing, extending validation time and limiting rapid quantum attack effectiveness~\cite{boneh2018verifiable}; and
(v) deploying quantum-resistant sidechains for high-value transactions, confining critical operations to quantum-secure environments while maintaining main chain interoperability~\cite{allende2023quantum,chauhan2023towards}.

\item \textit{Transaction Malleability:}  
The potential to exploit transaction malleability
is one of the significant threats posed by \gls{QC} to the security of \gls{BC}. This vulnerability allows attackers to manipulate weaknesses in transaction formatting to alter specific details (e.g., transaction fees or recipient addresses) without invalidating the signature~\cite{ramananandro2019everparse, decker2014bitcoin}. By accelerating the identification of cryptographic vulnerabilities, quantum computers could enable attackers to modify transaction details, potentially leading to double-spending, altered transaction amounts, or other fraudulent activities~\cite{porambage2021roadmap}.

To address transaction malleability in the quantum era, the following mitigation strategies are recommended:
(i) transitioning to quantum-resistant digital signature schemes to reduce the exploitability of malleability flaws under quantum attacks~\cite{nist2023postquantum, alagic2025status};
(ii) enforcing strict transaction formatting rules to eliminate inconsistencies and reduce the risk of field manipulation~\cite{ramananandro2019everparse};
(iii) adopting standardized canonical transaction format protocols to ensure consistency across all nodes, preventing exploitability due to format variability~\cite{decker2014bitcoin}; and
(iv) deploying protocol upgrades such as \gls{SegWit}, which separate signature data from the transaction ID calculation. By decoupling these elements, \gls{SegWit} directly addresses transaction malleability and enhances the overall robustness of legacy \gls{BC} systems~\cite{singh2020public}.

\item \textit{Transaction Reordering Attacks:}  
Transaction reordering attacks pose a significant threat to the security and integrity of \gls{BC} networks, especially those reliant on precise transaction order, such as \gls{DeFi} protocols or auction-based systems. \gls{QC}, with its exponential computational power, can exacerbate these threats by breaking cryptographic primitives (e.g., RSA, ECC) and accelerating the identification of vulnerabilities in consensus mechanisms. By exploiting compromised digital signatures or manipulating consensus processes, quantum attackers could reorder transactions within blocks, leading to malicious activities such as double-spending, front-running, and smart contract manipulation~\cite{leng2020blockchain}.

To mitigate these threats, \gls{BC} systems must adopt a layered and quantum-resistant approach that addresses both the computational and structural aspects of transaction reordering:  
(i) introducing cryptographically secured sequence numbers or timestamps to enforce tamper-proof transaction order and preserve consistency~\cite{ruan2020transactional}; 
(ii) embedding unspendable inputs to prevent the reuse of transaction outputs, thereby obstructing attempts to manipulate transaction flow or perform order-dependent double-spending~\cite{zaghloul2020bitcoin};  
(iii) implementing \glspl{POSet} within consensus protocols to define flexible but verifiable transaction orderings that maintain consistency in distributed and asynchronous environments~\cite{yanagita2022space,belchior2021survey};   
(iv) deploying \glspl{VDF} to introduce time-bound computational barriers, ensuring transaction ordering is validated through provable delays that are resistant to quantum speedup~\cite{boneh2018verifiable}; and  
(v) adopt quantum-resistant consensus protocols (e.g., voting or staking-based), which reduce reliance on quantum-vulnerable computations and enhance validation integrity via economic or fault-tolerant guarantees~\cite{BunzCampenGunther2018}.

\item \textit{Transaction Timestamp Manipulation:}  
Quantum attacks can exploit vulnerabilities in \gls{BC} timestamp generation, leading to inconsistencies in transaction history and potential disruptions in the consensus process. Quantum computers can undermine the cryptographic algorithms that secure timestamp data, allowing attackers to alter transaction times and sequences. This manipulation can result in out-of-order transactions, enabling malicious activities such as double-spending and hindering the network's ability to reach consensus~\cite{liu2023secure}.

To counter these timestamp manipulation vulnerabilities, several mitigation strategies can be employed: (i) using \glspl{VRF} to generate unpredictable and verifiable timestamps through cryptographic randomness, thereby enhancing security, preserving transaction order integrity, and preventing timestamp manipulation~\cite{gilad2017algorand,boneh2018verifiable}; (ii) synchronizing clocks across validator nodes to ensure consistent timestamp generation, preventing timing discrepancies and maintaining accurate transaction sequencing~\cite{chefrour2022evolution}; and (iii) deploying \gls{BFT} consensus protocols that preserve agreement among honest nodes even in the presence of timestamp manipulation attempts, ensuring network robustness against adversarial behaviors~\cite{zhong2023byzantine}.





 \item \textit{Transaction ID Collisions:}   
Transaction IDs, critical for ensuring the uniqueness and integrity of \gls{BC} transactions, may become vulnerable to quantum attacks. These identifiers are fundamental to maintaining consensus and preventing fraud within the network. Quantum algorithms like Grover's can significantly reduce the computational complexity of finding hash collisions, potentially enabling malicious actors to create multiple transactions with the same ID. Such collisions can lead to network confusion, disruption of consensus mechanisms, and exploitation scenarios, including double-spending or interference with smart contract execution~\cite{aggarwal2021blockchain}.

To address these vulnerabilities, several mitigation strategies are proposed: (i) migrating to quantum-resistant hashing algorithms, while actively contributing to the development of dedicated standards—particularly since NIST has standardized post-quantum public-key cryptography but not yet hash functions~\cite{wu2021quantum,ananth2023plausibility}; and (ii) extending transaction ID lengths to increase resistance against collision attacks. While this approach statistically lowers the probability of hash collisions under Grover's algorithm, it may introduce trade-offs in terms of storage and processing overhead~\cite{wu2021quantum,ananth2023plausibility}.




\item \textit{Classic Signature Vulnerabilities:}  
Classic digital signature algorithms, such as the \gls{ECDSA} and RSA, are foundational to \gls{BC} transaction verification and are widely used for signing transactions and ensuring authenticity. However, these algorithms are vulnerable to 
Shor's Algorithm~\cite{shor1999polynomial}.
A successful quantum attack would allow adversaries to forge digital signatures, inject unauthorized transactions, and compromise the integrity and trust of the \gls{BC} network~\cite{kearney2021vulnerability}. This vulnerability is especially critical for major \gls{BC} platforms like Bitcoin and Ethereum, which rely on ECDSA for transaction security.

To mitigate these risks, the following strategy is essential: (i) migration to quantum-resistant signature algorithms mentioned in Table~\ref{tab:Pre-Migration-Alg}. These algorithms are  designed to withstand attacks from quantum computers and are being standardized through initiatives like \gls{NIST} \gls{PQC}~\cite{nist2023postquantum}.


\item \textit{Quantum-Resistant Oracles:}   
In the world of \gls{BC}, smart contracts rely on oracles to bridge the gap between their internal logic and external data sources. These oracles fetch crucial information from the outside world, feeding it into the smart contract for transaction verification. 
However, traditional oracles pose a significant vulnerability to quantum attacks, as quantum computers could potentially manipulate the data they provide, compromising transaction verification integrity.

Key strategies to mitigate this challenge and proactively enhance security include: (i) adopting decentralized oracles to distribute data retrieval across independent nodes, creating redundancy that makes it considerably more difficult for attackers to manipulate data streams and disrupt transaction verification~\cite{basile2021enhancing,dong2023daon}; (ii) collaborating with quantum-secure data providers that use post-quantum protections for data transmission and storage, ensuring data integrity for smart contracts and strengthening overall network resilience against quantum threats~\cite{lewis2022secure}; and (iii) implementing essential on-chain data validation mechanisms that thoroughly authenticate and verify oracle inputs before contract execution, significantly mitigating risks of data tampering and transaction outcome manipulation~\cite{zhao2022toward}.

\begin{table*}[!hbpt]
\caption{Analysis of Transaction Verification  Component: Quantum Threats, Vulnerabilities, Mitigations, and Risk Assessment}
\small
\resizebox{\textwidth}{!}{%
\begin{tabular}{|m{0.09\linewidth}|m{0.19\linewidth}|m{0.17\linewidth}|m{0.18\linewidth}|m{0.29\linewidth}|m{0.30\linewidth}|m{0.14\linewidth}|c|c|c|}
\hline
\textbf{Layer} & \textbf{Exploited Vulnerabilities} & \textbf{Attack Vector} & \textbf{Potential Impacts} & \textbf{STRIDE Threats}& \textbf{Mitigation Strategies} & \textbf{Actionable Parties} & \textbf{L} & \textbf{I} & \textbf{R} 
\\ \hline
\multirow{30}{*}{\begin{minipage}[c]{\linewidth}
 {Transaction Verification Mechanism} 
\end{minipage}}
& 
Double-Spending Vulnerabilities 
& Grover's algorithm, Race Condition Attack, Finney Attack 
&\begin{myBullets} \item Financial Losses, \item Network Instability \vspace{-2ex} \end{myBullets} 
 &\begin{myBullets} 
 \item {Spoofing:} An attacker could impersonate a legitimate spender to conduct fraudulent transactions.
 \item {Elevation of Privilege:} Exploiting quantum efficiency, attackers may manipulate transaction validation to execute unauthorized double-spends. 
 \vspace{-2ex} \end{myBullets} 
 &\begin{myBullets} 
 \item Transition to Quantum-Resistant Signatures, 
 \item Enhanced Consensus Mechanisms, 
 \item Sharding and \glspl{DAG},
 \item Verifiable Delay Functions,
 \item Quantum-Resistant Sidechains.
 \vspace{-3ex} \end{myBullets} 
 & \begin{myBullets} 
 \item Miners, 
 \item Developers, 
 \item Auditors, 
 \item Community Participants.
 \vspace{-2ex} \end{myBullets} 
 & \high & \high & \high \\ \cline{2-10}
& Transaction Malleability 
& Exploiting weaknesses in transaction formatting to create a malleated transaction that retains a valid signature 
&\begin{myBullets} \item Compromised Transaction Integrity \vspace{-2ex} \end{myBullets} 
 &\begin{myBullets} 
 \item {Tampering:} Attackers alter transaction content without invalidating its signature.
 \vspace{-2ex} \end{myBullets} 
 &\begin{myBullets} 
 \item Adoption of Quantum-Resistant Cryptography,
 \item Strict Transaction Format Enforcement, 
 \item Transaction Standardization Protocols,
 \item Use of Segregated Witness.
 \vspace{-2ex} \end{myBullets} 
 & \begin{myBullets} 
 \item Developers, 
 \item Service Providers, 
 \item Auditors.
 \vspace{-2ex} \end{myBullets} 
 & \high & \high & \high \\ \cline{2-10}
& Transaction Reordering Attacks 
& Manipulating the order of transactions within a block 
& \begin{myBullets} 
\item Transaction Censorship, 
\item Reordering Attacks
\vspace{-2ex} \end{myBullets} 
 & \begin{myBullets} 
 \item {Repudiation:} Malicious nodes deny the authenticity of the correct transaction order.
 \item {Tampering:} Transaction sequences are intentionally altered to facilitate attacks.
 \vspace{-2ex} \end{myBullets} 
& \begin{myBullets} 
\item Sequence Numbers, 
\item Unspendable Inputs, 
\item Partially Ordered Sets,
\item Verifiable Delay Functions,
\item Enhanced Consensus Mechanism.
\vspace{-2ex} \end{myBullets} & 
\begin{myBullets} 
\item Developers, 
\item Node Operators, 
\item Auditors.
\vspace{-2ex} \end{myBullets} & \high & \high & \high \\ \cline{2-10}
& Transaction Timestamp Manipulation 
& Manipulating timestamps to disrupt the consensus process 
& \begin{myBullets} 
\item Disruption of Consensus Process, 
\item Denial of service 
\vspace{-2ex} \end{myBullets} 
 &\begin{myBullets} 
 \item {Tampering:} Altered timestamps impact transaction order and consensus stability.
 \vspace{-2ex} \end{myBullets} 
 &\begin{myBullets} 
 \item \glspl{VRF}, 
 \item Synchronized Clocks, 
 \item \gls{BFT} Protocols 
 \vspace{-2ex} \end{myBullets} 
 &\begin{myBullets} 
 \item Developers, 
 \item Node Operators, 
 \item Service Providers, 
 \item Auditors.
 \vspace{-2ex} \end{myBullets} 
 & \low & \high & \med \\ \cline{2-10}
& Transaction ID Collisions 
& Creating multiple transactions with the same transaction ID 
&\begin{myBullets} 
\item Double spending 
\vspace{-2ex} \end{myBullets} 
 &\begin{myBullets} 
 \item {Elevation of Privilege:} Quantum attackers exploit hash collisions to execute unauthorized transactions.
 \vspace{-2ex} \end{myBullets} 
 &\begin{myBullets} 
 \item Migration to Quantum-Resistant Hashing Functions, 
 \item Extended Transaction IDs 
 \vspace{-2ex} \end{myBullets} 
 & \begin{myBullets} 
 \item Developers, 
 \item Auditors, 
 \item Community Participants.
 \vspace{-2ex} \end{myBullets} 
 & \med & \high & \high \\ \cline{2-10}
& Classic Signature Vulnerabilities 
& Exploiting weaknesses in ECDSA, RSA, or similar algorithms 
&\begin{myBullets} 
\item Compromised Security, 
\item Loss of Funds, 
\item Disruption of Entire \gls{BC} 
\vspace{-2ex} \end{myBullets} 
 &\begin{myBullets} 
 \item {Spoofing:} Forged signatures allow malicious impersonation.
 \item {Tampering:} Attackers modify transaction data (e.g., recipient addresses or amounts) and forge corresponding signatures, compromising data integrity.
 \vspace{-2ex} \end{myBullets} 
 &\begin{myBullets} 
 \item Migration to Quantum-Resistant Signatures.
 \vspace{-2ex} \end{myBullets} 
 & \begin{myBullets} 
 \item Developers, 
 \item Service Providers, 
 \item Auditors, 
 \item Regulators
 \vspace{-2ex} \end{myBullets} 
 & \high & \high & \high \\ \cline{2-10}
& Quantum-Resistant Oracles 
& Manipulating data fed into smart contracts through oracles
& \begin{myBullets} 
\item Smart contract manipulation, 
\item Data Integrity Attacks, 
\item Potential Loss of Funds 
\vspace{-2ex} \end{myBullets} 
&\begin{myBullets} 
\item {Tampering:} Data inputs to smart contracts are manipulated.
\vspace{-2ex} \end{myBullets} 
 &\begin{myBullets} 
 \item Decentralized Oracles, 
 \item Quantum-Secure Data Providers, 
 \item Data Validation Mechanisms 
 \vspace{-2ex} \end{myBullets} 
 & \begin{myBullets} 
\item Developers, 
\item Service Providers, 
\item Node Operators.
\vspace{-2ex} \end{myBullets} 
& \high & \high & \high \\ \hline
 \end{tabular}%
}  \vspace{-0.3cm}
\label{tab:transaction_verification_mechanism_analysis}
\end{table*}
\item \textit{{Risk Assessment for Quantum Threats to Transaction Verification  Component}:}  
To assess the risks posed by \gls{QC} threats to \gls{BC} transaction verification mechanisms, we evaluate their likelihood and impact based on the criteria established in Tables~\ref{table:likelihood} and~\ref{table:impact}. 
Here is a breakdown of likelihood categories for transaction verification vulnerabilities:

\begin{enumerate}[itemsep=0ex]
 \item \textit{High:} This category includes double-spending, transaction malleability, transaction reordering, classic signature vulnerabilities, and quantum-resistant oracles. Double-spending represents one of the most critical threats, as Grover’s algorithm enables attackers to expedite the search for valid transaction signatures, especially affecting \glspl{BC} with slow confirmation times. Transaction malleability and reordering attacks are highly susceptible to quantum systems, which can exploit cryptographic weaknesses or formatting inconsistencies to alter transaction data or sequence. Classic signature vulnerabilities are particularly pressing, with Shor’s algorithm capable of breaking elliptic curve cryptography, exposing digital signatures to forgery and manipulation. Similarly, oracles, which serve as external data providers for \gls{BC} systems, are vulnerable to quantum-enabled manipulation, further amplifying these risks.
 
 \item \textit{Medium:} This category covers transaction ID collisions, where Grover’s algorithm reduces the effort required to find hash collisions. While these attacks are not yet practical under current conditions, employing quantum-resistant hashing algorithms and sufficiently long hash functions can effectively mitigate this risk. Nonetheless, proactive adoption of quantum-resistant measures is necessary to safeguard against future exploitation as \gls{QC} evolves.
 
 \item \textit{Low:} This category includes transaction timestamp manipulation. Advanced defenses such as \glspl{VRF}, synchronized clocks across nodes, and \gls{BFT} protocols significantly mitigate this threat. These measures ensure accuracy and consistency in transaction timestamps, making such attacks highly infeasible, even as quantum technologies mature.
\end{enumerate}
All transaction verification vulnerabilities carry a \textit{High} impact rating, mirroring the assessments for \gls{BC} networks and mining pools. This reflects the critical role transaction verification plays in safeguarding \gls{BC} transactions. A successful attack could wreak havoc, causing financial losses from double-spending or compromised integrity, network instability due to disrupted consensus mechanisms or manipulated transaction ordering, and even a complete compromise of the \gls{BC}'s security, shattering user trust and confidence. 
Overall risk levels are derived by combining likelihood and impact (Figure~\ref{fig:risk-matrix}). Table~\ref{tab:transaction_verification_mechanism_analysis} provides detailed analysis of vulnerabilities, likelihood, impact, risk ratings, and mitigation strategies. Adopting these countermeasures is essential for maintaining the security and stability of the \textit{transaction verification mechanism component} in the quantum era.
\end{enumerate}

\subsection{Smart Contract Attacks and Quantum Computing Threats}\label{Classic-Smart-Contract}

Smart contracts, self-executing programs stored on the \gls{BC}, offer immense potential for automating agreements and facilitating trustless interactions. However, the emergence of \gls{QC} introduces significant vulnerabilities that could compromise the security of these crucial components. Here is a deeper look at the potential threats posed by quantum computers to smart contracts:

 \begin{enumerate}[topsep=1ex, itemsep=1ex, wide, font=\itshape, labelwidth=!, labelindent=0pt, label*=D.\arabic*.]

\item \textit{Cryptographic Algorithm Vulnerabilities:} 
Cryptographic algorithms like ECDSA and RSA, vital for smart contract security, are vulnerable to \gls{QC}. Shor's algorithm 
allows attackers to forge signatures, steal funds, alter contract logic, and introduce malicious code, jeopardizing the integrity and trust of \gls{BC} systems~\cite{dasgupta2019survey, allende2023quantum}.
These vulnerabilities not only threaten financial stability by enabling fund theft but also compromise the functionality of \gls{BC} systems by allowing tampering with contract logic, leading to unintended behavior, malicious code execution, and disrupted contract execution.

To mitigate these vulnerabilities and secure smart contracts in the quantum era,  the following strategies are recommended: (i) conducting quantum-aware code audits to identify cryptographic weaknesses within smart contracts and prioritize remediation of components susceptible to quantum attacks~\cite{he2020smart,chen2022cross}; (ii) applying formal verification techniques under quantum-safe assumptions to mathematically prove the correctness and resilience of contract logic against cryptographic compromise~\cite{ethereum_formal_verification,lewis2023formal,tolmach2023securing}; and (iii) initiating phased migration plans that progressively transition smart contracts from quantum-vulnerable algorithms to standardized quantum-resistant alternatives as they mature and gain adoption~\cite{aydeger2024towards,nather2024migrating}.



\item \textit{Integer Overflow and Underflow Vulnerabilities:}  
Integer overflow and underflow vulnerabilities are common programming errors that can lead to unexpected behavior in smart contracts. These occur when arithmetic operations exceed the maximum or minimum storage capacity of a variable, causing unpredictable outcomes. Quantum computers, with their ability to perform calculations significantly faster, could accelerate the identification and exploitation of such vulnerabilities, making traditional detection methods less effective~\cite{baseri2024cybersecurityquantumeraassessing,baseri2024navigating}. For instance, a poorly designed smart contract may allow a large number to overflow the intended variable size, resulting in unintended behavior or even enabling malicious actors to manipulate funds. Exploitation of such vulnerabilities could destabilize not only individual smart contracts but also broader \gls{BC} ecosystems, leading to cascading failures across interconnected decentralized applications.

Addressing these vulnerabilities and strengthening defenses against quantum-accelerated exploitation requires the following strategies: (i) utilizing advanced static code analysis tools to proactively identify potential integer overflow and underflow vulnerabilities during the development phase~\cite{ghaleb2020effective,fontein2018comparison}; (ii) integrating safe math libraries into smart contract development that enforce strict runtime checks, to provide secure arithmetic operations that prevent overflows and underflows  even when handling large numbers~\cite{khor2020improved,al2024defi}; (iii) employing formal verification techniques with bounded arithmetic to mathematically prove the absence of these vulnerabilities, thereby ensuring robust security guarantees~\cite{tolmach2023securing}; and (iv) promoting proactive development practices, such as testing boundary conditions and adhering to strict validation rules, to minimize the risk of introducing arithmetic vulnerabilities from the outset~\cite{9402082}.

 

\item \textit{\gls{DoS} Attacks on Smart Contracts:} 
\gls{DoS} attacks target smart contracts by overwhelming them with a high volume of requests, far exceeding normal traffic, and impeding legitimate user access. While \gls{QC} does not directly enable \gls{DoS} attacks, its immense computational power could accelerate aspects such as attack generation and execution, enabling attackers to overwhelm smart contracts more rapidly and efficiently~\cite{hameed2022taxonomy}. This surge in malicious traffic disrupts contract operations, causing delays in transactions or even preventing their processing entirely.

To enhance smart contract security against DoS attacks and improve the overall resilience of the BC ecosystem, the following mitigation strategies are recommended: (i) enforcing resource limits to cap the number of interactions or computational operations per user over a given period, preventing abuse, resource exhaustion, and overload while maintaining overall system integrity~\cite{alladi2022comprehensive}; (ii) applying circuit breaker patterns that automatically halt contract execution during traffic surges to prevent system overload, allowing recovery after predefined downtime and minimizing DoS attack impact~\cite{quincozes2021survey}; (iii) implementing rate-limiting mechanisms that throttle excessive requests, ensure fair access for all users, and integrate transaction or gas fee mechanisms that help regulate network load and provide additional deterrents to malicious actors~\cite{chaganti2022comprehensive,riskhan2023adaptive}; (iv) employing network optimization techniques such as off-chain computation, batching, and caching to reduce on-chain load, minimize delays, and alleviate congestion during high-traffic periods while mitigating DoS attack effects~\cite{zou2019smart,das2019fastkitten}; and (v) optimizing gas usage by grouping multiple operations into a single transaction and offloading non-critical computations, reducing overall \gls{BC} load and enhancing system resilience during attacks~\cite{Ding2024Genuine,liu2022extending,wang2023towards}.

\begin{table*}[!h]
\caption{Analysis of Smart Contract Component: Quantum Threats, Vulnerabilities, Mitigations, and Risk Assessment}
\small
\resizebox{\textwidth}{!}{%
\begin{tabular}{|m{0.07\linewidth}|m{0.19\linewidth}|m{0.18\linewidth}|m{0.16\linewidth}|m{0.28\linewidth}|m{0.34\linewidth}|m{0.14\linewidth}|c|c|c|}
\hline
\textbf{Layer} & \textbf{Exploited Vulnerabilities} & \textbf{Attack Vector} & \textbf{Potential Impacts} & \textbf{STRIDE Threats} & \textbf{Mitigation Strategies} & \textbf{Actionable Parties} & \textbf{L} & \textbf{I} & \textbf{R} \\ \hline
\multirow{23}{*}{\begin{minipage}[c]{\linewidth}
 Smart Contract
\end{minipage}}
 
& Cryptographic Algorithm Vulnerabilities 
& Exploiting weaknesses in cryptographic primitives 
&\begin{myBullets} 
\item Steal Funds
\item Manipulate Contract Logic 
\item Disrupt Contract Execution 
\vspace{-2ex} \end{myBullets} 
&\begin{myBullets} 
\item Spoofing: Forged signatures allow impersonation. 
\item Tampering: Attackers modify contract logic or execution flow. 
\vspace{-2ex} \end{myBullets} 
&\begin{myBullets} 
\item Quantum-Resistant Code Audits 
\item Formal Verification with Quantum-Safe Assumptions 
\item Phased Migration to Quantum-Resistant Cryptography 
\vspace{-2ex} \end{myBullets} 
& \begin{myBullets}
\item Developers,
\item Auditors,
\item Regulators,
\item Community Participants.
\vspace{-2ex} \end{myBullets}
& \high & \high & \high \\ \cline{2-10}

& Integer Overflow and Underflow Vulnerabilities 
& Programming errors leading to unexpected behavior 
&\begin{myBullets} 
\item Financial Loss 
\item Data Corruption 
\vspace{-2ex} \end{myBullets} 
&\begin{myBullets} 
\item Tampering: Attackers exploit vulnerabilities to modify contract behavior. 
\vspace{-2ex} \end{myBullets} 
&\begin{myBullets} 
\item Static Code Analysis Tools 
\item Safe Math Libraries 
\item Formal Verification with Bounded Arithmetic 
\item Proactive Development Practices\vspace{-2ex} \end{myBullets} 
& \begin{myBullets}
\item Developers,
\item Auditors.
\vspace{-2ex} \end{myBullets}
& \med & \high & \high \\ \cline{2-10}

& {\gls{DoS} Attacks} 
& Overloading smart contracts with excessive requests, potentially accelerated by \gls{QC}'s computational power 
&\begin{myBullets} 
\item Service Disruption \item Financial Loss \item Transaction Failures or Logic Corruption
\vspace{-2ex} \end{myBullets} 
&\begin{myBullets} 
\item \gls{DoS}: Attackers overwhelm the contract with excessive requests, potentially accelerated by \gls{QC}. 
\vspace{-2ex} \end{myBullets} 
&\begin{myBullets} 
\item Resource Limits 
\item Circuit Breaker Patterns 
\item Rate Limiting Mechanisms
\item Network Optimization Techniques
\item Gas Optimization and Batching
\vspace{-2ex} \end{myBullets} 
& \begin{myBullets}
\item Developers,
\item Auditor,
\item Node Operators,
\item Service Providers.
\vspace{-2ex} \end{myBullets}
& \med & \high & \high \\ \cline{2-10}

& Inter-Contract Communication Vulnerabilities 
& Exploiting weaknesses in communication channels between contracts 
&\begin{myBullets} 
\item Unauthorized Access to Sensitive Data 
\item Manipulated Contract Execution 
\vspace{-2ex} \end{myBullets} 
&\begin{myBullets} 
\item Information Disclosure: Data interception reveals sensitive information. 
\item Tampering: Injected data alters inter-contract communication flows. 
\vspace{-2ex} \end{myBullets} 
&\begin{myBullets} 
\item Standardized Communication Protocols 
\item Access Control Mechanisms 
\item Quantum-Resistant Serialization Mechanisms 
\vspace{-2ex} \end{myBullets} 
& \begin{myBullets}
\item Developers,
\item Service Providers,
\item Auditors.
\vspace{-2ex} \end{myBullets}
& \high & \high & \high \\ \cline{2-10}

& Front-Running Attacks 
& Strategically placing transactions before others for unfair advantage 
&\begin{myBullets} 
\item Financial Loss 
\item Market Manipulation 
\vspace{-2ex} \end{myBullets} 
&\begin{myBullets} 
\item Elevation of Privilege: Attackers exploit quantum efficiency for prioritization. 
\item Information Disclosure: Attackers analyze pending transactions to gain advantage. 
\vspace{-2ex} \end{myBullets} 
&\begin{myBullets} 
\item Quantum-Resistant Encryption for Transaction Concealment,
 \item Commit-Reveal with Post-Quantum Security,
 \item Threshold Cryptography for Transaction Processing,
 \item Fair Transaction Ordering,
 \item Delayed Execution Mechanisms.\vspace{-2ex} \end{myBullets} 
& \begin{myBullets}
\item Service Providers,
\item Node Operators.
\vspace{-2ex} \end{myBullets}
& \high & \high & \high \\ \hline

\end{tabular}%
}  \vspace{-0.3cm}
\label{tab:smart_contract_analysis}
\end{table*}

\item \textit{Inter-Contract Communication Vulnerabilities:}  
In the intricate network of smart contracts, the pathways facilitating communication between these digital entities serve as critical conduits for data exchange and the execution of complex workflows. However, these channels are susceptible to vulnerabilities, offering potential entry points for exploitation by quantum attackers. These vulnerabilities may manifest in various forms, including the interception of sensitive data transmitted between contracts, the injection of malicious data into communication streams, or the disruption of interaction flows among contracts. Such actions could lead to severe consequences, ranging from the exposure of confidential information to the manipulation of data flows and the disruption of essential contract~\cite{parida2023post,ren2023building}. Consequently, implementing robust mitigation strategies becomes paramount to safeguarding against these risks. 

Mitigation strategies and proactive measures to immunize against inter-contract communication vulnerabilities include:  (i) developing and adopting standardized communication protocols that enforce security and support built-in data integrity verification~\cite{baseri2024evaluation,baseri2024navigating}; (ii) enforcing access control mechanisms that restrict unauthorized access to data and functionality during contract interactions~\cite{bellaj2023transpilation}; and (iii) integrating quantum-resistant serialization schemes to protect the integrity of exchanged data, even if intercepted by quantum-enabled adversaries~\cite{joshi2022scrutiny}.


\item \textit{Front-Running Attacks:} 
In some \gls{BC} networks, transaction fees are used to prioritize transaction processing. Quantum computers could potentially exploit this mechanism by analyzing pending transactions and strategically placing their own transactions before others (front-running). This could give attackers an unfair advantage in scenarios where transaction order is crucial.

To effectively mitigate front-running attacks, especially in the context of potential quantum threats, the following mitigation strategies can be implemented: 
(i) employing post-quantum encryption to conceal transaction contents prior to confirmation, preventing premature access even under quantum decryption capabilities and preserving transactional confidentiality~\cite{NISTPQC};  
(ii) adopting commit-reveal schemes enhanced with quantum-resistant cryptography, where transaction data is committed in encrypted form and revealed only after block inclusion, reducing early disclosure risks~\cite{fernandez2020towards};  
(iii) applying threshold cryptography, which requires multiple nodes to jointly decrypt transaction data, preventing unilateral access and mitigating front-running opportunities~\cite{zhang2022f3b,10.1007/978-3-030-77870-5_16};  
(iv) enforcing fair transaction ordering protocols—such as \gls{FIFO} queues—secured with quantum-resistant verification to ensure submission-order processing regardless of fee manipulation~\cite{zhang2022f3b}; and  
(v) implementing delayed execution mechanisms that defer transaction visibility until validation is complete, limiting real-time exploitability by quantum-equipped attackers~\cite{burdges2021delay,boneh2018verifiable}.

\item \textit{{Risk Assessment for Quantum Threats to Smart Contract Component}:} 
To assess the risks posed by \gls{QC} advancements to smart contracts, we evaluate their likelihood and impact using the criteria outlined in Tables \ref{table:likelihood} and \ref{table:impact}. The likelihood of these smart contract vulnerabilities is classified as follows:

\begin{enumerate}[itemsep=0ex]
\item \textit{High:} Vulnerabilities include cryptographic algorithm weaknesses, inter-contract communication vulnerabilities, and front-running attacks. Cryptographic algorithm vulnerabilities are the most critical due to the susceptibility of widely used schemes like ECDSA and RSA to Shor’s algorithm. If compromised, these algorithms would allow attackers to forge digital signatures, alter contract logic, or steal funds. Inter-contract communication vulnerabilities also pose a significant risk, as \gls{QC} could intercept or manipulate sensitive data exchanged between contracts, leading to unauthorized access or disruption of workflows. Similarly, front-running attacks gain feasibility as quantum systems can analyze pending transactions more quickly, enabling attackers to exploit priority mechanisms and gain unfair advantages in financial or auction-based systems.

\item \textit{Medium:} Vulnerabilities include integer overflow and underflow issues and \gls{DoS} attacks on smart contracts. Integer overflow and underflow vulnerabilities depend on specific coding flaws within the smart contract, which \gls{QC} could exploit more efficiently by accelerating vulnerability detection. However, these are less likely than cryptographic issues due to the specific conditions required for exploitation. \gls{DoS} attacks, while potentially amplified by quantum computational power, are application-layer threats reliant on specific design flaws in contract execution or resource management rather than systemic cryptographic weaknesses.


\end{enumerate}

A successful attack on any smart contract vulnerability carries a \textit{High} impact rating across the board. This highlights the potential for catastrophic consequences. Exploiting these vulnerabilities could lead to financial ruin for users and businesses through stolen funds, disrupted transactions, or manipulated contract logic. Furthermore, attackers could gain access to sensitive data or inject malicious content through inter-contract communication breaches. Additionally, \gls{DoS} attacks and communication vulnerabilities can cripple smart contract functionality, disrupting workflows and causing widespread disruptions. Even seemingly less severe attacks like front-running can manipulate markets by giving attackers an unfair advantage. This \textit{High} impact rating reflects the critical role smart contracts play in \gls{BC} ecosystems. A compromised smart contract can destroy trust and security, posing significant risks to all participants.  The overall risk rating is obtained by combining likelihood and impact using the risk matrix (Figure~\ref{fig:risk-matrix}). Table~\ref{tab:smart_contract_analysis} details vulnerabilities, likelihood, impact, overall risk, and mitigation strategies. Proactive deployment of quantum-resistant cryptography, secure contract design practices, and runtime verification are essential to preserve trust in the \textit{smart contract component} under emerging quantum threats.
\end{enumerate}

\subsection{User Wallet Attacks and Quantum Computing Threats}\label{Classic-User-Wallet}
The security of user wallets is paramount for ensuring trust and confidence in \gls{BC} technology. However, the emergence of \gls{QC} poses a significant threat to user wallets, potentially leading to stolen cryptocurrency and compromised financial security~\cite{thanalakshmi2023quantum,allende2023quantum}. Here is a breakdown of the vulnerabilities user wallets face in the quantum era:

 \begin{enumerate}[topsep=1ex, itemsep=1ex, wide, font=\itshape, labelwidth=!, labelindent=0pt, label*=E.\arabic*.]
\item \textit{Private Key Exposure:} 
User wallets rely on robust cryptographic algorithms to protect private keys, which grant access to the funds within the wallet. Popular algorithms like \gls{ECDSA} are currently used for key generation and signing transactions. 
However, vulnerability of these schemes to Shor's Algorithm
could potentially render the wallet susceptible to theft, allowing attackers to gain unauthorized access, derive the private key from the public key, and siphon off associated funds. Furthermore, it could facilitate the forging of seemingly legitimate transactions without the user's knowledge or consent, resulting in unauthorized transfers.

To mitigate private key exposure in the quantum era, the following strategies are recommended: (i) adopting quantum-resistant algorithms for private key generation in wallets—research on these algorithms is ongoing with promising options emerging—ensuring that keys remain secure even under quantum adversaries~\cite{sahu2024quantum}; and (ii) deploying \gls{MPC}-based wallets integrated with post-quantum cryptographic primitives, which enables distributed key generation and signing without revealing private key material, thereby combining the benefits of decentralized trust with quantum-resistant security guarantees~\cite{10.1007/978-3-030-77870-5_16}.


\item \textit{Random Number Generators Manipulation:}  
Many wallets rely on \glspl{RNG} to create strong cryptographic keys. While traditional \glspl{RNG} might suffice currently, they could be vulnerable to manipulation by quantum computers~\cite{mannalatha2023comprehensive,jacak2021quantum}. Attackers could exploit weaknesses in \glspl{RNG} to generate predictable keys, compromising the security of user wallets.

To mitigate quantum-specific attacks on \glspl{RNG}, the following strategies can be employed: (i) deploying \glspl{DI-QRNG} that leverage quantum nonlocality to certify randomness independent of device trust assumptions, ensuring unpredictability even under quantum adversaries~\cite{liu2018device};  
(ii) applying post-processing with quantum-proof randomness extractors to eliminate statistical biases in raw entropy and produce uniformly random outputs resilient to quantum inference~\cite{berta2017quantum};  
(iii) adopting entropy sources grounded in quantum physical processes—such as photon measurements or nuclear decay—which offer intrinsic resistance to quantum tampering and enhance randomness integrity~\cite{marangon2017source}; and  
(iv) implementing continuous entropy monitoring systems capable of detecting anomalies or manipulation patterns, thereby preserving the robustness of key generation under potential quantum interference~\cite{ma2013postprocessing}.

 
 
 

\item \textit{Password Hashing Vulnerabilities:} 
User accounts are typically protected with passwords, which are hashed (one-way encrypted) for secure storage. Popular hashing algorithms like SHA 256 are currently used in password storage. However, some quantum algorithms could potentially accelerate brute-force attacks, making it easier for attackers to guess passwords by trying a large number of combinations~\cite{li2020quantum, durmuth2021towards}.

To mitigate password hashing vulnerabilities in user wallets under quantum threats, the following strategies are recommended:  
(i) transitioning to cryptographic hash functions with larger output sizes—such as SHA-3 or SHA-512—that provide sufficient bit security against Grover’s algorithm~\cite{nist2024transition};  
(ii) adopting memory-hard hashing algorithms like Argon2, which increase resource requirements through computational and memory constraints, limiting large-scale brute-force feasibility for both classical and quantum adversaries~\cite{biryukov2021rfc};  
(iii) configuring adaptive password hashing algorithms with elevated iteration counts, ensuring that computational cost scales with attacker capabilities, including quantum speedups~\cite{durmuth2021towards}; and  
(iv) enforcing strong password policies that mandate longer, more complex passwords to increase entropy and expand the effective search space, aligning with both classical and quantum-era security practices~\cite{baseri2024cybersecurityquantumeraassessing}.

\begin{table*}[!htbp]
\caption{Analysis of User Wallet Component: Quantum Threats, Vulnerabilities, Mitigations, and Risk Assessment}
\small
\resizebox{\textwidth}{!}{%
\begin{tabular}{|m{0.06\linewidth}|m{0.19\linewidth}|m{0.2\linewidth}|m{0.16\linewidth}|m{0.33\linewidth}|m{0.28\linewidth}|m{0.14\linewidth}|c|c|c|}
\hline
\textbf{Layer} & \textbf{Exploited Vulnerabilities} & \textbf{Attack Vector} & \textbf{Potential Impacts} & \textbf{STRIDE Threats with Reasoning} & \textbf{Mitigation Strategies} & \textbf{Actionable Parties} & \textbf{L} & \textbf{I} & \textbf{R} \\ \hline
\multirow{23}{*}{\begin{minipage}[c]{\linewidth}
 {User Wallet} 
\end{minipage}}
& Private Key Exposure 
& Deriving private keys from public keys using Shor's algorithm 
& \begin{myBullets} 
\item Loss of Funds 
\item Identity Theft 
\item Compromised Transactions 
\vspace{-2ex} \end{myBullets} 
& \begin{myBullets} 
\item {Spoofing:} Impersonation of legitimate users by deriving private keys and signing unauthorized transactions.
\item {Tampering:} Forging transactions using compromised private keys allows attackers to manipulate the ledger. 
 
\vspace{-2ex} \end{myBullets} 
& \begin{myBullets} 
\item Quantum-Resistant Key Generation 
\item \gls{MPC} Wallets with Post-
Quantum Cryptography~\cite{10.1007/978-3-030-77870-5_16}
 
\vspace{-2ex} \end{myBullets} 
& \begin{myBullets} 
\item Developers 
\item Service Providers 
\item End Users 
\item Auditors 
\vspace{-2ex} \end{myBullets} 
& \high & \high & \high \\ \cline{2-10}

& \glspl{RNG} Manipulation
& Manipulation of \glspl{RNG} to generate predictable keys 
& \begin{myBullets} 
\item Compromised Security 
\item Unauthorized Access 
 
\vspace{-2ex} \end{myBullets} 
& \begin{myBullets} 
\item {Tampering:} Alter or manipulation of \glspl{RNG} process to produce predictable keys, compromising wallet security.
\item {Information Disclosure:} Weak \glspl{RNG} could reveal key generation patterns, exposing cryptographic material to attackers. 
 
\vspace{-2ex} \end{myBullets} 
& \begin{myBullets} 
 \item \glspl{DI-QRNG}
 \item Post-Processing with Quantum-Proof Extractors
 \item Adoption of Entropy Sources Resistant to Quantum Tampering
 \item Continuous Entropy Monitoring 
 
\vspace{-2ex} \end{myBullets} 
& \begin{myBullets} 
\item Developers 
\item Service Providers 
\item Auditors 
\item Regulators 
\vspace{-2ex} \end{myBullets} 
& \med & \high & \high \\ \cline{2-10}

& Password Hashing Vulnerabilities 
& Accelerated brute-force attacks on hashed passwords 
& \begin{myBullets} 
\item Data Breach 
\item Unauthorized Transactions 
 
\vspace{-2ex} \end{myBullets} 
& \begin{myBullets} 
\item {Repudiation:} Attackers or users could deny responsibility for compromised accounts, especially in cases of weak password policies.
\item {Information Disclosure:} Quantum-powered brute-force attacks could reveal hashed passwords, granting unauthorized access. 
 
\vspace{-2ex} \end{myBullets} 
& \begin{myBullets} 
 \item Cryptographic Hash Functions with Adequate Bit Security,
 \item Memory-Hard Functions,
 \item Increased Hash Iterations with Adaptive Difficulty,
 \item Password Policies with Enhanced Complexity. 
\vspace{-2ex} \end{myBullets} 
& \begin{myBullets} 
\item Developers 
\item Service Providers 
\item End Users 
\item Auditors 
\vspace{-2ex} \end{myBullets} 
& \med & \high & \high \\ \cline{2-10}

& Transaction Interception
& Altering transaction data leading to theft or redirection of funds 
& \begin{myBullets} 
\item Financial Loss 
\item Data Manipulation 
 
\vspace{-2ex} \end{myBullets} 
& \begin{myBullets} 
\item {Tampering:} Quantum attackers could alter transaction data in transit, redirecting funds or changing transaction details.
\item {Information Disclosure:} Intercepted transactions could expose sensitive data, such as recipient addresses and transaction amounts. 
 
\vspace{-2ex} \end{myBullets} 
& \begin{myBullets} 
\item Quantum-Resistant Encryption Protocols
\item Enhanced Transaction Security 
\item \gls{BC} Integrity Measures 
\item Quantum-Resistant Transaction Verification
\item Secure Transmission Channels
 
\vspace{-2ex} \end{myBullets} 
& \begin{myBullets} 
\item Developers 
\item Service Providers 
\item Auditors 
\item Node Operators 
\item Regulators 
\vspace{-2ex} \end{myBullets} 
& \high & \high & \high \\ \hline

\end{tabular}%
}  \vspace{-0.2cm}
\label{tab:user_wallet_analysis}
\end{table*}

\item \textit{Transaction Interception:}  
Quantum attackers leveraging advanced quantum computational capabilities could exploit vulnerabilities in traditional cryptographic protocols to intercept or tamper with transaction data during transmission. This poses significant risks, including theft or redirection of funds, unauthorized modifications, and exposure of sensitive transactional details. Quantum-specific attacks focus on breaking encryption or exploiting flaws in transaction validation mechanisms.

Mitigation strategies against transaction interception vulnerabilities in quantum-capable threat environments include:
(i) integrating quantum-resistant encryption protocols to secure data in transit, ensuring intercepted transaction content remains indecipherable and tamper-proof even under quantum computation~\cite{chen2016report};
(ii) reinforcing \gls{BC} integrity using cryptographic constructs such as Merkle trees, threshold cryptography, and secure multiparty computation, which safeguard transaction authenticity and integrity and make it computationally infeasible for quantum adversaries to tamper with \gls{BC} data~\cite{truong2018strengthening};
(iii) deploying robust quantum-safe verification mechanisms—such as hash-based digital signatures and quantum-resistant \glspl{ZKP}—to maintain transaction validity and ensure they remain untampered under quantum adversarial conditions~\cite{allende2023quantum}; and
(iv) establishing secure transmission channels with post-quantum TLS protocols between users and \gls{BC} nodes, preventing eavesdropping and mitigating interception and injection threats during transaction propagation~\cite{refc17}.







\item \textit{{Risk Assessment for Quantum Threats to User Wallet Component}:} 
To assess the risks posed by \gls{QC} advancements to user wallets, we evaluate the likelihood and impact of potential attacks using criteria established in Tables~\ref{table:likelihood} and~\ref{table:impact}. 
The following breakdown summarizes the likelihood categories for user wallet vulnerabilities:
\begin{enumerate}[itemsep=0ex]
 \item \textit{High:} Private key exposure represents a critical vulnerability with a high likelihood of exploitation as \gls{QC} advancements progress. Shor’s algorithm makes it feasible to derive private keys from public keys, posing a severe threat to wallet security. Additionally, transaction interception could also fall into the high-likelihood category. Advanced quantum systems might compromise encryption protocols, allowing attackers to intercept or alter transaction data. While current secure communication protocols mitigate this risk, insufficiently updated encryption methods will be vulnerable to future quantum threats.
 \item \textit{Medium:} \glspl{RNG} manipulation is a potential threat, with quantum systems capable of exploiting weaknesses in traditional \glspl{RNG} to generate predictable keys. Although robust quantum-resistant random number generation techniques significantly reduce this risk, poorly implemented \glspl{RNG} remain vulnerable. Password hashing vulnerabilities are also categorized as medium likelihood, as Grover’s algorithm could accelerate brute-force attacks, making it easier for attackers to exploit weak password hashing implementations. However, the adoption of strong password policies, memory-hard hashing algorithms, and cryptographic schemes with higher security margins can help mitigate this threat.

\end{enumerate}
 A successful attack on a user wallet vulnerability carries a \textit{High} impact rating. Exploiting these vulnerabilities could lead to catastrophic consequences for users. This includes loss of funds, identity theft, and compromised transactions. Attackers could gain unauthorized access to user accounts, steal funds, or manipulate transaction data for personal gain. Overall risk is determined by combining likelihood and impact using the matrix in Figure~\ref{fig:risk-matrix}. Table~\ref{tab:user_wallet_analysis} details  vulnerabilities, likelihood, impact, overall risk, and mitigation strategies. Ensuring resilience of the \textit{user wallet component} in the quantum era requires migration to \gls{PQC} primitives, robust key management, and the adoption of secure entropy sources.
\end{enumerate}

\section{Roles and Responsibilities in Mitigating \gls{QC} Impacts on \gls{BC}}\label{sec:roles}

\begin{table*}[!hbpt]
\caption{Roles and Responsibilities in Mitigating \gls{QC} Impacts on \gls{BC}}\label{table:roles}
\small
\renewcommand{\arraystretch}{1.2}
\resizebox{\textwidth}{!}{%
\begin{tabular}{|m{0.1\linewidth}|m{0.12\linewidth}|m{0.52\linewidth}|m{0.27\linewidth}|m{0.25\linewidth}|m{0.32\linewidth}|}
\hline
\multirow{1}{*}{\textbf{Role}} & \multirow{1}{*}{\textbf{Attack Vector}} & \multirow{1}{*}{\textbf{STRIDE Threats}} & \multirow{1}{*}{\textbf{Quantum Implications}} & \multirow{1}{*}{\textbf{Possible Vulnerabilities}} & \multirow{1}{*}{\textbf{Quantum-Resistance Measures}} \\
\hline
\multirow{12}{*}{Miners} & Quantum Accelerated Mining & \begin{myBullets} 
\item Spoofing: Quantum speed allows impersonation within mining pools.
\item Tampering: Quantum-accelerated hashing enables modification of blocks or mining history.
\vspace{-2ex} \end{myBullets} & Quantum computers may significantly speed up mining processes, especially in \gls{PoW}, potentially disrupting consensus by enabling faster hash solving~\cite{bard2022quantum,scharer2023quantum}. & Higher vulnerability to 51\% attacks if mining power is concentrated among quantum-equipped miners~\cite{fernandez2020towards,sayeed2019assessing}. & Adopting longer hash outputs, quantum-resistant hashing algorithms, and hybrid cryptographic techniques can help mitigate mining centralization~\cite{shekhawat2024quantum,khodaiemehr2023navigating}. \\
\cline{2-6}
 & Threat to Consensus Mechanisms & \begin{myBullets} 
\item Spoofing: Quantum power enables impersonation of network nodes.
\item Tampering: Quantum attacks allow altering transaction data.
\item Repudiation: Attackers may deny responsibility for malicious transactions.
\item Info. Disclosure: Decryption powered by quantum technology could expose private transaction data.
\item \gls{DoS}: High-speed processing enables flooding of network nodes.
\item Elevation of Privilege: Quantum power could give attackers control over nodes.
\vspace{-2ex} \end{myBullets} & \gls{QC}'s ability to break cryptographic primitives could undermine \gls{PoW} and \gls{PoS} security and affect scalability in \gls{BFT} systems~\cite{fernandez2020towards,gomes2023fortifying}. & Cryptographic vulnerabilities in consensus algorithms and scalability issues in \gls{BFT} protocols~\cite{sinai2024performance,chen2021construction}. & Development of quantum-resistant cryptographic algorithms, hybrid cryptographic systems, and restructuring \gls{BFT} protocols for quantum resilience~\cite{baseri2024evaluation,khodaiemehr2023navigating}. \\
\cline{2-6}
 & Energy Consumption & \begin{myBullets} 
\item Elevation of Privilege: Quantum efficiency enables unauthorized operations at lower energy costs.
\vspace{-2ex} \end{myBullets} & \gls{QC} may improve mining efficiency, potentially reducing energy consumption, though security implications remain critical~\cite{khodaiemehr2023navigating,chen2021construction}. & Trade-offs between energy efficiency and security as quantum capabilities increase~\cite{bard2022quantum}. & Design of low-energy quantum-resistant protocols for sustainable \gls{BC} mining~\cite{scharer2023quantum}. \\
\hline
\multirow{13}{*}{\begin{minipage}[c]{\linewidth}
 Service Providers
\end{minipage}}& Quantum Cryptanalysis & \begin{myBullets} 
\item Spoofing: Quantum attacks could enable impersonation of legitimate service providers.
\item Tampering: Decryption allows unauthorized data modifications.
\item Info. Disclosure: Breaking encryption exposes sensitive data.
\vspace{-2ex} \end{myBullets} & Quantum attacks on encryption expose sensitive data and allow unauthorized access~\cite{shor1999polynomial,grover1996fast}. & Encryption vulnerabilities compromise data confidentiality and integrity~\cite{shor1999polynomial,grover1996fast}. & Implementation of quantum-resistant and hybrid cryptographic approaches~\cite{shor1999polynomial,grover1996fast}. \\
\cline{2-6}

 & Smart Contract Vulnerabilities & \begin{myBullets} 
\item Tampering: Quantum attacks enable unauthorized changes to contract states.
\item Repudiation: Attackers may deny malicious actions within contracts.
\item Info. Disclosure: Quantum attacks could expose confidential contract data.
\item Elevation of Privilege: Quantum bypasses enable unauthorized access to contract functions.
\vspace{-2ex} \end{myBullets} & Quantum threats could compromise contract integrity, allowing unauthorized control or data manipulation~\cite{akhai2024quantum,gharavi2024post}. & Increased risk of smart contract exploitation due to compromised cryptographic primitives~\cite{akhai2024quantum,gharavi2024post}. & Adoption of quantum-resistant cryptographic protocols, formal verification~\cite{ethereum_formal_verification,lewis2023formal,tolmach2023securing}, and regular auditing~\cite{akhai2024quantum,gharavi2024post}. \\
\cline{2-6}

 & Data Confidentiality & \begin{myBullets} 
\item Spoofing: Quantum attacks enable impersonation of authorized data handlers.
\item Tampering: Unauthorized modifications to stored data due to broken encryption.
\item Info. Disclosure: Quantum decryption reveals sensitive stored data.
\vspace{-2ex} \end{myBullets} & Quantum attacks threaten data confidentiality by breaking encryption~\cite{shuaib2022effect}. & Risk of sensitive data exposure and data manipulation~\cite{shuaib2022effect}. & Post-quantum cryptographic algorithms for data protection~\cite{shuaib2022effect}. \\
\cline{2-6}

 & Access Control Compromise & \begin{myBullets} 
\item Spoofing: Quantum-assisted impersonation bypasses access controls.
\item Elevation of Privilege: Breaking cryptography grants unauthorized access.
\vspace{-2ex} \end{myBullets} & Quantum attacks could undermine role-based access and multi-signature wallets~\cite{siddiqui2023smart, wang2019blockchain}. & Unauthorized access to sensitive functions or data~\cite{siddiqui2023smart, wang2019blockchain}. & Quantum-resistant multi-signature schemes and hybrid authentication protocols~\cite{siddiqui2023smart, wang2019blockchain}. \\
\hline

\multirow{12}{*}{End Users} & Quantum Key Extraction (Wallet Security Compromise) & \begin{myBullets} 
\item {Spoofing}: Quantum decryption enables impersonation of wallets.
\item {Tampering}: Unauthorized transactions using compromised private keys.
\item {Info. Disclosure}: Quantum attacks reveal private keys.
\vspace{-2ex} \end{myBullets} & Quantum attacks on ECC can expose private keys, compromising wallet security~\cite{kethepalli2023reinforcing}. & Private keys and wallet data are vulnerable, risking asset security~\cite{kethepalli2023reinforcing}. & Quantum-resistant wallets with advanced cryptography and hardware wallets~\cite{kethepalli2023reinforcing}. \\
\cline{2-6}
& Transaction Verification & \begin{myBullets} 
\item {Tampering}: Quantum attacks enable unauthorized transaction modifications.
\item {Repudiation}: Attackers may deny responsibility for fraudulent transactions.
\item {Elevation of Privilege}: Unauthorized transaction approvals via compromised signatures.
\vspace{-2ex} \end{myBullets} & Quantum threats may compromise digital signatures, affecting transaction integrity~\cite{gomes2023fortifying}. & Risk of fraud and financial loss due to compromised transaction verification~\cite{gomes2023fortifying}. & Quantum-resistant digital signatures, multi-signature protocols, and threshold cryptography~\cite{gomes2023fortifying}. \\
\cline{2-6}

& Phishing Attacks & \begin{myBullets} 
\item {Spoofing}: Quantum attacks enable impersonation of trusted sources.
\item {Info. Disclosure}: Quantum attacks expose credentials by breaking secure channels.
\vspace{-2ex} \end{myBullets} & Quantum threats to secure channels increase phishing risks and exposure of credentials~\cite{mohanty2023comprehensive,fahim2022comparative}. & Higher risk of credential theft and unauthorized access~\cite{mohanty2023comprehensive}. & Quantum-resistant communication protocols and user education on phishing risks~\cite{mohanty2023comprehensive}. \\
\cline{2-6}

& Privacy Breaches & \begin{myBullets} 
\item {Info. Disclosure}: Quantum decryption reveals user identities and transaction history.
\item {Spoofing}: Quantum-assisted impersonation could lead to data leaks.
\vspace{-2ex} \end{myBullets} & Quantum decryption could reveal user identities and transaction history, risking privacy~\cite{fernandez2020towards,malina2021post}. & Compromised anonymity and data confidentiality~\cite{zhang2019security}. & Quantum-resistant privacy technologies like \glspl{ZKP} and data anonymization~\cite{fernandez2020towards}. \\
\hline
\multirow{13}{*}{\begin{minipage}[c]{\linewidth}
 Developers
\end{minipage}} & Quantum Cryptanalysis & \begin{myBullets} 
\item Spoofing: Impersonation of smart contract owners or developers to deploy malicious updates.
\item Tampering: Unauthorized modifications to cryptographic protocols or contract logic using quantum-powered attacks.
\vspace{-2ex} \end{myBullets} & \gls{QC} undermines cryptographic primitives (e.g., RSA, ECC) used in \gls{BC} protocols and smart contracts~\cite{shor1999polynomial}. & Risk of protocol compromise, unauthorized updates, and systemic security failures if cryptographic primitives are not updated. & Adoption of lattice-based and hash-based cryptography, hybrid cryptographic schemes, and integration of post-quantum cryptographic libraries~\cite{vasani2024embracing}. \\
\cline{2-6}

 & Digital Signature Vulnerabilities & \begin{myBullets} 
\item Spoofing: Impersonation of valid signers in multi-signature schemes.
\item Repudiation: Attackers deny responsibility for signed malicious transactions.
\item Info. Disclosure: Exposure of private keys due to quantum decryption.
\vspace{-2ex} \end{myBullets} & Digital signatures, critical for authentication and authorization, are rendered insecure by quantum attacks~\cite{fernandez2020towards,karbasi2020post}. & Compromised transaction authenticity, multi-signature wallets, and node authorization in consensus protocols. & Adoption of quantum-resistant digital signatures, such as lattice-based schemes, and regular validation of security mechanisms~\cite{karbasi2020post}. \\
\cline{2-6}

& Transition to Quantum-Resistant Solutions & \begin{myBullets} 
\item Tampering: Exploitation of legacy cryptographic systems during the transition phase.
\item Elevation of Privilege: Attackers exploit unpatched vulnerabilities in transitioning protocols.
\vspace{-2ex} \end{myBullets} & Updating smart contracts and protocols to quantum-resistant algorithms is necessary but may introduce temporary security gaps~\cite{vasani2024embracing}. & Outdated cryptographic primitives, compatibility issues, and risks of incomplete migration. & Phased updates to quantum-resistant solutions, fallback mechanisms, testing in simulated environments, and collaboration with standardization bodies~\cite{vasani2024embracing}. \\
\cline{2-6}

 & Testing and Validation & \begin{myBullets} 
\item Info. Disclosure: Weaknesses in quantum-resistant algorithms could expose vulnerabilities.
\item Tampering: Exploitation of untested or improperly validated quantum-resistant cryptographic schemes.
\vspace{-2ex} \end{myBullets} & Quantum-resistant algorithms require rigorous testing to ensure practicality, resilience, and scalability in \gls{BC} environments. & Vulnerabilities due to improperly tested algorithms and reduced system performance during integration. & Rigorous testing in controlled environments, performance benchmarking, and phased deployment~\cite{vasani2024embracing}. \\
\hline
\multirow{12}{*}{\begin{minipage}[c]{\linewidth}
 Node Operators
\end{minipage}} & Quantum Hashing Attacks & \begin{myBullets}
\item Tampering: Breaking traditional hashes enables modification of transaction history.
\item Spoofing: Impersonation of valid nodes or miners using quantum-assisted hash manipulation.
\vspace{-2ex} \end{myBullets} & Grover’s algorithm halves the effective security of hashing algorithms (e.g., SHA 256), reducing resistance to brute-force attacks~\cite{kearney2021vulnerability}. & Vulnerability to tampering with \gls{BC} data, weakening of \gls{PoW} consensus mechanisms, and potential double-spending risks. & Adoption of quantum-resistant hashing algorithms, such as sponge constructions and hash-based designs, and updates to \gls{PoW} systems~\cite{yang2022decentralization}. \\
\cline{2-6}

 & Data Integrity Risks & \begin{myBullets}
\item Info. Disclosure: Breaking encryption exposes sensitive data, including private keys.
\item Tampering: Quantum attacks enable unauthorized modifications to \gls{BC} or node data.
\vspace{-2ex} \end{myBullets} & Quantum attacks on encryption (e.g., RSA, ECC) can compromise stored and transmitted \gls{BC} data, risking confidentiality and integrity~\cite{sahu2024quantum}. & Risks to private key security, data confidentiality, and historical transaction accuracy. & Adoption of lattice-based and hash-based cryptography for encryption, secure key management, and data storage protections~\cite{sahu2024quantum}. \\
\cline{2-6}

 & \gls{DoS} Attacks & \begin{myBullets}
\item \gls{DoS}: Quantum-assisted attacks enable large-scale traffic flooding or resource exhaustion.
\item Spoofing: Exploitation of compromised authentication to launch coordinated attacks.
\vspace{-2ex} \end{myBullets} & \gls{QC} enhances the scalability and sophistication of \gls{DoS} attacks by accelerating resource exhaustion techniques~\cite{mangla2023mitigating}. & Disruption of node operations, reduced service availability, and potential node isolation. & Adoption of decentralized \gls{DoS} mitigation systems, quantum-resistant authentication protocols, and enhanced network-layer protections~\cite{mangla2023mitigating}. \\
\cline{2-6}

 & Unauthorized Access & \begin{myBullets}
\item Elevation of Privilege: Breaking cryptographic protections to gain unauthorized access to nodes.
\item Tampering: Compromised nodes enable manipulation of network data and configurations.
\vspace{-2ex} \end{myBullets} & Quantum decryption of private keys or authentication data may allow attackers to bypass security measures and gain control of node infrastructure~\cite{putz2019secure}. & Risks of unauthorized access, data breaches, manipulation of network state, and disruption of node operations. & Implementation of quantum-resistant authentication protocols, multi-factor authentication, regular security audits, and proactive infrastructure monitoring~\cite{putz2019secure}. \\
\hline
\end{tabular}}
\end{table*}

\addtocounter{table}{-1}

\begin{table*}[!htbp]
\small
\caption{(Cont.) Roles and Responsibilities in Mitigating \gls{QC} Impacts on \gls{BC}}
\renewcommand{\arraystretch}{1.2}
\resizebox{\textwidth}{!}{%
\begin{tabular}{|m{0.1\linewidth}|m{0.12\linewidth}|m{0.52\linewidth}|m{0.27\linewidth}|m{0.25\linewidth}|m{0.32\linewidth}|}
\hline
\multirow{1}{*}{\textbf{Role}} & \multirow{1}{*}{\textbf{Attack Vector}} & \multirow{1}{*}{\textbf{STRIDE Threats}} & \multirow{1}{*}{\textbf{Quantum Implications}} & \multirow{1}{*}{\textbf{Possible Vulnerabilities}} & \multirow{1}{*}{\textbf{Quantum-Resistance Measures}} \\ \hline
\multirow{16}{*}{\begin{minipage}[c]{\linewidth}
 Regulators \& Compliance Authorities
\end{minipage}} & Encryption Vulnerabilities & \begin{myBullets}
\item Info. Disclosure: Quantum attacks expose encrypted communications between regulators and stakeholders.
\item Tampering: Compromised encryption allows alteration of regulatory data or directives.
\item Repudiation: Attackers deny responsibility for breaches or altered communications.
\vspace{-2ex} \end{myBullets} & Quantum computers can break cryptographic standards like RSA and ECC, requiring regulators to transition to quantum-resistant encryption~\cite{li2019toward}. & Risks to secure communications, data integrity, and enforcement of cryptographic compliance standards. & Adoption of post-quantum encryption standards, collaboration with global standardization bodies, and secure cryptographic audits~\cite{li2019toward}. \\
\cline{2-6}

 & Data Security Risks & \begin{myBullets}
\item Info. Disclosure: Breaking encryption exposes sensitive regulatory and compliance data, including \gls{KYC} records.
\item Tampering: Quantum attacks enable unauthorized modification of stored or transmitted data.
\item Spoofing: Impersonation of regulatory authorities to issue fraudulent compliance directives.
\vspace{-2ex} \end{myBullets} & Quantum attacks compromise encryption, creating vulnerabilities in regulatory systems and sensitive data~\cite{sahu2024quantum}. & Breaches of \gls{KYC} records, manipulation of compliance records, and impersonation of regulatory entities, leading to compliance failures. & Implementation of quantum-resistant encryption, secure storage mechanisms, and periodic audits of regulatory systems~\cite{sahu2024quantum}. \\
\cline{2-6}

 & Legal Framework Updates & \begin{myBullets}
\item Spoofing: Impersonation of regulators to issue fraudulent directives or policies.
\item Tampering: Unauthorized changes to compliance or legal frameworks using quantum-enhanced tools.
\item Elevation of Privilege: Exploiting outdated regulatory frameworks to bypass enforcement mechanisms.
\vspace{-2ex} \end{myBullets} & Quantum threats necessitate updates to cybersecurity and data protection laws to address cryptographic obsolescence and emerging quantum vulnerabilities~\cite{kong2024realizing}. & Outdated legal frameworks that fail to mandate quantum-resistant systems and enforce compliance standards. & Regular updates to legal frameworks, phased mandates for quantum-resistant cryptographic systems, and post-quantum readiness audits~\cite{kong2024realizing}. \\
\cline{2-6}

 & Privacy Mechanism Vulnerabilities & \begin{myBullets}
\item Info. Disclosure: Quantum attacks expose anonymized or privacy-protected data.
\item Spoofing: Impersonation of entities to exploit privacy-preserving tools.
\item Repudiation: Attackers deny responsibility for breaches of privacy-preserving systems.
\vspace{-2ex} \end{myBullets} & Quantum capabilities weaken privacy-preserving mechanisms, such as \glspl{ZKP} and mixers, potentially undermining \gls{AML} and \gls{KYC} compliance~\cite{derose2022establishing}. & Non-compliance of privacy tools with regulatory standards, exposure of sensitive user data, and challenges balancing privacy and \gls{AML} requirements. & Enhanced oversight of privacy mechanisms, integration of quantum-resistant privacy tools, and regular technical audits to ensure compliance~\cite{derose2022establishing}. \\
\hline
\multirow{18}{*}{\begin{minipage}[c]{\linewidth}
 Auditors
\end{minipage}}
 & Quantum-Accelerated Cryptanalysis & \begin{myBullets}
\item Tampering: Manipulation of \gls{BC} protocols or audit processes using quantum decryption.
\item Info. Disclosure: Exposure of cryptographic weaknesses and sensitive audit findings.
\vspace{-2ex} \end{myBullets} & Quantum computers can break cryptographic primitives (e.g., RSA, ECC), requiring auditors to assess quantum-resistant solutions~\cite{white2020blockchain}. & Ineffective detection of cryptographic vulnerabilities and outdated audit practices that fail to address quantum risks. & Development of quantum-aware audit frameworks, adoption of \gls{NIST} post-quantum standards, and regular audits for cryptographic resilience~\cite{kahyaouglu2023evaluation}. \\
\cline{2-6}

 & Transition Planning & \begin{myBullets}
\item Spoofing: Impersonation of auditors to exploit migration vulnerabilities.
\item Info. Disclosure: Quantum decryption exposes sensitive transition strategies.
\item Tampering: Exploitation of unpatched or legacy systems during migration phases.
\vspace{-2ex} \end{myBullets} & Auditors guide \gls{BC} projects through transitions to quantum-resistant cryptography, mitigating migration-specific risks~\cite{ciulei2022preparation}. & Flawed migration plans, incomplete updates, and temporary compatibility issues during transitions. & Comprehensive evaluation of transition strategies, phased migration plans, and adoption of hybrid cryptographic systems to ensure smooth transitions~\cite{kahyaouglu2023evaluation}. \\
\cline{2-6}

 & Post-Quantum Readiness Assessments & \begin{myBullets}
\item Tampering: Exploitation of improperly deployed quantum-resistant implementations.
\item Spoofing: Impersonation of auditors during readiness testing or compliance audits.
\item Info. Disclosure: Exposure of vulnerabilities during assessments or simulated attacks.
\vspace{-2ex} \end{myBullets} & Auditors test quantum-resistant solutions for effectiveness and compliance with evolving standards~\cite{kahyaouglu2023evaluation}. & Improperly deployed quantum-resistant systems and inadequate alignment with post-quantum standards. & Simulating quantum-based attacks, validating system alignment with regulatory frameworks, and conducting ongoing compliance audits~\cite{white2020blockchain}. \\
\cline{2-6}

 & Collaborative Security Reviews & \begin{myBullets}
\item Tampering: Misaligned collaboration leaves vulnerabilities unaddressed.
\item Info. Disclosure: Exposure of sensitive findings during collaborative processes.
\item Repudiation: Stakeholders deny responsibility for implementing quantum-resistant solutions.
\vspace{-2ex} \end{myBullets} & Collaboration ensures robust quantum-resistant designs and alignment with technical and regulatory standards~\cite{kahyaouglu2023evaluation}. & Miscommunication, unaligned standards, or incomplete reviews can weaken system resilience. & Joint reviews with developers, cryptographers, and regulators to validate quantum-resistant designs and enforce global standards~\cite{kahyaouglu2023evaluation}. \\
\cline{2-6}

 & Skill Development and Training & \begin{myBullets}
\item Inadequate training leads to ineffective audits and overlooked vulnerabilities.
\item Exploitation of outdated tools and methodologies by attackers.
\vspace{-2ex} \end{myBullets} & Auditors must acquire expertise in \gls{PQC}, quantum-specific auditing tools, and advanced security frameworks~\cite{kahyaouglu2023evaluation}. & Insufficient expertise compromises audits and weakens assessments of post-quantum readiness. & Continuous skill development through workshops, certifications, and training in quantum-resistant practices~\cite{kahyaouglu2023evaluation}. \\
\hline
\multirow{18}{*}{\begin{minipage}[c]{\linewidth}
 Governance Participants
\end{minipage}}
 & Quantum-Supported Manipulation & \begin{myBullets}
\item Spoofing: Forgery of governance participant identities to disrupt decision-making.
\item Tampering: Manipulation of governance decisions or proposals.
\vspace{-2ex} \end{myBullets} & \gls{QC} undermines cryptographic protections, enabling identity forgery and manipulation of decision-making processes~\cite{johnson2019governance}. & Risk of unauthorized governance decisions, identity spoofing, and governance manipulation affecting system trust. & Adoption of quantum-resistant identity verification, secure quorum mechanisms, and tamper-proof governance protocols~\cite{johnson2019governance}. \\
\cline{2-6}

 & Decentralized Governance Systems & \begin{myBullets}
\item Spoofing: Quantum attackers forge participant identities to influence decision-making.
\item Tampering: Alteration of governance-related data, outcomes, or rules.
\item Repudiation: Denial of responsibility for fraudulent governance actions.
\vspace{-2ex} \end{myBullets} & Quantum threats could compromise decentralized governance systems, disrupting transparency, accountability, and trust~\cite{malik2023building}. & Risk of governance manipulation, unauthorized access, and reduced confidence in decentralized systems. & Implementation of quantum-resistant cryptographic protocols, tamper-proof governance mechanisms, and enhanced audit trails~\cite{malik2023building}. \\
\cline{2-6}

 & Voting Mechanisms & \begin{myBullets}
\item Info. Disclosure: Quantum attacks expose confidential votes or voter identities.
\item Tampering: Unauthorized alteration of cast votes or voting outcomes.
\item Spoofing: Impersonation of voters to skew results or manipulate governance.
\vspace{-2ex} \end{myBullets} & \gls{QC} compromises vote confidentiality and integrity, enabling manipulation of governance outcomes~\cite{aurangzeb2024enhancing}. & Privacy breaches, vote manipulation, vote-buying, coercion, and loss of confidence in governance outcomes. & Adoption of quantum-resistant cryptographic techniques and advanced privacy-preserving mechanisms like quantum-resistant \glspl{ZKP}~\cite{aurangzeb2024enhancing}. \\
\cline{2-6}

 & Smart Contract Execution & \begin{myBullets}
\item Tampering: Quantum attackers disrupt governance-related transactions or smart contract execution.
\item Elevation of Privilege: Unauthorized execution of governance-related actions.
\vspace{-2ex} \end{myBullets} & Quantum threats compromise governance-related smart contracts, enabling transaction manipulation, unauthorized actions, or financial tampering~\cite{ma2024blockchain}. & Risk of disrupted decision-making processes, treasury manipulation, and governance system failures. & Implementation of quantum-resistant smart contracts, formal verification practices, and secure governance transaction frameworks~\cite{ma2024blockchain}. \\
\cline{2-6}

 & Quorum Formation and Proposal Verification & \begin{myBullets}
\item Spoofing: Forged identities disrupt quorum formation, enabling Sybil attacks.
\item Tampering: Quantum attacks alter governance proposals or their verification.
\vspace{-2ex} \end{myBullets} & \gls{QC} enables identity forgery and tampering, undermining quorum legitimacy and proposal verification~\cite{kassen2023blockchain}. & Loss of governance integrity, fraudulent proposals, and weakened trust in decision-making structures. & Adoption of quantum-resistant identity verification systems, tamper-proof proposal mechanisms, and decentralized governance frameworks~\cite{kassen2023blockchain}. \\
\hline
\multirow{12}{*}{\begin{minipage}[c]{\linewidth}
 Oracles
\end{minipage}} & Quantum Tampering & \begin{myBullets} 
\item Tampering: Manipulation of external data feeds from multiple sources.
\item Info. Disclosure: Interception of sensitive oracle communications.
\item Spoofing: Impersonation of legitimate oracles to deliver fraudulent data.
\vspace{-2ex} \end{myBullets} & Quantum attacks compromise cryptographic protections, enabling manipulation or interception of oracle data, undermining trust in smart contract operations~\cite{leng2020blockchain,johnson2019governance}. & Risk of corrupted or falsified data feeds, compromised multi-oracle systems, and disrupted decision-making in oracle-driven applications. & Adoption of quantum-resistant cryptographic protocols, tamper-proof data verification mechanisms, and secure multi-source aggregation to prevent single-point failures~\cite{chawla2023roadmap,li2020quantum}. \\
\cline{2-6}

 & Quantum Eavesdropping & \begin{myBullets} 
\item Info. Disclosure: Quantum-enabled interception of data during transmission.
\vspace{-2ex} \end{myBullets} & Secure communication channels between oracles and \gls{BC} systems are vulnerable to quantum interception, exposing sensitive data or disrupting workflows~\cite{chawla2023roadmap}. & Exposure of confidential data, leading to breaches of contract logic and manipulation of \gls{BC} processes reliant on oracle inputs. & Transition to quantum-safe communication protocols and continuous monitoring of data flow integrity~\cite{li2020quantum}. \\
\cline{2-6}

 & Quantum-Compromised Smart Contracts & \begin{myBullets} 
\item Tampering: Manipulation of oracle-driven smart contract executions.
\item Elevation of Privilege: Unauthorized execution of high-priority actions through compromised inputs.
\vspace{-2ex} \end{myBullets} & Quantum attacks undermine the cryptographic foundations of smart contracts that depend on oracle inputs, disrupting contract execution and financial operations~\cite{karbasi2020post,fernandez2020towards}. & Disruption of contract reliability, unauthorized fund transfers, and manipulation of automated workflows in oracle-driven systems. & Implementing quantum-resistant smart contracts, regularly auditing oracle integrations, incorporating fail-safe mechanisms for data integrity, and fostering collaborative development with \gls{BC} stakeholders~\cite{karbasi2020post}. \\
\hline
\multirow{12}{*}{\begin{minipage}[c]{\linewidth}
 Community Participants
\end{minipage}} & Cryptographic Breaks & \begin{myBullets}
\item Tampering: Manipulation of transactions or digital signatures.
\item Spoofing: Forgery of participant identities.
\vspace{-2ex} \end{myBullets} & Quantum attacks on cryptography could enable identity forgery and transaction manipulation~\cite{gharavi2024post}. & Loss of trust in \gls{BC} systems, compromised digital signatures, and stolen participant identities or private keys. & Advocacy for the adoption of quantum-resistant cryptography by developers, active participation in governance discussions, and promotion of secure digital identity systems~\cite{khodaiemehr2023navigating}. \\
\cline{2-6}

 & Privacy Breaches & \begin{myBullets}
\item Info. Disclosure: Exposure of transaction data and identities.
\item Spoofing: Impersonation of users or community members.
\vspace{-2ex} \end{myBullets} & Quantum-enabled decryption threatens confidentiality of \gls{BC} transactions~\cite{malina2021post}. & Loss of user anonymity, exposure of sensitive data, and reduced trust in public \glspl{BC}. & Promotion of privacy-preserving technologies like \glspl{ZKP}, secure \glspl{MPC}, and community-driven awareness campaigns on privacy risks~\cite{ikeda2018security}. \\
\cline{2-6}

 & Misinformation & \begin{myBullets}
\item Tampering: Manipulation of educational materials or \gls{BC} narratives.
\item Spoofing: Creation of false sources to spread quantum-related misinformation.
\vspace{-2ex} \end{myBullets} & Misinformation about quantum vulnerabilities can cause panic and distrust in \gls{BC} ecosystems~\cite{radanliev2024cyber}. & Erosion of trust, spread of incorrect security measures, and delayed adoption of quantum-safe technologies. & Transparent communication, fact-checking initiatives, collaboration with trusted experts, and leveraging social media to dispel myths~\cite{radanliev2024cyber}. \\
\cline{2-6}

 & Educational Gaps & \begin{myBullets}
\item Info. Disclosure: Lack of knowledge delays the adoption of protective measures.
\item Repudiation: Failure to provide accurate and accessible educational resources.
\vspace{-2ex} \end{myBullets} & Limited awareness of quantum risks leaves the community unprepared for emerging threats~\cite{ku2019awareness}. & Vulnerabilities due to delayed adoption of quantum-safe practices, especially among non-technical participants. & Community-led educational initiatives like webinars, forums, outreach programs, and collaborative workshops to spread awareness~\cite{radanliev2024cyber}. \\
\hline
\end{tabular}}
\end{table*}

\gls{BC} technology relies on cryptographic techniques to ensure security and integrity. Each role within the ecosystem—miners, service providers, end users, developers, regulators, auditors, and community participants—faces unique quantum-related challenges, necessitating tailored strategies to address potential vulnerabilities and ensure resilience.

Miners need to adapt their mining algorithms to incorporate quantum-resistant designs, effectively addressing the risks posed by quantum-accelerated mining and the potential for centralization.
Service providers play a critical role in implementing quantum-resistant cryptography, securing smart contracts, and ensuring data confidentiality and integrity. End users should enhance wallet security, rigorously verify transactions, and adopt measures to mitigate quantum-assisted phishing attacks.
Developers bear the responsibility of transitioning \gls{BC} protocols and smart contracts to quantum-resistant cryptographic algorithms. They must conduct rigorous testing of these solutions and collaborate with cryptographic experts to ensure scalability, security, and resilience. Regulators need to update legal frameworks to address the implications of \gls{QC}, enforce quantum-resistant standards, and monitor \gls{BC} ecosystems for emerging vulnerabilities.
Auditors play a vital role in assessing \gls{BC} systems for quantum readiness. By evolving audit methodologies, conducting quantum readiness assessments, and guiding the transition to secure solutions, auditors help maintain system integrity.
Community participants contribute by raising awareness, advocating for quantum-resistant practices, educating users, and supporting open-source development, conducting security audits, and promoting the adoption of best practices.

Table~\ref{table:roles} details the challenges and mitigation strategies associated with each role, emphasizing the necessity of quantum-resistant cryptography, advanced security protocols, and proactive risk management. By addressing these challenges and collaborating with standardization bodies such as \gls{NIST}, the \gls{BC} ecosystem can maintain its security and resilience in the face of \gls{QC} advancements.


\section{Challenges and Strategies After Transitioning to Quantum-Resistant \gls{BC} Systems}\label{sec:navigate}
 While the integration of quantum-resistant algorithms seems like a natural solution, the transition itself introduces a new wave of challenges \cite{mashatan2021complex}. This section delves into the complexities organizations face after transitioning their \gls{BC} systems to a quantum-resistant future. We will explore the multifaceted attack vectors that extend beyond just encryption vulnerabilities, and the strategic considerations required to navigate this critical shift. By examining key components like the \gls{BC} network, mining pools, and user wallets, we will identify the challenges associated with adapting each element to quantum-resistant cryptography. We will also explore potential solutions that organizations can implement to ensure a smooth and secure transition, safeguarding the long-term viability of their \gls{BC} deployments in the face of this evolving technological landscape.

\subsection{\gls{BC} Network}\label{Post-BC-Network}
The emergence of \gls{QC} necessitates a paradigm shift in securing \gls{BC} networks. While the integration of quantum-resistant algorithms offers a solution, it demands a comprehensive re-evaluation of the underlying network architecture. Unlike a straightforward cryptographic substitution, these new algorithms often necessitate larger key sizes and impose significantly higher computational workloads. Consequently, adjustments to existing protocols, data structures, and even the fundamental infrastructure of the network become paramount \cite{wicaksana2025survey}. Additionally, these changes can affect network latency and transaction throughput, further complicating the transition. 
This section explores the multifaceted challenges associated with this crucial transition, including potential incompatibilities with current systems, the need to accommodate larger keys and increased workloads, the imperative for network infrastructure upgrades, and the critical task of maintaining seamless interoperability during the migration process. By addressing these challenges and implementing effective solutions, organizations can ensure a smooth migration to a quantum-resistant \gls{BC} network, safeguarding its long-term security and scalability within this evolving technological landscape.
Here is a breakdown of the key challenges and potential solutions for \gls{BC} networks in this context:
\begin{enumerate}[topsep=1ex, itemsep=1ex, wide, font=\itshape, labelwidth=!, labelindent=0pt, label*=A.\arabic*.]
\item \textit{Compatibility with Existing Protocols and Data Structures:} 
New cryptographic algorithms might not be readily compatible with existing \gls{BC} protocols and data structures~\cite{banaeian2021blockchain}. This could lead to issues with data validation, consensus mechanisms, and overall network functionality. Potential solutions to address these compatibility challenges include: 
(i) developing adaptations to existing protocols that accommodate the specific requirements and constraints of \gls{PQC}, ensuring seamless integration while preserving network functionality; 
(ii) exploring alternative data structures inherently compatible with \gls{PQC}, with redesigns focused on optimizing both performance and security; and 
(iii) implementing a phased migration approach—initially transitioning critical functionalities, followed by gradual adoption across remaining network components—to minimize disruption and maintain system stability.

 
\item \textit{Accommodating Larger Key Sizes and Increased Computational Demands:} 
\gls{PQC} algorithms typically require significantly larger key sizes than classical cryptographic schemes~\cite{openquantumsafe_benchmarking,baseri2024navigating,baseri2024evaluation}. This leads to increased storage needs and computational overhead for tasks such as transaction verification and block validation, potentially impacting network latency and throughput. To address these challenges, the following strategies are recommended:
(i) analyzing trade-offs between security levels and key sizes to adopt optimal configurations that balance robustness and efficiency;
(ii) upgrading \gls{BC} node infrastructure—including processing power, memory, and storage—to support increased computational demands; and
(iii) exploring alternative consensus mechanisms with lower computational complexity that remain secure in a \gls{QC} environment.

 \item \textit{Network Infrastructure Upgrades:} 
Upgrading the underlying infrastructure of the \gls{BC} network is essential to ensure seamless integration with \gls{PQC}. This encompasses software updates, hardware replacements, and maintaining compatibility with existing development and deployment tools~\cite{allen2024governance,marchsreiter2024towards}. To address these infrastructure challenges, the following strategies are recommended: 
(i) developing clear and well-documented upgrade paths for \gls{BC} nodes, including compatibility testing tools and step-by-step migration guidelines; 
(ii) fostering ecosystem-wide collaboration to coordinate adoption of new software versions and maintain operational consistency across diverse network participants; and 
(iii) incorporating backward compatibility features during the transition phase to reduce disruption for users and applications that depend on legacy infrastructure.

\item \textit{Maintaining Interoperability:} 
During the transition to quantum-resistant cryptography, some nodes may still operate under legacy algorithms while others adopt \gls{PQC}~\cite{yang2024survey,edwards2020review,liu2023secure}. This coexistence can lead to network fragmentation or disruptions if interoperability is not properly ensured.  To overcome this challenge, the following strategies are recommended:
(i) deploying robust hybrid \gls{BC} architectures—such as composite designs, which streamline integration through unified ledgers, and non-composite designs, which use independent dual-network strategies, as detailed in Section~\ref{sec:hybrid}—to ensure seamless interoperability between classical and post-quantum cryptographic \gls{BC} systems;
(ii) designing a staged migration process with defined milestones and rollback mechanisms to address failures and compatibility issues; and
(iii) fostering collaboration and transparency within the \gls{BC} community to ensure a smooth and coordinated transition.

 
\end{enumerate}

\subsection{Mining Pool}\label{Post-Mining-Pool}
Mining pools face significant challenges during the transition to quantum-secure cryptography. Increased computational overheads associated with post-quantum algorithms can strain resources and potentially exacerbate centralization risks. Furthermore, new attack vectors, such as quantum-based \gls{DoS} attacks or cryptanalysis of quantum-secure primitives, may emerge. To ensure continued success, mining pools must implement robust security measures, optimize resource utilization, and actively mitigate centralization risks~\cite{chen2020survey}. The following explores these challenges and proposes strategies for successful operation in a quantum-secure environment.
\begin{enumerate}[topsep=1ex, itemsep=1ex, wide, font=\itshape, labelwidth=!, labelindent=0pt, label*=B.\arabic*.]
\item \textit{Enhanced Security Measures:} 
The transition to quantum-resistant algorithms introduces new attack vectors—including quantum cryptanalysis and side-channel vulnerabilities—that can compromise mining operations. To mitigate these evolving threats, the following defense strategies are recommended:
(i) designing and implementing key management systems capable of operating with both classical and quantum-resistant cryptographic keys for critical infrastructure components~\cite{giron2023post};
(ii) deploying intrusion detection and prevention systems (IDS/IPS) to monitor and block anomalous activities within the mining pool infrastructure~\cite{quincozes2021survey};
(iii) conducting regular security audits to identify and remediate vulnerabilities in mining pool software, configurations, and exposed network interfaces~\cite{white2020blockchain,kahyaouglu2023evaluation}; and
(iv) enforcing stricter access controls and multi-factor authentication to secure administrative interfaces, mining resources, and user accounts~\cite{sinigaglia2020survey}.


\item \textit{Resource Optimization:} 
Quantum-resistant algorithms often require larger key sizes and greater computational effort, leading to increased energy consumption and hardware demands~\cite{yang2024survey,edwards2020review,liu2023secure}. To address these resource-intensive requirements, the following optimization strategies are recommended: 
(i) utilizing specialized hardware accelerators—such as GPUs, FPGAs, or ASICs—designed to efficiently execute quantum-resistant cryptographic operations; 
(ii) deploying dynamic resource allocation systems through cloud computing to elastically manage fluctuating mining workloads; 
(iii) developing energy-efficient mining algorithms tailored to post-quantum cryptography and integrating renewable energy sources to minimize environmental impact; and 
(iv) quantifying resource overheads—such as the percentage increase in energy consumption per transaction due to larger key sizes—and addressing thermal challenges with advanced cooling solutions.



\item \textit{Maintaining Competitiveness and Fairness:} 
Increased hardware and operational costs may exacerbate centralization risks, as only well-funded mining pools can afford the necessary infrastructure~\cite{allende2023quantum,baseri2024evaluation,baseri2024navigating}. To promote fairness and competitiveness, the following strategies are recommended:
(i) encouraging collaboration among mining pools to share computational resources and collectively strengthen network security;
(ii) advocating for the development of resource-efficient quantum-resistant algorithms to lower entry barriers for smaller participants;
(iii) fostering transparency and fair competition within the mining ecosystem to ensure equitable access to advanced technologies and software; and
(iv) engaging with policymakers to incentivize equitable mining practices and renewable energy adoption through subsidies, tax credits, or grants.



\item \textit{Phased Transition and Continuous Adaptation:} 
A phased roadmap ensures smooth transition to quantum-resistant operations while maintaining security, trust, and efficiency~\cite{aydeger2024towards, baseri2024evaluation}. Key considerations include:
(i) \textit{Short-term:} assess infrastructure, upgrade hardware/networking, implement baseline quantum-resistant algorithms for registration/communications, and pilot test for vulnerabilities;
(ii) \textit{Medium-term:} deploy advanced quantum-resistant algorithms for rewards/verification, optimize mining software for post-quantum efficiency, implement quantum-secure consensus, optimize energy consumption, and establish quantum-resistant/legacy interoperability;
(iii) \textit{Long-term:} achieve complete quantum resistance via standardized post-quantum cryptography, align with quantum-resistant consensus, implement secure inter-pool communication, and establish continuous adaptation;
(iv) \textit{Continuous monitoring:} deploy predictive analytics for hash rate, difficulty, computational, and energy changes, enabling proactive allocation and threat positioning.

\item \textit{Governance and Economic Considerations:}  
The transition to quantum-secure \gls{BC} systems requires coordination and financial support to prevent smaller mining pools from being marginalized~\cite{chen2020decentralized,times2023quantum}. To promote fairness and long-term competitiveness, the following strategies are recommended:
(i) establishing a decentralized governance structure to coordinate upgrades across mining pools and enforce compliance with quantum-resistant standards;
(ii) implementing shared resource models—such as pooled access to quantum-resistant hardware, secure cloud infrastructure, or open-source platforms—to reduce entry costs for smaller participants; and
(iii) engaging with policymakers to provide financial incentives, including grants, tax credits, or subsidies, to accelerate the adoption of quantum-resistant technologies and renewable energy sources.

\end{enumerate}

\subsection{Transaction Verification Mechanism}\label{Post-Transaction-Verification} The transition to quantum-resistant algorithms presents significant challenges for verifying transactions within \gls{BC} networks. Careful consideration and adaptation are crucial to maintain network integrity and security. By proactively addressing these challenges, organizations can ensure a smooth transition and safeguard the enduring security and efficiency of their \gls{BC} networks.
 \begin{enumerate}[topsep=1ex, itemsep=1ex, wide, font=\itshape, labelwidth=!, labelindent=0pt, label*=C.\arabic*.]


\item \textit{Consensus Protocol Adaptation:}  
 Existing consensus mechanisms, which ensure network integrity by verifying transactions and reaching agreement on the \gls{BC} state, might require adjustments to mitigate the risk of unauthorized control after the switch to quantum-resistant cryptography~\cite{gomes2023fortifying}. Quantum-resistant algorithms can alter the computational power dynamics within the network (e.g., some algorithms might be more efficient for specific hardware), potentially creating vulnerabilities for malicious actors~\cite{zhang2024reaching,sinai2024performance}. To address these challenges, the following proactive strategies are recommended: 
(i) analyzing existing consensus mechanisms to identify vulnerabilities introduced by the adoption of quantum-resistant algorithms; 
(ii) modifying consensus parameters, such as block validation rules, voting weights, and dispute resolution processes, to preserve balanced participation and network security; and 
(iii) exploring or designing alternative consensus protocols specifically tailored for quantum-resistant environments, including post-quantum voting-based variants that inherently resist unauthorized control.

\item \textit{Increased Computational {and Communication} Overheads:}  
 \begin{table*}[!ht]
\caption{{Comparison of NIST-Standardized Post-Quantum Cryptographic Algorithms}}
\label{tab:qresistant}
\scriptsize
\resizebox{\linewidth}{!}{{\begin{tabular}{|l|c|c|c|c|c|c|c|c|}
\hline
  \textbf{Algorithm} & \textbf{Security Level\textsuperscript{*}} &\textbf{Type}& \textbf{KeyGen (s)} & \textbf{Sign/Enc (s)} & \textbf{Verify/Dec (s)} & \textbf{Public Key (B)} & \textbf{Secret Key (B)} & \textbf{Ciphertext/Sig. (B)} \\
\hline
Kyber512 & L1 & \gls{KEM/ENC}& 0.032 & 0.032 & 0.022 & 800 & 1632 & 768 \\
Kyber768 & L3 & \gls{KEM/ENC}& 0.214 & 0.046 & 0.032 & 1184 & 2400 & 1088 \\
Kyber1024 & L5 & \gls{KEM/ENC}& 0.052 & 0.053 & 0.046 & 1568 & 3168 & 1568 \\
\hline
HQC-128 & L1 &\gls{KEM/ENC}&  0.108 & 0.145 & 0.232 & 2249 & 2289 & 4481 \\
HQC-192 & L3 &\gls{KEM/ENC}&  0.407 & 0.332 & 0.511 & 4522 & 4562 & 9026 \\
 HQC-256 & L5 &\gls{KEM/ENC}&  0.542 & 0.600 & 0.926 & 7245 & 7285 & 14469 \\
\hline
Dilithium2 & L2 &Signature&  0.495 & 0.179 & 0.073 & 1312 & 2528 & 2420 \\
Dilithium3 & L3 &Signature& 0.130 & 0.260 & 0.107 & 1952 & 4000 & 3293 \\
Dilithium5 & L5 &Signature& 0.145 & 0.250 & 0.128 & 2592 & 4864 & 4595 \\ \hline
SPHINCS+ (SHA-256-128f) & L1 &Signature& 1.155 & 28.111 & 3.093 & 32 & 64 & 17088 \\
SPHINCS+ (SHA-256-192f) & L3 &Signature& 1.562 & 45.656 & 4.596 & 48 & 96 & 35664 \\
SPHINCS+ (SHA-256-256f) & L5 &Signature& 4.203 & 92.525 & 4.788 & 64 & 128 & 49856 \\ \hline
Falcon-512 & L1 &Signature& 12.686 & 0.525 & 0.110 & 897 & 1281 & 690 \\
Falcon-1024 & L5 &Signature& 34.212 & 1.003 & 0.199 & 1793 & 2305 & 1330 \\
\hline
\end{tabular}}}
\vspace{3pt} \newline
 {\scriptsize{$^*$  Security Levels: L1 = 128-bit, L3 = 192-bit, L5 = 256-bit quantum security.\\
$^{**}$ Performance was evaluated using \textit{liboqs} \cite{oqs-liboqs} with reference implementations from \textit{PQClean} \cite{pqclean} on Intel Xeon E5-2670 v3 (Ubuntu 20.04, 1,000 TPS).
}}
  \vspace{-0.3cm}
\end{table*}
 The complex nature of quantum-resistant cryptographic operations might introduce   {scalability challenges and} performance slowdowns~\cite{openquantumsafe_benchmarking,baseri2024navigating,baseri2024evaluation} due to increased computational {and communicational} demands for transaction verification. {Specifically, NIST-standardized PQC algorithms introduce significant performance trade-offs compared to classical cryptography, as detailed in Table~\ref{tab:qresistant}. For key establishment, ML-KEM (CRYSTALS-Kyber) public keys (800-1,568 bytes) and ciphertexts (768-1,568 bytes) are considerably larger than classical ECDH public keys (e.g., 64 bytes) \cite{FIPS203}. While ML-KEM's computational speed for key encapsulation is often comparable to or even faster than RSA-KEM, it is generally slower than ECDH. For digital signatures, ML-DSA (CRYSTALS-Dilithium) signatures are 38-52x larger than 64-byte ECDSA signatures (2,420-3,309 bytes) and its verification is typically 2-5x slower~\cite{FIPS204}. While FN-DSA (FALCON)~\cite{FalconWebsite} offers smaller signatures (690-1,330 bytes), SLH-DSA (SPHINCS+), chosen for its conservative hash-based security, presents the largest overheads: signatures can be 123-779x larger (7,856-49,856 bytes) and signing 10-100x slower \cite{FIPS205}. This overhead may increase block sizes and transmission delays, leading to slower block propagation across the network. In bandwidth-constrained environments, this can exacerbate transaction backlogs in the mempool and raise the risk of forks, thereby requiring infrastructure upgrades in bandwidth, memory, and processing capacity to sustain \gls{BC} scalability.}

To mitigate this challenge, the following strategies are recommended: (i) optimizing the implementation of quantum-resistant algorithms within the transaction verification process through code refactoring and hardware acceleration;
(ii) selecting PQC schemes that offer a favorable scalability-security trade-off
for transaction verification;
(iii) considering scalability solutions like sharding~\cite{singh2020public,dang2019towards} and increased block size to distribute the workload of transaction verification across the network;
(iv) leveraging interoperability protocols to facilitate load distribution and network communication between different \glspl{BC}~\cite{LI2025100286,Deng2025SFPoW,belchior2021survey}; and
(v) utilizing quantum-resistant off-chain scaling solutions, such as state channels~\cite{9627997}, sidechains~\cite{Deng2025SFPoW}, and rollups~\cite{ethereum_rollups}, to augment scalability by processing transactions externally and finalizing only crucial state changes on-chain. Ensuring the integrity of off-chain operations under this approach relies on the development of efficient, quantum-resistant \glspl{ZKP}~\cite{refc31,zhou2024leveraging}.

\item \textit{Implementation Complexity:}  
\gls{PQC} integration introduces substantial complexity to existing \gls{BC} systems, particularly in transaction verification, consensus protocols, and key operations~\cite{joseph2022transitioning,mashatan2021complex}. {Such complexity can introduce new security vulnerabilities, performance bottlenecks, and integration challenges, which may further compound scalability and interoperability issues as network adoption grows. To mitigate these risks, the following proactive strategies are recommended: 
(i) conducting comprehensive analysis and testing of \gls{PQC} integration workflows, including component-level validation, interface compatibility checks, targeted vulnerability assessments, and the use of abstraction layers to isolate algorithmic intricacies from core \gls{BC} operations;
(ii) adopting secure coding practices and adhering to established cryptographic standards to reduce the risk of misconfiguration or flawed implementation;
(iii) providing specialized training and development tools to help engineers manage the intricacies of \gls{PQC} integration, particularly within verification mechanisms; and
(iv) collaborating with quantum cryptography experts to audit and validate \gls{PQC} components across the \gls{BC} stack, ensuring correctness, resilience, and long-term maintainability.}


 \item \textit{Maintaining Decentralization:} 
 The increased computational demands of quantum-resistant algorithms may inadvertently drive centralization, as only entities with access to powerful hardware can efficiently participate in transaction verification~\cite{yang2022decentralization}. To preserve decentralization, the following strategies are recommended:
(i) promoting the development of resource-efficient quantum-resistant algorithms compatible with a broad range of hardware;
(ii) exploring consensus mechanisms that enable equitable participation despite high computational costs; and
(iii) designing incentive structures that reward verification efficiency independently of hardware capabilities.
\end{enumerate}
\subsection{Smart Contract}\label{Post-Smart-Contract} 
The shift to quantum-resistant algorithms within \gls{BC} networks introduces a unique set of challenges for smart contracts. These challenges include potential security vulnerabilities arising from interactions with new algorithms, compatibility issues with existing smart contracts, and the need for user education regarding the transition. Addressing these challenges is crucial for ensuring the seamless integration and continued functionality of smart contracts in a quantum-secure environment.
 \begin{enumerate}[topsep=1ex, itemsep=1ex, wide, font=\itshape, labelwidth=!, labelindent=0pt, label*=D.\arabic*.]
 \item \textit{Compatibility with Existing Smart Contracts:} 
 Upgrading existing smart contracts to utilize quantum-resistant algorithms might not be straightforward. This could lead to a situation where some contracts operate with the old cryptography while others adopt the new, potentially hindering interoperability and causing disruptions. To mitigate this, the following strategies are recommended: 
(i) developing migration tools and frameworks to enable seamless transitions~\cite{cisa2023quantum}; 
(ii) designing standardized wrappers or compatibility layers to facilitate interaction between legacy and quantum-resistant contracts, while managing potential performance overhead~\cite{nist2023migration}; and 
(iii) promoting modular and future-proof contract design to ease adaptation to evolving cryptographic standards~\cite{fernandez2020towards}.
\item \textit{User Education and Awareness:} 
Users interacting with smart contracts must be aware of the transition to quantum-resistant algorithms and the potential risks of engaging with outdated, vulnerable contracts. To promote secure participation, the following strategies are recommended:
(i) developing educational materials and awareness campaigns to encourage users to upgrade their wallets and tools for compatibility with quantum-resistant contracts~\cite{naz2024surveying};
(ii) integrating warnings and notifications within applications to alert users about the risks of engaging with legacy contracts~\cite{naz2024surveying}; and
(iii) promoting best practices for secure interaction, including key management and contract verification prior to execution~\cite{tsankov2018securify}.

\item \textit{{Zero-Day and Emergent Security Vulnerabilities}:} {The integration of \gls{PQC} primitives into smart contract code introduces a new class of risks: zero-day vulnerabilities (previously unknown exploitable flaws)~\cite{11028033,ahmad2023zero} and emergent issues arising from fundamental shifts in cryptographic assumptions. Unlike conventional smart contract vulnerabilities, these stem from the novel challenges of integrating \gls{PQC} with existing \gls{BC} infrastructure. These risks are amplified by the distinct computational and structural characteristics of post-quantum algorithms, the complexity of maintaining backward compatibility during migration, and the limited adversarial testing to date. While specific vulnerabilities cannot be anticipated, the integration process creates conditions conducive to unforeseen attack vectors, particularly in hybrid environments.} {To mitigate these risks, the following proactive measures are recommended:} (i) conducting thorough audits and validation processes to identify vulnerabilities emerging from the interaction between smart contracts and quantum-resistant primitives~\cite{kahyaouglu2023evaluation,white2020blockchain}; (ii) developing and applying formal verification techniques to mathematically guarantee contract correctness and security in post-quantum environments~\cite{ethereum_formal_verification,lewis2023formal,tolmach2023securing}; and (iii) establishing best practices for secure smart contract development, with an emphasis on minimizing attack surfaces during \gls{PQC} integration.
 

\item \textit{Performance Implications:}  
Integrating quantum-resistant algorithms may degrade smart contract performance due to increased computational demands, potentially resulting in higher execution costs and longer processing times~\cite{schierbauer2024performance}. To mitigate these issues, the following strategies are recommended:
(i) optimizing smart contract code by reducing complexity, minimizing redundant computations, and improving data structures;
(ii) batching multiple transactions into single operations to reduce on-chain activity and alleviate congestion~\cite{wang2023towards};
(iii) refining quantum-resistant algorithms to reduce computational overhead and enhance runtime efficiency;
(iv) offloading computations off-chain where feasible to minimize on-chain processing~\cite{Ding2024Genuine,liu2022extending};
(v) enabling modular and asynchronous execution of smart contracts to allow independent task processing and increase throughput; and
(vi) benchmarking and monitoring smart contract performance regularly to identify and resolve bottlenecks.\\

\item \textit{Regulatory Considerations}

The transition to quantum-resistant cryptography introduces regulatory challenges that require proactive compliance. Organizations must align with evolving legal standards to mitigate operational and reputational risks~\cite{gao2023blockchain,nist2023migration}. Recommended actions include:
(i) staying informed about regulatory developments in quantum-resistant cryptography and \gls{BC} technologies;
(ii) engaging with policymakers and industry groups to help shape practical, forward-looking regulations; and
(iii) implementing compliance frameworks that anticipate and align with emerging regulatory standards.

\end{enumerate}

\subsection{User Wallet}\label{Post-User-Wallet} 
The transition to quantum-resistant \gls{BC} presents specific challenges for user wallets, which are critical interfaces for users to access and manage their digital assets. Ensuring a seamless and secure transition is paramount to maintain user trust and the integrity of the \gls{BC} ecosystem.
In the following, we examine the primary challenges and propose solutions to facilitate this transition.

\begin{enumerate}[topsep=1ex, itemsep=1ex, wide, font=\itshape, labelwidth=!, labelindent=0pt, label*=E.\arabic*.]




 \item \textit{Wallet Compatibility and Security Upgrades:}  
 Existing user wallets might not be compatible with the new quantum-resistant cryptography. This leaves users vulnerable to potential attacks that exploit weaknesses in older cryptographic algorithms~\cite{allende2023quantum,kearney2021vulnerability,sinai2024performance}.  To ensure secure and seamless transitions, the following strategies are recommended:
(i) developing clear upgrade paths and detailed guidance to support users migrating to quantum-resistant wallets;
(ii) implementing backward-compatible solutions where feasible to maintain continuity during migration;
(iii) promoting adoption of standardized quantum-resistant protocols among wallet developers to preserve interoperability; and
(iv) enhancing security through multi-signature wallet configurations that require multiple keys to authorize transactions.

\item \textit{User Education and Awareness:} 
In the context of user adoption~\cite{alnahawi2021state}, individuals might be hesitant or unaware of the need to upgrade their wallets, potentially leaving them vulnerable even after the transition to quantum-resistant algorithms. To promote informed adoption, the following strategies are recommended: 
(i) launching educational campaigns to highlight the importance of transitioning to quantum-resistant wallets and their security benefits;
(ii) integrating intuitive prompts and automated update mechanisms into wallet applications to support seamless user migration; and
(iii) providing accessible resources, such as tutorials and FAQs, to assist users throughout the upgrade process.


\item \textit{Maintaining User Experience:} 
Upgrading wallets and potentially adopting new functionalities related to quantum-resistant cryptography might introduce complexities that could hinder user experience~\cite{de2022software}. To maintain a seamless user experience, the following strategies are recommended: 
(i) designing wallet interfaces that integrate enhanced security features while maintaining simplicity and user-friendliness; 
(ii) conducting usability testing to identify and resolve issues stemming from the integration of quantum-resistant elements; and 
(iii) implementing security enhancements in a transparent manner to minimize disruptions to users’ typical interactions with the wallet.



\end{enumerate}

\section{{Hybrid \gls{BC} Architectures}}\label{sec:hybrid}

\begin{table*}[!h]
\caption{{Hybrid Signature Strategies}}
\label{tab:hybrid-signature-strategy}
\scriptsize
\resizebox{\linewidth}{!}{%
\begin{tabular}{|m{0.15\linewidth}|m{0.25\linewidth}|m{0.43\linewidth}|m{0.3\linewidth}|}
\hline
\textbf{Combiners} &\textbf{Mathematical Definition} & \textbf{Pros} & \textbf{Cons} \\
\hline
Concatenation \cite{bindel2017transitioning} & 
{\begin{minipage}[c]{\linewidth}
\vspace{-0.1cm}$Sign_{hybrid} = \sigma_1 \| \sigma_2$, where
\begin{myBullets}
\item $\sigma_i = \text{Sign}_i(sk_i, m)$
 \end{myBullets}
\end{minipage}}&
\begin{myBullets}
\item Supports lightweight operations, simple logic, and easy implementation.
\item Retains unforgeability when both signature algorithms are unforgeable.
 \vspace{-2ex} \end{myBullets} &
\begin{myBullets}
\item Does not support non-separability property for both signature algorithms.
 \vspace{-2ex} \end{myBullets} \\
\hline
Weak Nesting~\cite{bindel2017transitioning} & 
{\begin{minipage}[c]{\linewidth}

\vspace{-0.1cm}$Sign_{hybrid} = \text{Sign}_2(sk_2, \sigma_1)$, where
\begin{myBullets}\vspace{-0.1cm}
\item  $\sigma_1 = \text{Sign}_1(sk_1, m)$
 \end{myBullets}
\end{minipage}}
&
\begin{myBullets}
\item Preserves unforgeability when the first signature algorithm is unforgeable.
\item Supports non-separability property for the second signature algorithm.
 \vspace{-2ex} \end{myBullets}&
\begin{myBullets}
\item Unforgeability of weak nesting depends crucially on the unforgeability of the first signature scheme.
 \vspace{-2ex} \end{myBullets} \\
\hline
Strong Nesting~\cite{bindel2017transitioning,ghinea2022hybrid} & {\begin{minipage}[c]{\linewidth}
\vspace{-0.1cm}$Sign_{hybrid} = (\sigma_1, \sigma_2)$, where
\begin{myBullets}\vspace{-0.1cm}
\item $\sigma_1 =$ $\text{Sign}_1(sk_1, m)$ and
\item $\sigma_2 =$ $ \text{Sign}_2(sk_2, m \mathbin{\|} \sigma_1)$
 \end{myBullets}
\end{minipage}}
&
\begin{myBullets}
\item Retains unforgeability when both signature algorithms are unforgeable.
\item Preserves non-separability property for the second signature algorithm.
 \vspace{-2ex} \end{myBullets} &
\begin{myBullets}
\item Caution needed with Strong Nesting due to potential signature leaks from one of its underlying schemes.
 \end{myBullets} \\
\hline
Dual Nesting~\cite{bindel2017transitioning} & {\begin{minipage}[c]{\linewidth}
\vspace{-0.1cm}$Sign_{hybrid}=(\sigma_1, \sigma_{outer})$, where
\begin{myBullets}\vspace{-0.1cm}
\item $\sigma_1 = \text{Sign}_1(sk_1, m_1)$ and
\item $\sigma_{outer} = \text{Sign}_1(sk_1, \text{Sign}_2(sk_2, m_2))$
 \end{myBullets}
\end{minipage}}&
\begin{myBullets}
\item Preserves unforgeability of each message under its corresponding signature scheme.
\item Retains unforgeability of both messages when the outer signature scheme is unforgeable.
 \vspace{-2ex} \end{myBullets} &
\begin{myBullets}
\item Not designed to provide the unforgeability of both messages under either signature scheme.
 \vspace{-2ex} \end{myBullets} \\
\hline
\end{tabular}%
 }  \vspace{-0.3cm}
\end{table*}
{The transition to \gls{PQC} poses a fundamental dilemma for \gls{BC} networks. An immediate and comprehensive migration is technically demanding and financially costly, with the potential to degrade the performance of high-throughput financial \glspl{BC} that process millions of transactions per day. Conversely, postponing migration exposes networks to \gls{HNDL} attacks~\cite{barenkamp2022steal}, whereby adversaries intercept and archive encrypted data with the objective of future decryption once large-scale \gls{QC} becomes viable.
Hybrid \gls{BC} architectures have emerged as a pragmatic solution to this challenge~\cite{bindel2017transitioning,giron2023post}, combining classical and \gls{PQC} primitives such that the compromise of one cryptographic scheme does not immediately undermine overall system security. By selectively integrating \gls{PQC} into critical \gls{BC} components, hybrid approaches preserve operational performance and ecosystem compatibility while incrementally strengthening resilience throughout the quantum transition~\cite{joseph2022transitioning,allende2023quantum,driscoll-pqt-hybrid-terminology-02}.}

{This section develops a comprehensive framework for hybrid \gls{BC} architectures, systematically categorizing them into \emph{composite} and \emph{non-composite} paradigms based on their cryptographic integration strategies. We analyze their respective trade-offs, implementation complexity, and scalability implications, with emphasis on cryptographic combiners, selective verification, interoperability, and consensus integrity. The analysis examines practical deployment factors including performance overhead, migration pathways, and backward compatibility requirements. By synthesizing theoretical foundations with implementation challenges, this framework provides a structured reference for researchers and practitioners navigating \gls{BC} security decisions in the quantum era.}

\subsection{{Hybrid Strategies for Cryptographic Primitives}}\label{Hybrid-Strategies}

{Hybrid cryptographic strategies facilitate secure \gls{BC} migration by combining classical and \gls{PQC} schemes through mathematically rigorous \textit{combiners}, providing both crypto-agility~\cite{alnahawi2023state,petrenko2019assessing} and defense-in-depth~\cite{giron2023post}. Hybrid strategies for \textit{digital signatures}, the most critical primitives in the \gls{BC} ecosystem, integrate multiple independent signatures via structured combiners to preserve unforgeability against adaptive chosen-message attacks~\cite{kwon2024compact,ghinea2022hybrid}. The principal approaches (Table~\ref{tab:hybrid-signature-strategy}) include: \textit{Concatenation}, a direct combination with simple implementation, which preserves security as long as at least one scheme remains secure; \textit{Weak Nesting}~\cite{bindel2017transitioning}, a sequential composition that reduces implementation cost but inherits dependence on the inner scheme; \textit{Strong Nesting}~\cite{bindel2017transitioning,ghinea2022hybrid}, which dual-signs each message to ensure unforgeability and non-separability as long as one scheme remains quantum-resistant; and \textit{Dual Nesting}~\cite{bindel2017transitioning}, which extends nesting to multi-message and heterogeneous \gls{BC} architectures at the expense of added complexity.}
{Combiner selection must balance security guarantees, computational overhead, and deployment feasibility. Strong Nesting delivers the strongest resilience but increases verification costs. Concatenation provides implementation simplicity and robustness as long as one scheme remains secure, though at the cost of signature size. Weak Nesting simplifies deployment yet inherits vulnerabilities from the inner scheme. Dual Nesting offers flexibility for heterogeneous or multi-message contexts, though with greater complexity. These trade-offs highlight the importance of aligning combiner selection with \gls{BC}-specific requirements for performance, scalability, and interoperability.}

\subsection{{Architectural Design Approaches}}\label{Architectural-Design}
{Building on the hybrid cryptographic strategies outlined above, two principal hybrid \gls{BC} paradigms have emerged: \textit{composite} and \textit{non-composite}. These paradigms differ in how classical and post-quantum primitives are integrated, either within unified structures or across distinct operational domains, each offering trade-offs tailored to specific migration strategies, performance constraints, and security models.}

\begin{enumerate}[topsep=1ex, itemsep=1ex, wide, font=\itshape, labelwidth=!, labelindent=0pt, label*=B.\arabic*.]
\item \textit{{Non-Composite Architecture:}}\label{Non-Composite} 
{The non-composite or dual \gls{BC} architecture employs a \textit{dual-network strategy}, maintaining two independent \gls{BC} networks with distinct cryptographic implementations and optimized consensus mechanisms. This approach provides clear security boundaries, enables specialized optimization for different cryptographic paradigms, and facilitates a structured migration path to quantum resistance while preserving complete decentralization. The architecture addresses the fundamental trade-off between performance and quantum security by allowing users to choose between networks based on their security requirements and risk tolerance.}

\begin{enumerate}[topsep=1ex, itemsep=1ex, wide, font=\itshape, labelwidth=!, labelindent=0pt, label*=\arabic*.]

\item \textit{{Architecture Components and Design Principles:} }
{The non-composite architecture comprises two independent \gls{BC} networks: (a) a \textit{classical \gls{BC} network} and (b) a \textit{post-quantum \gls{BC} network}. These networks are bridged through a dedicated \textit{cross-chain interoperability framework} that enables secure coordination and asset transfers between the classical and post-quantum environments.}

\begin{figure}[htbp]
\centering

\resizebox{\linewidth}{!}{%
\begin{tikzpicture}[
  ledger/.style={
    rectangle,
    rounded corners=5pt,
    minimum width=5cm,
    minimum height=5.5cm,
    draw,
    line width=1.5pt
  },
  badge/.style={
    circle,
    minimum size=2.2cm,
    draw,
    line width=2.5pt,
    fill=white,
  },
  inner_badge/.style={
    circle,
    minimum size=1.6cm,
    draw,
    line width=1.5pt,
    dashed,
    fill=none
  },
  coordination/.style={
    rectangle,
    rounded corners=3pt,
    minimum width=5.3cm,
    minimum height=3.8cm,
    draw,
    line width=1.5pt,
    dashed
  },
  tier/.style={
    rectangle,
    rounded corners=2pt,
    minimum width=4.5cm,
    minimum height=0.5cm,
    draw,
    line width=0.8pt 
  },
  arrow/.style={
    ->,
    line width=2pt,
    >=Stealth
  },
  dashed_arrow/.style={
    ->,
    line width=2pt,
    >=Stealth,
    dashed
  }
]

\definecolor{classicalblue}{RGB}{25,118,210}
\definecolor{quantumpurple}{RGB}{123,31,162}
\definecolor{coordinationorange}{RGB}{255,102,0}
\definecolor{highvaluepink}{RGB}{233,30,99}
\definecolor{routinegreen}{RGB}{76,175,80}

\node[ledger, fill=blue!10, draw=classicalblue] (classical) at (-4.5,0.5) {};

\node[ledger, fill=purple!10, draw=quantumpurple] (quantum) at (4.5,0.5) {};

\node[coordination, fill=orange!10, draw=coordinationorange] (coord) at (0,-0.1) {};
\node[text width=6cm, align=center, font=\bfseries\color{coordinationorange}] at (0,1.3) {Quantum-Safe Cross-Chain Coordination Layer};
\node[tier, fill=yellow!20, draw=orange!80!black] at (0,0.5) {};
\node (Tier-1) at (0,0.5) {\color{orange!80!black}\scriptsize Tier 1: PQ-HTLCs};

\node[tier, fill=orange!20, draw=orange!80!black] at (0,-0.3) {};
\node (Tier-2) at (0,-0.3) {\color{orange!80!black}\scriptsize Tier 2: Commitment Protocols};

\node[tier, fill=red!20, draw=red!70] at (0,-1.1) {};
\node (Tier-3) at (0,-1.1) {\color{orange!80!black}\scriptsize Tier 3: Decentralized PQ Oracles};

\draw[arrow, coordinationorange] ($(Tier-1.south)+(-1,0)$) -> ($(Tier-2.north)+(-1,0)$);
\draw[arrow, coordinationorange] ($(Tier-2.north)+(1,0)$) -> ($(Tier-1.south)+(1,0)$);
\draw[arrow, coordinationorange] ($(Tier-2.south)+(-1,0)$) -> ($(Tier-3.north)+(-1,0)$);
\draw[arrow, coordinationorange] ($(Tier-3.north)+(1,0)$) -> ($(Tier-2.south)+(1,0)$);

\node at (0,-1.7) {\color{coordinationorange}\scriptsize Cross-Ledger Operations};

\node[badge, draw=classicalblue] at (-4.5,5) {};
\node[inner_badge, draw=classicalblue] at (-4.5,5) {};
\node[badge, draw=quantumpurple] at (4.5,5) {};
\node[inner_badge, draw=quantumpurple] at (4.5,5) {};

\node[classicalblue] at (-4.5,5.3) {\scriptsize\textbf{SECURITY ASSURANCE}};
\node[classicalblue] at (-4.5,4.5) {\tiny CLASSICAL};
\draw[classicalblue, line width=1pt] (-4.2,4.7) rectangle (-4.8,5.1);
\draw[classicalblue, line width=1pt] (-4.5,4.92) circle (0.1);
\fill[classicalblue] (-4.3,4.7) rectangle (-4.7,4.9);

\node[quantumpurple] at (4.5,5.3) {\scriptsize\textbf{SECURITY ASSURANCE}};
\node[quantumpurple] at (4.5,4.5) {\tiny QUANTUM-SAFE};
\draw[quantumpurple, line width=1pt] (4.2,4.7) rectangle (4.8,5.1);
\draw[quantumpurple, line width=1pt] (4.5,4.92) circle (0.1);
\fill[quantumpurple] (4.3,4.7) rectangle (4.7,4.9);

\draw[arrow, classicalblue] (-4.5,4) -- (classical.north);
\draw[arrow, quantumpurple] (4.5,4) -- (quantum.north);

\node[classicalblue] at (-4.5,2.6) {\textbf{Classical Cryptography}};
\node at (-4.5,2.3) {\textbf{\color{classicalblue} Blockchain}};

\fill[classicalblue, opacity=0.7] (-5.2,1.5) rectangle (-3.8,2);
\foreach \y in {1.6,1.7,1.8,1.9} {
  \draw[white, line width=1pt] (-5.1,\y) -- (-3.9,\y);
}
\node at (-4.5,1) {\textbf{\color{classicalblue}Ledger Node}};
\node at (-4.5,0) {\color{classicalblue}\small • ECDSA Signatures};
\node at (-4.5,-0.4) {\color{classicalblue}\small • SHA-256 Hashing};
\node at (-4.5,-0.8) {\color{classicalblue}\small • ~3,000 TPS};

\node[quantumpurple] at (4.5,2.6) {\textbf{Post-Quantum Cryptography}};
\node at (4.5,2.3) {\textbf{\color{quantumpurple} Blockchain}};

\fill[quantumpurple, opacity=0.7] (3.8,1.5) rectangle (5.2,2);
\foreach \y in {1.6,1.7,1.8,1.9} {
  \draw[white, line width=1pt] (3.9,\y) -- (5.1,\y);
}
\node at (4.5,1) {\textbf{\color{quantumpurple}Ledger Node}};
\node at (4.5,0) {\color{quantumpurple}\small • Dilithium Signatures};
\node at (4.5,-0.4) {\color{quantumpurple}\small • SHA-3 Hashing};
\node at (4.5,-0.8) {\color{quantumpurple}\small • ~10–16 TPS};


\end{tikzpicture}}
\caption{{Non-Composite \gls{BC} Architecture: A Dual-Public Ledger Design with Quantum-Safe Cross-Chain Interoperability.}}
\label{fig:Non-Composite}
\vspace{-0.1cm}
\end{figure}
{The \textit{classical \gls{BC} network} handles existing applications, high-frequency transactions, and users prioritizing performance and compatibility. It utilizes established classical cryptographic standards such as ECDSA for digital signatures, SHA-256 for hashing, and proven consensus mechanisms like \gls{PoW} or \gls{PoS}. This network maintains full compatibility with existing tooling, wallets, and applications while providing optimal throughput and minimal computational overhead. However, its reliance on classical cryptography makes it increasingly vulnerable to quantum attacks as \gls{QC} technology advances. The \textit{post-quantum \gls{BC} network} serves security-conscious users, high-value transactions, and long-term asset storage requiring quantum resistance. It leverages \gls{PQC} algorithms such as CRYSTALS-Dilithium for digital signatures, quantum-resistant hash functions like SHA-3 with extended output lengths, as suggested by \gls{NIST}~\cite{nist2024transition}, and consensus mechanisms adapted for \gls{PQC}. This network implements quantum-resistant address generation, transaction formats, and state management optimized for long-term security against quantum adversaries.}

{The {cross-chain interoperability framework} enables seamless asset and data transfer between networks through quantum-resistant bridge protocols~\cite{LI2025100286}. The interoperability layer can be implemented through three core mechanisms: (1) \textit{Quantum-Resistant Atomic Swaps}  utilizing post-quantum \glspl{HTLC}, which retain the structure of classical \gls{HTLC}~\cite{bip199} but replace standard hash functions with longer variants resistant to quantum (e.g. SHA-3 with 512 bit, as suggested in~\cite{nist2024transition}) to withstand Grover's algorithm~\cite{hu2025quantum}. (2) \textit{Cross-Chain Bridge Validators} operating quantum-resistant multi-signature schemes where validator sets can verify transactions from both classical ECDSA and post-quantum signature schemes, ensuring atomic execution across networks~\cite{Herlihy2018Atomic,Jia2023Cross-Chain}; (3) \textit{Hybrid Oracle Networks} providing cross-chain state verification through decentralized oracle networks using post-quantum digital signatures and threshold cryptography to prevent single points of cryptographic failure.}

{In the above approach, both networks maintain complete decentralization, and there's no performance degradation from hybrid cryptography integration. Users can migrate between networks based on their evolving security needs, as each network is optimized specifically for its cryptographic approach. This also facilitates market-driven adoption without mandatory transitions and provides effective risk isolation between classical and quantum-resistant operations.}

{Bridge security is paramount, as the interoperability layer represents the primary attack surface. Cross-chain transactions inherit the security properties of both underlying networks and the bridge protocol, necessitating careful design of atomic swap mechanisms, validator incentives, and dispute resolution protocols. This architecture supports gradual migration as quantum threats emerge, all while preserving full backward compatibility and user choice throughout the transition period. Figure~\ref{fig:Non-Composite} illustrates this dual-network architecture, detailing the cross-chain transaction flow and interoperability mechanisms between the classical and post-quantum \gls{BC} networks.}
\item \textit{{Case Study - Dual-\gls{BC} CBDC Architecture for Quantum Transition:}}
{To demonstrate the viability of non-composite \gls{BC} architectures in quantum-safe financial infrastructures, we present a conceptual case study, EuroChain, inspired by the coexistence and migration strategies analyzed by Weinberg et al.~\cite{weinberg2025will}. While their work does not propose a specific implementation, it outlines essential cryptographic primitives (e.g., MPC, OT) and architectural requirements for parallel classical and post-quantum digital currencies. Building on this foundation, EuroChain models a dual-ledger CBDC architecture comprising two fully decentralized public \glspl{BC}: EuroChain-Classical, optimized for high-throughput retail payments (3,000–5,000 TPS) using ECDSA and SHA-256; and EuroChain-PQC, a quantum-resilient ledger secured with CRYSTALS-Dilithium-3 and SHA3-384, tailored for high-security use cases such as interbank settlements and long-term digital asset custody (50–100 TPS with 128-bit security).
This architecture reflects the non-composite paradigm, where classical and post-quantum operations are handled by separate chains, preserving cryptographic isolation and enabling performance-security optimization without hybridization. Users can dynamically select the appropriate ledger based on risk, latency, or compliance needs, supporting market-driven migration without centralized enforcement. A quantum-safe coordination layer ensures secure interoperability between chains via three mechanisms: (1) quantum-safe \glspl{HTLC} for atomic swaps with calibrated timeouts, utilizing NIST-recommended quantum-safe hash functions, such as SHA3-512, as suggested in~\cite{nist2024transition}; (2) hybrid bridge validators supporting both ECDSA and Dilithium with slashing/staking guarantees; and (3) decentralized oracle networks leveraging BLS aggregation and verifiable delay functions for state synchronization.
While EuroChain introduces implementation-level assumptions beyond those in~\cite{weinberg2025will}, it operationalizes their proposed principles into a dual-chain, non-composite framework suitable for gradual, decentralized quantum transition in digital currency ecosystems.}

\end{enumerate}
\item \textit{{Composite Architecture:}}\label{Composite:} 
{The \textit{composite \gls{BC}} integrates classical and \gls{PQC} functionalities within a single, unified ledger. This approach prioritizes simplicity and seamless integration, offering robust security through cryptographic redundancy, facilitating gradual migration, and maintaining a consolidated ledger structure.}
\begin{enumerate}[topsep=1ex, itemsep=1ex, wide, font=\itshape, labelwidth=!, labelindent=0pt, label*=\arabic*.]
  \item \textit{{Architecture Design and Implementation:} }
  {This unified ledger strategy employs three key design principles: (a) integrated block structure, (b) dual cryptographic operations within each block, and (c) a unified consensus mechanism.
  Each block incorporates \textit{integrated sections} dedicated to both classical and post-quantum cryptography. This structure enables simultaneous support for multiple cryptographic schemes, allowing transactions to select appropriate security levels based on their requirements, thereby eliminating the need for separate ledgers. }

  {To ensure robustness against both classical and quantum adversaries, the architecture applies \textit{dual cryptographic operations}. Following the hybrid strategies outlined in Subsection~\ref{Hybrid-Strategies}, both classical and post-quantum cryptographic methods are applied in parallel within each transaction block and combined using the cryptographic combiners described therein. \textit{Composite architectures may, for instance, employ co-signing techniques} where each transaction includes both a classical (e.g., ECDSA) and a post-quantum (e.g., Dilithium) signature, validated in parallel. This ensures resilience against both threat models and facilitates a seamless transition to quantum-safe infrastructures without compromising interoperability or performance predictability. During the transition period, a relaxed hybrid verification policy may be adopted, accepting a transaction if either the classical or post-quantum signature is valid. This facilitates backward compatibility with legacy systems and enables incremental adoption of \gls{PQC}. However, such a strategy inherently limits security to the weaker of the two schemes. If classical signatures are rendered forgeable (e.g., by quantum adversaries), the hybrid signature becomes insecure unless post-quantum verification is mandated and classical-only acceptance is gradually deprecated.}

\begin{figure}[!h]
\centering

\begin{subfigure}[b]{\linewidth}
\centering
\resizebox{\linewidth}{!}{%
\begin{tikzpicture}[
badge/.style={
  circle,
  minimum size=2.2cm,
  draw,
  line width=2.5pt,
  fill=white,
  inner sep=0pt
},
  inner_badge/.style={
  circle,
  minimum size=2.2cm,
  draw,
  line width=1.9pt,
  dashed,
  fill=none
  inner sep=0pt
},
  pqc_section/.style={
    rectangle,
    rounded corners=2pt,
    minimum width=1.4cm,
    minimum height=2.6cm,
    draw=pqcorange,
    line width=1.5pt,
    fill=orange!15
  },
  classic_section/.style={
    rectangle,
    rounded corners=2pt,
    minimum width=1.4cm,
    minimum height=2.6cm,
    draw=classicblue,
    line width=1.5pt,
    fill=blue!15
  },
  security_badge/.style={
    circle,
    minimum size=1.5cm,
    draw,
    line width=2.5pt,
    fill=white,  
  },
  inner_badge/.style={
    circle,
    minimum size=1.1cm,
    draw,
    line width=1pt,
    dashed,
    fill=none
  },
    crypto_link_node/.style={ 
        circle,
        minimum size=1cm,
        draw=gray!70!black,
        line width=1.5pt,
        fill=gray!10,
        font=\small\bfseries,
        text=gray!80!black,
    },
        block_link_arrow/.style={ 
        ->,
        line width=2pt,
        gray!70!black,
        >=Stealth,
        shorten >=2pt,
        shorten <=2pt
    },
  chain_link/.style={
    ellipse,
    minimum width=0.8cm,
    minimum height=0.4cm,
    draw=blue,
    line width=2pt,
    fill=white!30
  },
 combiner_circle/.style={
  circle,
  draw=orange!90,
  fill=orange!15,
  line width=2pt,
  minimum size=1.1cm,
  inner sep=0pt,
  text=orange!90!black,
  align=center
 },
  security_system/.style={
    rectangle,
    rounded corners=5pt,
    minimum width=5cm,
    minimum height=1.2cm,
    draw=orange!90,
    line width=2.0pt,
    fill=orange!15,
     text=orange!90!black,
  },
  connection_line/.style={
    line width=2pt,
    purple!70
  },
  arrow/.style={
    ->,
    line width=2.5pt,
    >=Stealth,
    orange!80
  },
  detail_block_style/.style={
    draw=gray!70,
    thick,
    rectangle,
    minimum width=2.8cm,
    minimum height=0.5cm,
    fill=gray!5,
    rounded corners=2pt
  },
  detail_tx_style/.style={
    rectangle,
    minimum width=2.6cm,
    minimum height=0.6cm,
    fill=blue!10,
    draw=blue!30,
    rounded corners=1pt,
    font=\footnotesize\ttfamily,
    align=center,
    text=black
  },
  detail_separator_style/.style={
    draw=gray!100,
    thick,
    dashed
  },
  detail_arrow_style/.style={
    ->,
    thick,
    gray!70,
    >=stealth
  },
  detail_label_style/.style={
    font=\scriptsize,
    anchor=west,
    text=gray!100
  }
]

\definecolor{classicalblue}{RGB}{25,118,210}
\definecolor{quantumpurple}{RGB}{123,31,162}

\definecolor{pqcorange}{RGB}{255,152,0}
\definecolor{classicblue}{RGB}{33,150,243}
\definecolor{facefrontL}{RGB}{102,153,255}
\definecolor{facefrontR}{RGB}{72,124,200}
\definecolor{faceleft}{RGB}{72,124,200}
\definecolor{facetopL}{RGB}{187,212,255}
\definecolor{facetopR}{RGB}{160,190,255}

\tikzset{
 cube/.pic={
  \def\cubex{3}
  \def\cubey{3}
  \def\cubez{3}
  \def\half{\cubex/2}
  \def\epsilon{0.015} 

  \fill[facetopL!50!white] (0,\cubey,0) -- ++(0,0,\cubez) -- ++(\half,0,0) -- ++(0,0,-\cubez) -- cycle;
  \fill[facetopR!50!white] (\half,\cubey,0) -- ++(0,0,\cubez) -- ++(\half,0,0) -- ++(0,0,-\cubez) -- cycle;

  \fill[facetopL] (0,0,\cubez) -- ++(\half,0,0) -- ++(0,\cubey,0) -- ++(-\half,0,0) -- cycle;
  \fill[facetopR] (\half,0,\cubez) -- ++(\half,0,0) -- ++(0,\cubey,0) -- ++(-\half,0,0) -- cycle;

  \fill[faceleft] (0,0,0) -- ++(0,0,\cubez) -- ++(0,\cubey,0) -- ++(0,0,-\cubez) -- cycle;

  \fill[facefrontL] (0,0,0) -- ++(\half,0,0) -- ++(0,0,\cubez) -- ++(-\half,0,0) -- cycle;
  \fill[facefrontR] (\half,0,0) -- ++(\half,0,0) -- ++(0,0,\cubez) -- ++(-\half,0,0) -- cycle;

  \draw[black, thick] (0,0,0) -- ++(\cubex,0,0) -- ++(0,\cubey,0) -- ++(-\cubex,0,0) -- cycle;
  \draw[black, thick] (0,0,0) -- ++(0,0,\cubez);
  \draw[black, thick] (\cubex,0,0) -- ++(0,0,\cubez);
  \draw[black, thick] (0,\cubey,0) -- ++(0,0,\cubez);
  \draw[black, thick] (\cubex,\cubey,0) -- ++(0,0,\cubez);
  \draw[black, thick] (0,0,\cubez) -- ++(\cubex,0,0) -- ++(0,\cubey,0) -- ++(-\cubex,0,0) -- cycle;

\draw[black, thick, dashed] (\half,0,0) -- (\half,0,\cubez);

\draw[black, thick, dashed] (\half,0,\cubez) -- (\half,\cubey,\cubez);

\draw[black, thick, dashed] (\half,\cubey,0) -- (\half,\cubey,\cubez);

\draw[black, thick, dashed] (\half,0,0) -- (\half,\cubey,0);

\coordinate (-center) at (\cubex/2,\cubey/2,\cubez/2); 
  \coordinate (-bottom) at (\cubex/2,\cubey/2,0);     

 }
}

\pic (block1) at (-6,0) {cube};
\node at (-6,1) {\textbf{Block N-1}};
\pic (block2) at (0,0) {cube};
\node at (0,1) {\textbf{Block N}};
\pic (block3) at (6,0) {cube};
\node at (6,1) {\textbf{Block N+1}};


\node[crypto_link_node] (hash_node1) at ($(block1-center)!0.5!(block2-center)+(0,-0.2)$) {\faHashtag}; 
\draw[block_link_arrow] ($(block1-center)+(0,-0.2)$) -- (hash_node1);
\draw[block_link_arrow] (hash_node1) -- ($(block2-center)+(0,-0.2)$);

\node[crypto_link_node] (hash_node2) at ($(block2-center)!0.5!(block3-center)+(0,-0.2)$) {\faHashtag}; 
\draw[block_link_arrow] ($(block2-center)+(0,-0.2)$) -- (hash_node2);
\draw[block_link_arrow] (hash_node2) -- ($(block3-center)+(0,-0.2)$);


\tikzset{
 classicalbadge/.pic={

  \node[badge, draw=classicalblue, inner sep=0pt] at (0,0) {};

  \draw[classicalblue, dashed, line width=1pt] (0,0) circle [radius=0.8cm];

  \node[classicalblue] at (0,0.4) {\tiny\textbf{SECURITY ASSURANCE}};
  \node[classicalblue] at (0,-0.4) {\tiny \textbf{CLASSICAL}};

\draw[classicalblue, line width=1pt] (0.3,-0.2) rectangle (-0.3,0.2);
\draw[classicalblue, line width=1pt] (0,0.02) circle (0.1);
\fill[classicalblue] (0.2,-0.2) rectangle (-0.2,0);
 }
}
\tikzset{
 quantumbadge/.pic={
  \node[badge, draw=quantumpurple, inner sep=0pt] at (0,0) {};

  \draw[quantumpurple, dashed, line width=1pt] (0,0) circle [radius=0.8cm];

  \node[quantumpurple] at (0,0.4) {\tiny\textbf{SECURITY ASSURANCE}};
  \node[quantumpurple] at (0,-0.4) {\tiny \textbf{QUANTUM-SAFE}};

\draw[quantumpurple, line width=1pt] (0.3,-0.2) rectangle (-0.3,0.2);
\draw[quantumpurple, line width=1pt] (0,0.02) circle (0.1);
\fill[quantumpurple] (0.2,-0.2) rectangle (-0.2,0);

 }
}

\pic[scale=0.8, transform shape] at (0.3,3.5) {quantumbadge};

\pic[scale=0.8, transform shape] at (2.4,3.5) {classicalbadge};

\node[combiner_circle] (combiner1) at (1.35,3.5) {\scriptsize Combiner};

\pic[scale=0.8, transform shape] at (-5.5,3.5) {quantumbadge};

\pic[scale=0.8, transform shape] at (-3.4,3.5) {classicalbadge};

\node[combiner_circle] (combiner2) at (-4.45,3.5) {\scriptsize Combiner};

\pic[scale=0.8, transform shape] at (6.3,3.5) {quantumbadge};

\pic[scale=0.8, transform shape] at (8.4,3.5) {classicalbadge};

\node[combiner_circle] (combiner3) at (7.35,3.5) {\scriptsize Combiner};

\node[security_system,below of=combiner1, yshift=-5.1cm] (security_system){\textbf{Overall System Security}};
\draw[arrow] (combiner1) -- (security_system.north);
\draw[arrow] (combiner2) |- (security_system.west);
\draw[arrow] (combiner3) |- (security_system.east);

\begin{scope}[shift={(10cm,1.2cm)}, scale=2]

   \node[font=\normalsize\bfseries, color=gray!80] at (1.4,2.3) {Transaction-Level};

   \draw[detail_block_style, line width=1.5pt] (0,-0.1) rectangle (2.8,2.2);

   \node[font=\small, color=gray!80] at (1.4,2.05) {\textbf{Block Header}};
   \draw[detail_separator_style, line width=0.8pt] (0.1,1.8) -- (2.7,1.8);

\node[detail_tx_style,scale=1.2] (tx1) at (1.4,1.55) { $\text{Tx}_1 : (\text{msg}_1, \text{Sig}_1^{\text{PQC}} \circ \text{Sig}_1^{\text{Classic}})$};

   \draw[detail_separator_style, line width=0.8pt] (0.1,1.3) -- (2.7,1.3);

\node[detail_tx_style,scale=1.2] (tx2) at (1.4,1.05)   
{ $\text{Tx}_2 : (\text{msg}_2, \text{Sig}_2^{\text{PQC}} \circ \text{Sig}_2^{\text{Classic}})$};

   \draw[detail_separator_style, line width=0.8pt] (0.1,0.8) -- (2.7,0.8);

   \node[detail_tx_style,scale=1.2] (tx3) at (1.4,0.55) { $\text{Tx}_3 : (\text{msg}_3, \text{Sig}_3^{\text{PQC}} \circ \text{Sig}_3^{\text{Classic}})$};

   \draw[detail_separator_style, line width=0.8pt] (0.1,0.3) -- (2.7,0.3);
   \node[font=\small, color=gray!80] at (1.4,0.1) {\textbf{Block Footer}};

   \draw[dashed, gray!50, line width=2pt] ($(block3-center)+(1, 0.5)$) .. controls +(right:0.6cm) and +(left:0.6cm) .. (0,1.2); 

   \node[font=\small, color=gray!80] at (0.65,-0.2) {{$\circ$: Hybrid Combiner}};
  \node[font=\normalsize, color=blue!70] at (1.4,-0.4) {\textit{Detail of Block N+1 }};

\end{scope}

\end{tikzpicture}}
\caption{}
\label{fig:composite_block_structure}
\end{subfigure}

\begin{subfigure}[b]{\linewidth}
\centering

\resizebox{\textwidth}{!}{%

\definecolor{compositecolor}{RGB}{102,51,153} 
\definecolor{accentcolor}{RGB}{255,87,51} 
\definecolor{consensusgrey}{RGB}{52,73,94} 
\definecolor{blockcolor}{RGB}{236,240,241} 
\definecolor{systemcolor}{RGB}{46,134,193} 
\definecolor{signaturecolor}{RGB}{26,188,156} 

\resizebox{\linewidth}{!}{%
\begin{tikzpicture}[
 node distance=4cm,
 composite_system_block/.style={
  draw=systemcolor,
  fill=white,
  rounded corners=8pt,
  inner sep=10pt,
  text width=3.5cm,
  align=center,
  drop shadow={shadow xshift=0.2ex, shadow yshift=-0.2ex, fill=black!15},
  line width=2pt
 },
 participant/.style={
  draw=compositecolor,
  fill=compositecolor!15,
  circle,
  minimum size=1.4cm,
  inner sep=4pt,
  font=\bfseries\LARGE,
  drop shadow={shadow xshift=0.15ex, shadow yshift=-0.15ex, fill=black!12}
 },
 blockchain_block/.style={
  draw=consensusgrey,
  fill=white,
  rectangle,
  rounded corners=8pt,
  minimum width=4.2cm,
  minimum height=1.2cm,
  align=center,
  font=\small\bfseries,
  line width=1.8pt,
  drop shadow={shadow xshift=0.15ex, shadow yshift=-0.15ex, fill=black!12}
 },
 transaction/.style={
  draw=accentcolor!80,
  fill=accentcolor!5,
  rectangle,
  rounded corners=4pt,
  minimum width=3.4cm,
  minimum height=0.6cm,
  font=\footnotesize\bfseries,
  line width=1pt,
  align=center
 },
 composite_sig/.style={
  draw=signaturecolor!80,
  fill=signaturecolor!10,
  rectangle,
  rounded corners=3pt,
  minimum width=3.4cm,
  minimum height=0.4cm,
  font=\scriptsize,
  line width=1pt,
  align=center
 },
 consensus_layer/.style={
  draw=consensusgrey,
  fill=consensusgrey!3,
  rectangle,
  rounded corners=12pt,
  minimum width=8cm,
  minimum height=2cm,
  align=center,
  font=\large\bfseries,
  line width=2pt,
  drop shadow={shadow xshift=0.2ex, shadow yshift=-0.2ex, fill=black!15}
 },
 arrcomposite/.style={
  draw=compositecolor,
  -{Stealth[length=4mm]},
  line width=2.5pt,
  shorten >=2pt,
  shorten <=2pt
 },
 arrtx/.style={
  draw=accentcolor,
  -{Stealth[length=7mm]},
  line width=2pt,
  shorten >=2pt,
  shorten <=2pt
 },
]

 \node (A) [participant] at (-3,6) {\color{white}\textbf{A}};
 \node (B) [participant] at (-6,2) {\color{white}\textbf{B}};
 \node (C) [participant] at (3,6) {\color{white}\textbf{C}};
 \node (D) [participant] at (6,2) {\color{white}\textbf{D}};

 \node (sysA) [composite_system_block] at (-6,8) {
  \textbf{\color{systemcolor}\large Composite} \\
  \textbf{\color{systemcolor}\large Cryptographic System} \\[2mm]
  \color{black}\footnotesize
  • Hybrid Signatures \\
  • Quantum-safe Hashing \\
  • Integrated Security Layer
 };
 
 \node (sysB) [composite_system_block] at (-9,2) {
  \textbf{\color{systemcolor}\large Composite} \\
  \textbf{\color{systemcolor}\large Cryptographic System} \\[2mm]
  \color{black}\footnotesize
  • Hybrid Signatures \\
  • Quantum-safe Hashing \\
  • Integrated Security Layer
 };

 \node (sysC) [composite_system_block] at (6,8) {
  \textbf{\color{systemcolor}\large Composite} \\
  \textbf{\color{systemcolor}\large Cryptographic System} \\[2mm]
  \color{black}\footnotesize
  • Hybrid Signatures \\
  • Quantum-safe Hashing \\
  • Integrated Security Layer
 };
 
 \node (sysD) [composite_system_block] at (9,2) {
  \textbf{\color{systemcolor}\large Composite} \\
  \textbf{\color{systemcolor}\large Cryptographic System} \\[2mm]
  \color{black}\footnotesize
  • Hybrid Signatures \\
  • Quantum-safe Hashing \\
  • Integrated Security Layer
 };

 \node (consensus_container) [consensus_layer] at (0,1) {
   \color{consensusgrey}\textbf{Unified Consensus Mechanism} \\[1mm]
   \color{consensusgrey}\small\text{Validates, Orders \& Commits} \\
   \color{consensusgrey}\small\text{Creates Proposed/Validated Blocks for \gls{BC}}
 };

 \node[
  draw=compositecolor!60,
  fill=compositecolor!3,
  rectangle,
  rounded corners=12pt,
  minimum width=12cm,
  minimum height=4cm,
  below=2cm of consensus_container,
  line width=2pt,
  dashed
 ] (blockchain_container) {};
 
 \node[above=2mm of blockchain_container.north, font=\large\bfseries, color=compositecolor] {\gls{BC} Ledger (Output of Consensus)};

 \node (block1) [blockchain_block, scale=0.7] at (-4,-4) {
  \color{consensusgrey}\textbf{\large Block N} \\[4mm]
  \begin{tikzpicture}
    \node (tx1) [transaction] at (0,0.9) {\color{accentcolor!80}\textbf{Tx1: A→B}};
    \node (sig1) [composite_sig] at (0,0.3) {\color{signaturecolor}\textbf{Hybrid: RSA+Dilithium}};
    \node [font=\scriptsize, color=black!60] at (0,-0.1) {Hash: SHA-3};
    
    \node (tx2) [transaction] at (0,-0.7) {\color{accentcolor!80}\textbf{Tx2: C→D}};
    \node (sig2) [composite_sig] at (0,-1.3) {\color{signaturecolor}\textbf{Hybrid: ECDSA+Falcon}};
    \node [font=\scriptsize, color=black!60] at (0,-1.7) {Hash: BLAKE3};
  \end{tikzpicture}
 };

 \node (block2) [blockchain_block, scale=0.7] at (0,-4) {
  \color{consensusgrey}\textbf{\large Block N+1} \\[4mm]
  \begin{tikzpicture}
    \node (tx3) [transaction] at (0,0.9) {\color{accentcolor!80}\textbf{Tx3: B→C}};
    \node (sig3) [composite_sig] at (0,0.3) {\color{signaturecolor}\textbf{Hybrid: Ed25519+SPHINCS+}};
    \node [font=\scriptsize, color=black!60] at (0,-0.1) {Hash: SHA-3};
    
    \node (tx4) [transaction] at (0,-0.7) {\color{accentcolor!80}\textbf{Tx4: D→A}};
    \node (sig4) [composite_sig] at (0,-1.3) {\color{signaturecolor}\textbf{Hybrid: P-256+Dilithium}};
    \node [font=\scriptsize, color=black!60] at (0,-1.7) {Hash: BLAKE3};
  \end{tikzpicture}
 };

 \node (block3) [blockchain_block, scale=0.7] at (4,-4) {
  \color{consensusgrey}\textbf{\large Block N+2} \\[4mm]
  \begin{tikzpicture}
    \node (tx5) [transaction] at (0,0.9) {\color{accentcolor!80}\textbf{Tx5: A→C}};
    \node (sig5) [composite_sig] at (0,0.3) {\color{signaturecolor}\textbf{Hybrid: RSA+Dilithium}};
    \node [font=\scriptsize, color=black!60] at (0,-0.1) {Hash: SHA-3};
    
    \node (tx6) [transaction] at (0,-0.7) {\color{accentcolor!80}\textbf{Tx6: B→D}};
    \node (sig6) [composite_sig] at (0,-1.3) {\color{signaturecolor}\textbf{Hybrid: P-384+Falcon}};
    \node [font=\scriptsize, color=black!60] at (0,-1.7) {Hash: BLAKE3};
  \end{tikzpicture}
 };

 \draw [line width=2.5pt, color=compositecolor, -Stealth] (block1.east) -- (block2.west);
 \draw [line width=2.5pt, color=compositecolor, -Stealth] (block2.east) -- (block3.west);

 \draw [arrcomposite, <->] (A) to[bend right=15] (B);
 \draw [arrcomposite, <->] (A) to[bend left=15] (C) ;
 \draw [arrcomposite, <->] (A) to[bend left=20] (D) ;
 \draw [arrcomposite, <->] (B) to[bend left=20] (C) ;
 \draw [arrcomposite, <->] (B) to[bend left=15] (D);
 \draw [arrcomposite, <->] (C) to[bend left=15] (D);

 \draw [arrtx] (consensus_container.south west) to[bend right=30] (block1.north);
 \draw [arrtx] (consensus_container.south) -- (block2.north);
 \draw [arrtx] (consensus_container.south east) to[bend left=45] (block3.north);
 
 \draw [arrtx] (A.south) to[bend right=20] (consensus_container.north);
 \draw [arrtx] (B.east) to[bend left=10] (consensus_container.west);
 \draw [arrtx] (C.south) to[bend left=20] (consensus_container.north);
 \draw [arrtx] (D.west) to[bend right=10] (consensus_container.east) ;

 \draw [dashed, gray!50, line width=1.2pt] (A) -- (sysA);
 \draw [dashed, gray!50, line width=1.2pt] (B) -- (sysB);
 \draw [dashed, gray!50, line width=1.2pt] (C) -- (sysC);
 \draw [dashed, gray!50, line width=1.2pt] (D) -- (sysD);

 \begin{scope}[yshift=-12cm, xshift=0cm]
 \draw [arrcomposite, <->] (-5.5,5) -- (-3.5,5) node [right, font=\small] {\textbf{Composite Channels (Data + Hybrid Signatures)}};
  \draw [arrtx] (-5.5,4.3) -- (-3.5,4.3) node [right, font=\small] {\textbf{Consensus Process (Input $\rightarrow$ Validation $\rightarrow$ Output)}};
  \draw [line width=2.5pt, color=compositecolor, -Stealth] (-5.5,3.6) -- (-3.5,3.6) node [right, font=\small] {\textbf{Blockchain (Immutable Ledger)}};
  
  \node [composite_sig, minimum width=2cm, minimum height=0.4cm] at (5,5) {\color{signaturecolor}\footnotesize \textbf{Hybrid Sig}};
  \node [transaction, minimum width=2cm, minimum height=0.5cm] at (5,4.3) {\color{accentcolor!80}\footnotesize \textbf{Transaction}};
  \node [blockchain_block, minimum width=2cm, minimum height=0.6cm] at (5,3.6) {\color{consensusgrey}\tiny \textbf{Block}};
 \end{scope}

\end{tikzpicture}%
}
}
\caption{}
\label{fig:composite_system_workflow}
\end{subfigure}
\caption{{Composite \gls{BC} Architectures. 
(a) \textit{Block Structure:} Each transaction incorporates both post-quantum and classical signatures, combined through cryptographic combiners for unified security assurance. 
(b) \textit{System Workflow:} Participants submit hybrid-signed transactions through a unified consensus mechanism that validates and records them on the ledger.}}
\label{fig:composite_architecture_combined}
\end{figure}

  {To accommodate smart contract platforms {(e.g., Ethereum)}, composite ledgers must \textit{extend virtual machine capabilities} to support dual-signature verification, cryptographic combiners, and runtime selection based on transaction-level security policies. For backward compatibility during transition, \textit{composite blocks can define optional PQC segments} interpretable only by upgraded nodes, allowing classical-only participants to process legacy transactions without disruption. \textit{Cryptographic agility} should also be supported via per-transaction algorithm negotiation, enabling future upgrades without hard forks or complete protocol redesign~\cite{yang2024survey}. While enabling robust security, the protocol-layer complexity introduced by signature verification redundancy and consensus logic must be managed through \textit{modular design and formal verification} to prevent new attack surfaces. Unlike the non-composite approach, the composite architecture maintains a \textit{single unified consensus pipeline} that validates both classical and post-quantum transactions. This simplification streamlines network operations and eliminates synchronization issues inherent in multi-ledger systems. Figure~\ref{fig:composite_architecture_combined} illustrates two complementary perspectives: (a) a structural view of block-level hybrid cryptographic segmentation and combiners; (b) a system-wide workflow of transaction flow, composite signature generation, consensus processing, and ledger formation. This dual-view highlights how hybrid strategies support post-quantum migration while ensuring backward compatibility and operational continuity.
\item \textit{Supply Chain Case Study:}  
Consider a hypothetical global supply chain consortium adopting the composite architecture to track products from manufacturing to retail. The system secures routine tracking updates with classical cryptography for efficiency, while critical events such as customs clearance or ownership transfers employ \gls{PQC} for enhanced protection.
The unified ledger simplifies integration with existing enterprise systems and seamlessly adapts for security-sensitive operations. Manufacturers can retain existing interfaces with minimal modifications, while customs authorities and financial institutions benefit from quantum-resistant safeguards for high-value transactions.
A key advantage of the composite design is its transparency. Unlike the non-composite approach, users interact with a single ledger, and cryptographic mechanisms are applied automatically based on predefined policies and transaction attributes. This abstraction reduces user complexity while maintaining strong, context-aware security guarantees.}
\end{enumerate}
\end{enumerate}

\subsection{{Risk Assessment and Security Posture}}\label{hybrid-Risk}
{Hybrid \gls{BC} architectures enhance resilience during the transition to post-quantum security by leveraging cryptographic diversity. These architectures require transaction-level risk assessments based on the specific cryptographic primitives used for signing and verification.
In \emph{composite} architectures, each transaction is simultaneously protected by both classical and post-quantum cryptographic primitives, typically through dual signatures or dual-layer encryption. This design provides a \textit{defense-in-depth} security posture: as long as at least one primitive remains secure, the transaction maintains its integrity. Accordingly, the effective transaction-level risk is determined by the stronger (i.e., lower-risk) of the two primitives:}
\begin{eqnarray*}
\text{Risk} = \min(\text{Risk}_{\text{Classic}}, \text{Risk}_{\text{PQC}}).
\end{eqnarray*}
{In  \emph{non-composite} architecture, risk evaluation must account for three distinct transaction types, each with different security properties and attack surfaces. The effective risk depends on the specific transaction type and cryptographic domains involved:}
\begin{equation*}
\text{Risk} =\hspace{-0.1cm}
\begin{cases*}
\text{Risk}_{\text{Classic}}, & if $tx\in $ classic ledger only, \\
\text{Risk}_{\text{PQC}}, & if $tx \in$ PQC ledger only, \\
\begin{aligned}
\max(\text{Risk}_{\text{Classic}}, \text{Risk}_{\text{PQC}}) \\
+\, \text{Risk}_{\text{Coord}},
\end{aligned}
\hspace{-0.2cm}& if $tx$ spans both ledgers,
\end{cases*}
\end{equation*}
{where $\text{Risk}_{\text{Coord}}$ captures additional risk arising from cross-ledger coordination, including quantum-safe \glspl{HTLC}, oracle dependencies, atomic swap mechanisms, and synchronization constraints. Formally, this coordination risk can be expressed as a function of its key contributing factors:}
\begin{equation*}
\text{Risk}_{\text{Coord}} = f(\text{Risk}_{\text{Oracle}}, \text{Risk}_{\text{HTLC}}, \text{Risk}_{\text{Sync}}, \text{Risk}_{\text{Timeout}}). 
\end{equation*}


{This architecture supports risk segregation, enabling gradual migration and tiered security based on sensitivity and risk tolerance. However, it does not ensure system-wide minimum risk. Cross-ledger operations are particularly vulnerable to coordination failures, oracle compromise, and downgrade attacks that may force reliance on weaker cryptographic paths. Additional failure modes include partial transaction execution, timeout exposures, and architectural complexity that expands the attack surface.}

\subsection{Hybrid \gls{BC} Architectures: Comparative Analysis, Design Challenges, and Strategic Pathways}\label{hybrid-comparative}

{Both non-composite and composite hybrid \gls{BC} architectures offer quantum-resistant pathways with distinct trade-offs. Non-composite architectures achieve resilience through cryptographic isolation but face significant coordination complexity. Key challenges include dual ledger synchronization requiring complex middleware for transactional atomicity~\cite{WANG2024103977}, interoperability mismatches in signature schemes and data models~\cite{LI2025100286}, and performance gaps causing finality delays where assets remain visible across chains~\cite{refc17}. 
Additional barriers include smart contract execution across heterogeneous domains~\cite{9631953,10261941}, amplified computational, networking, and storage demands due to dual ledger maintenance—particularly under \gls{PQC} key and signature workloads—and persistent cross-ledger synchronization challenges in non-composite chains.
Strategic solutions include optimistic execution with \gls{PQC} auditing, enabling rapid classical processing with provisional finality while \gls{PQC} chains act as asynchronous verifiers. Dual-layer finality mechanisms~\cite{refc25} provide fast provisional confirmation for routine transactions and rigorous \gls{PQC} verification for critical ones, with ML models dynamically setting thresholds~\cite{refml28,refml17,refml10,refml1}. Selective \gls{PQC} verification mirrors only high-risk transactions to \gls{PQC} chains, supported by state checkpointing, cross-chain locking~\cite{hu2025quantum}, and unified ledger anchoring. Migration protocols~\cite{Ding2024Genuine,liu2022extending} ensure continuity during quantum threats through immediate fallback and secure state-preserving transitions.}

{Composite architectures streamline integration through unified ledgers but incur scalability challenges. Block sizes increase 60–80\% from dual cryptographic operations, with ML-DSA signatures 38–52$\times$ larger than ECDSA (2,420–3,309 vs. 64 bytes)~\cite{FIPS204} and ML-KEM keys 12.5–24.5$\times$ larger than ECDH (800–1,568 vs. 64 bytes)~\cite{FIPS203}. These designs also depend on evolving \gls{PQC} standards~\cite{nist2024transition,NISTIR8545} and require real-time performance optimization.
Mitigation strategies include layered verification, where \gls{PQC} validation is applied to high-value transactions while routine operations use classical cryptography, guided by ML-based risk assessment~\cite{refml27}. Robust auditing uses tamper-proof logs with adaptive criteria adjustment. 
Off-chain computation~\cite{Ding2024Genuine,liu2022extending} with recursive \glspl{ZKP}~\cite{refc14,refc33} mitigates the computational overhead of \gls{PQC} verification, while time-lock mechanisms~\cite{bip199} ensure secure synchronization between off-chain computation and on-chain verification, and sidechain–mainchain coordination~\cite{Deng2025SFPoW} preserves ledger consistency.
Backward-compatible  blocks~\cite{petrenko2019assessing,alnahawi2023state} embed PQC sections for upgraded nodes, validated by unified consensus across legacy and \gls{PQC} schemes, with per-transaction algorithm negotiation enabling crypto-agility and upgrades without hard forks.
Figure~\ref{fig:hybrid_bc_complete_framework} summarizes the key comparative findings, maps design challenges to strategic pathways for both composite and non-composite hybrid \gls{BC} architectures, and provides a roadmap for future research priorities.}
\begin{figure}[!t] \vspace{-0.1cm}
\definecolor{compblue}{RGB}{198,219,239}   
\definecolor{challorange}{RGB}{251,180,174} 
\definecolor{solugreen}{RGB}{204,235,197}  
\definecolor{arrowgray}{RGB}{80,80,80}     
\centering
\resizebox{\linewidth}{!}{
\begin{tikzpicture}[
    font=\footnotesize,
        concept/.style={rectangle, draw,  align=center, minimum height=1.2cm, font=\normalsize\bfseries},
    header/.style={  align=center,    font=\bfseries},
        comparison/.style={rectangle, draw, text width=4cm, minimum height=16.95cm, align=justify, fill=blue!15},
    challenge/.style={rectangle, draw, text width=4.8cm, minimum height=1.3cm, align=justify, fill=red!15},
    solution/.style={rectangle, draw, text width=6.9cm, minimum height=1.3cm, align=justify, fill=green!15},
    arrow/.style={->, thick, >=stealth, color=arrowgray},
    node distance=0.2cm
]

\node[concept, fill=blue!15,text width=4cm] (comparative) {Comparative Analysis};
\node[concept, fill=red!15, ,text width=4.8cm, right=of comparative,xshift=0.2cm] (challenges) {Design Challenges};
\node[concept, fill=green!15, ,text width=6.9cm, right=of challenges,xshift=0.2cm] (pathways) {Strategic Pathways};

\node[comparison, below=of comparative,yshift=-0.4cm] (comp) {
   \textbf{Performance:}
    \begin{myBullets}
        \item PQC: 10-16 TPS~\cite{abelian}
        \item Classical: 1000+ TPS
        \item Cross-ledger sync latency~\cite{perez2023analyzing}
    \end{myBullets}
    \textbf{Block Size:}
    \begin{myBullets}
        \item 60-80\% increase~\cite{FIPS204,FIPS203}
        \item ML-DSA: 38-52$\times$ larger~\cite{FIPS204}
        \item ML-KEM: 12.5-24.5$\times$ larger~\cite{FIPS203}
    \end{myBullets}
    \textbf{Security Strength:}
    \begin{myBullets}
        \item Composite: 
        \begin{itemize}
            \item $\max$(Classical,PQC)
        \end{itemize}
        \item Non-composite: 
        \begin{itemize}
            \item         $\min$(Public,Private)
        \end{itemize}
    \end{myBullets}
    \textbf{Trade-offs:}
    \begin{myBullets}
        \item Implementation complexity
        \item Operational flexibility
        \item Risk management approaches
        \item Cost structures
        \item Future adaptability
    \end{myBullets}
};

\node[header,  below=of challenges, xshift=2.8cm,yshift=0.2cm] (nc_header) {Non-Composite};

\node[challenge, below=of challenges, yshift=-0.4cm] (nc_c1) {
    \textbf{Dual Ledger Coordination}: Complex middleware for atomicity across divergent crypto primitives~\cite{WANG2024103977} 
};
\node[solution, right=of nc_c1, xshift=0.2cm] (nc_s1) {
    \begin{myBullets}\vspace{-0.2cm}
  \item State checkpointing with crypto summaries
  \item Cross-chain transaction locking~\cite{hu2025quantum}
  \item Unified ledger anchoring
    \end{myBullets}
};

\node[challenge, below=of nc_c1] (nc_c2) {
    \textbf{Cross-ledger interoperability}: Signature schemes, key formats, data models, fee structure mismatches~\cite{LI2025100286} 
};
\node[solution, right=of nc_c2, xshift=0.2cm] (nc_s2) {
    \begin{myBullets}\vspace{-0.2cm}
  \item Advanced coordination logic with rollback/ compensation/ timeout~\cite{9631953,10261941}
  \item Asynchronous state updates with reconciliation
    \end{myBullets}
};

\node[challenge, below=of nc_c2] (nc_c3) {
     \textbf{Performance gaps}: Classical vs PQC finality delays, state discrepancies~\cite{refc17} 
};
\node[solution, right=of nc_c3, xshift=0.2cm] (nc_s3) {
    \begin{myBullets}\vspace{-0.2cm}
  \item Optimistic execution + PQC auditing
  \item Automated rollback with fork resolution
  \item Fraud-proof mechanisms with incentives
    \end{myBullets}
};

\node[challenge, below=of nc_c3] (nc_c4) {
    \textbf{Synchronization issues}: Smart contract execution across heterogeneous domains 
};
\node[solution, right=of nc_c4, xshift=0.2cm] (nc_s4) {
    \begin{myBullets}\vspace{-0.2cm}
  \item Dual-layer finality: fast provisional + rigorous PQC~\cite{refc25}
  \item ML-based dynamic thresholds using risk profiles~\cite{refml28,refml17,refml10,refml1}
    \end{myBullets}
};

\node[challenge, below=of nc_c4] (nc_c5) {
\textbf{Resource overhead}: Computational/networking/storage affecting resource-constrained participants 
};
\node[solution, right=of nc_c5, xshift=0.2cm] (nc_s5) {
    \begin{myBullets}\vspace{-0.2cm}
  \item Selective PQC verification (critical transactions only)
  \item Off-chain computation with Merkle anchoring~\cite{Ding2024Genuine,liu2022extending}
    \end{myBullets}
};

\node[challenge, below=of nc_c5] (nc_c6) {
\textbf{Migration complexity}: Quantum compromise scenarios requiring system continuity 
};
\node[solution, right=of nc_c6, xshift=0.2cm] (nc_s6) {
    \begin{myBullets}\vspace{-0.2cm}
  \item Dual-chain fallback to PQC ledger
  \item State snapshot preservation
  \item Comprehensive migration protocols
    \end{myBullets}
};

\node[header, below=of nc_c6, xshift=2.8cm,yshift=0.2cm]  (c_header) {Composite};

\node[challenge, below=of nc_c6, yshift=-0.4cm] (c_c1) {
\textbf{Block size inflation}: 60-80\% increase; ML-DSA 38-52$\times$ larger (2,420-3,309 vs 64 bytes); ML-KEM 12.5-24.5$\times$ larger (800-1,568 vs 64 bytes)~\cite{FIPS204,FIPS203} 
};
\node[solution, right=of c_c1, xshift=0.2cm] (c_s1) {
    \begin{myBullets}\vspace{-0.3cm}
  \item Layered verification: PQC for high-value, classical for routine
  \item ML-based classifiers for dynamic assignment
  \item Backward-compatible blocks with PQC sections
     \vspace{-0.1cm}\end{myBullets}
};

\node[challenge, below=of c_c1] (c_c2) {
     \textbf{PQC standards dependence}: Evolving standards volatility, protocol modification security risks 
};
\node[solution, right=of c_c2, xshift=0.2cm] (c_s2) {
    \begin{myBullets}\vspace{-0.2cm}
  \item Per-transaction algorithm negotiation~\cite{petrenko2019assessing,alnahawi2023state}
  \item Tamper-proof auditing with accountability
  \item Threat-adaptive algorithms
    \end{myBullets}
};

\node[challenge, below=of c_c2] (c_c3) {
\textbf{Dynamic performance optimization}: Real-time throughput-latency-security balancing for varying crypto demands~\cite{li2021close} 
};
\node[solution, right=of c_c3, xshift=0.2cm] (c_s3) {
    \begin{myBullets}\vspace{-0.3cm}
  \item Off-chain recursive \glspl{ZKP} with on-chain attestations~\cite{Ding2024Genuine,refc14,refc33}
  \item Time-lock synchronization~\cite{bip199}
  \item Cross-environment consensus protocols
    \vspace{-0.1cm}\end{myBullets}
};

\node[challenge, below=of c_c3] (c_c4) {
 \textbf{Performance degradation}: PQC 10-16 TPS vs 1000+ classical; cross-ledger sync latency~\cite{abelian,perez2023analyzing} 
};
\node[solution, right=of c_c4, xshift=0.2cm] (c_s4) {
    \begin{myBullets}\vspace{-0.3cm}
  \item Sidechain-mainchain coordination: classical routine, hybrid critical~\cite{Deng2025SFPoW}
  \item Merkle proof validation with periodic state submissions
  \item Efficient communication for rapid discrepancy detection
     \vspace{-0.1cm}\end{myBullets}
};

\node[challenge, below=of c_c4] (c_c5) {
\textbf{Migration complexity}: Quantum threats without transaction history loss 

};
\node[solution, right=of c_c5, xshift=0.2cm] (c_s5) {
    \begin{myBullets}\vspace{-0.2cm}
  \item Freeze-and-migrate protocols for compromised sidechains
  \item Atomic migration ensuring secure transitions~\cite{Herlihy2018Atomic}
  \item Unified consensus validating both crypto types
    \end{myBullets}
};

\draw[arrow] (nc_c1) -- (nc_s1);
\draw[arrow] (nc_c2) -- (nc_s2);
\draw[arrow] (nc_c3) -- (nc_s3);
\draw[arrow] (nc_c4) -- (nc_s4);
\draw[arrow] (nc_c5) -- (nc_s5);
\draw[arrow] (nc_c6) -- (nc_s6);
\draw[arrow] (c_c1) -- (c_s1);
\draw[arrow] (c_c2) -- (c_s2);
\draw[arrow] (c_c3) -- (c_s3);
\draw[arrow] (c_c4) -- (c_s4);
\draw[arrow] (c_c5) -- (c_s5);
\draw[arrow] (comparative) -- (challenges);
\draw[arrow] (challenges) -- (pathways);
\end{tikzpicture}
}
\caption{Hybrid \gls{BC} Architectures: Design Challenges, and Strategic Pathways}
\vspace{-0.3cm}
\label{fig:hybrid_bc_complete_framework}
\end{figure}
\begin{table}[!htbp]
\centering
\caption{Strategic Comparison of Composite and Non-Composite Hybrid \gls{BC} Architectures: A Decision Framework}
\label{tab:comprehensive-hybrid-architecture}
\scriptsize
\resizebox{\linewidth}{!}{
\begin{tabular}{|p{0.19\linewidth}|p{0.45\linewidth}|p{0.45\linewidth}|}
\hline
\textbf{Dimension} & \textbf{Composite Architecture} & \textbf{Non-Composite  Architecture} \\ \hline

{Performance Impact} 
& Block size overhead of {60--80\%} due to dual signatures and larger PQC keys. Requires \emph{protocol adjustments} (e.g., consensus and storage optimizations) but not a complete chain rebuild. 
& Cross-chain synchronization and bridge latency dominate performance costs. Overhead arises from \emph{coordination mechanisms} rather than native network congestion.
\\ \hline

{Scalability Considerations} 
& All nodes must process both PQC and classical cryptographic operations, introducing intra-chain computational bottlenecks. 
& Dual-network strategy partitions workload across classical and PQC ledgers, which {optimizes per-chain throughput} by specialization, avoiding intra-chain saturation. \\ \hline

{Operational Flexibility} 
& Fine-grained, \emph{per-transaction adaptability}: cryptographic primitives can be selected dynamically within a single ledger.
& Coarse-grained, \emph{per-ledger adaptability}: users select between classical and PQC chains depending on security requirements; flexibility is fixed at the ledger level.
\\ \hline

{Risk Mitigation Strategies} 
& Unified auditability and \emph{crypto-agility}: blocks inherit the strongest available primitive, ensuring consistent end-to-end resilience. 
& Ledger isolation \emph{ensures cryptographic isolation} between classical and PQC domains, but cross-ledger coordination (bridges/swaps) introduces residual systemic risks.
\\ \hline

{Deployment Complexity} 
& Requires protocol redesign for dual signatures, consensus modifications, and per-transaction algorithm negotiation (see Sec.~VII.D). 
& Requires middleware and interoperability frameworks to maintain atomicity and consistency across distinct ledgers.
\\ \hline

\end{tabular}}
\end{table}

{The selection between composite and non-composite hybrid \gls{BC} architectures hinges on cryptographic strategy, operational constraints, security posture, and adaptability throughout the quantum transition. Table~\ref{tab:comprehensive-hybrid-architecture} provides a structured decision framework to align architectural choices with performance, scalability, and compliance objectives. Composite architectures are suited to ecosystems prioritizing interoperability, gradual \gls{PQC} adoption, and unified auditing with cryptographic agility. They offer broad compatibility and per-transaction flexibility but face scalability limits from enlarged block sizes and dual-cryptographic processing. Non-composite architectures ensure strong security isolation and per-ledger performance optimization, but at the cost of higher complexity, cross-ledger latency, and operational overhead.}

\section{Security Evaluation of Major \gls{BC} Platforms}\label{sec:platform}
This section presents a detailed analysis of the potential impact of \gls{QC} on major \gls{BC} platforms, including Bitcoin~\cite{nakamoto2008bitcoin}, Ethereum~\cite{buterin2013ethereum}, Ripple~\cite{armknecht2015ripple}, Litecoin~\cite{jumaili2021comparison}, and Zcash~\cite{hopwood2016zcash}, a privacy-focused cryptocurrency. The analysis examines their vulnerable components, assesses potential impacts, analyzes associated STRIDE threats, and provides an evaluation of likelihood, impact, and overall risk levels under the assumption of a sufficiently advanced \gls{QC} environment. Table~\ref{tab:quantum_impact} offers a consolidated reference for understanding these implications.
\begin{table*}[!htbp]
\caption{Impact of Quantum Computing on Major \gls{BC} Platforms}
\label{tab:quantum_impact}
\small
\resizebox{\textwidth}{!}{%
\begin{tabular}{|m{0.07\linewidth}|m{0.24\linewidth}|m{0.25\linewidth}|m{0.44\linewidth}|l|l|l|m{0.4\linewidth}|}
\hline
\textbf{Platform} & \textbf{Vulnerable Components} & \textbf{Potential Impacts} & \textbf{STRIDE Threats} & \textbf{L} & \textbf{I} & \textbf{R} & \textbf{Mitigation Strategies} \\ \hline
Bitcoin &
\begin{myBullets} 
\item \gls{ECDSA}: Used for digital signatures.
\item SHA-256: Used for hashing blocks.
 \vspace{-2ex} \end{myBullets} &
\begin{myBullets} 
\item Forgery of digital signatures, enabling unauthorized spending.
\item Breaking collision resistance of SHA-256, disrupting mining and block verification.
 \vspace{-2ex} \end{myBullets} &
\begin{myBullets} 
\item Spoofing: Impersonation through compromised signatures
\item Tampering: Alteration of transaction data or blocks
\item Repudiation: Disputes arising from forged transactions
\item \gls{DoS}: Mining and consensus disruption
\item Elevation of Privilege: Unauthorized network control through compromised keys
 \vspace{-2ex} \end{myBullets} &
\high & \high & \high &
\begin{myBullets} 
\item Transition to quantum-resistant signature schemes
\item Transition to a stronger hash function
\item Develop quantum-resistant \gls{PoW} or use alternative approaches such as proof-of-space or memory-hard \gls{PoW} to mitigate quantum mining advantages.
 \vspace{-2ex} \end{myBullets} \\ \hline
Ethereum &
\begin{myBullets} 
\item \gls{ECDSA}: Used for digital signatures.
\item Keccak-256: Used for hashing transactions and blocks.
\item \gls{ECIES}: Encrypts communication (less common).
 \vspace{-2ex} \end{myBullets} &
\begin{myBullets} 
\item Signature forgery and potential disruption of consensus mechanism.
\item Breach of encrypted communication.
\item Tampering with smart contract states or inputs.
 \vspace{-2ex} \end{myBullets} &
\begin{myBullets} 
\item Spoofing: Identity impersonation via ECDSA compromise
\item Tampering: Block manipulation by breaking Keccak-256
\item Repudiation: Disputes from forged or altered transactions
\item Information Disclosure: Leaked data through broken ECIES encryption
\item \gls{DoS}: Quantum attacks degrading network operations
\item Elevation of Privilege: Unauthorized access to smart contracts
 \vspace{-2ex} \end{myBullets} &
\high & \high & \high &
\begin{myBullets} 
\item Migration to quantum-resistant signature schemes
\item Replace ECIES with a post-quantum KEM 
\item Use a higher-security hash function in block hashing
\item Formal verification of smart contracts to eliminate vulnerabilities
\item Develop quantum-resistant smart contract libraries and frameworks
 \vspace{-2ex} \end{myBullets} \\ \hline
Ripple &
\begin{myBullets} 
\item \gls{ECDSA}: Used for digital signatures.
\item SHA-256: Used for hashing transactions.
 \vspace{-2ex} \end{myBullets} &
\begin{myBullets} 
\item Unauthorized manipulation of transactions and account balances.
\item Limited relevance for Information Disclosure due to Ripple's simpler metadata structure.
 \vspace{-2ex} \end{myBullets} &
\begin{myBullets} 
\item Spoofing: Forged transactions through signature compromise
\item Tampering: Manipulation of transaction data
\item Repudiation: Disputes arising from altered balances
\item \gls{DoS}: Attacks exploiting consensus mechanisms
\item Elevation of Privilege: Misuse of validator roles via key compromise
 \vspace{-2ex} \end{myBullets} &
\high & \high & \high &
\begin{myBullets} 
\item Similar mitigation strategies as Bitcoin and Ethereum
\item Explore alternative consensus mechanisms less vulnerable to quantum attacks (e.g., voting-based consensus algorithms)
\item Strengthen validator key management to prevent privilege escalation.

\vspace{-2ex} \end{myBullets} \\ \hline
Litecoin &
\begin{myBullets} 
\item Scrypt: Memory-intensive \gls{PoW} hashing algorithm (more ASIC-resistant than SHA-256).
\item \gls{ECDSA}: Used for digital signatures.
\vspace{-2ex} \end{myBullets} &
\begin{myBullets} 
\item Scrypt provides some resistance, but \gls{QC} can break it.
\item ECDSA is vulnerable to signature forgery.
\item Enhanced quantum mining could enable 51\% attacks.
\item Mining disruptions may cause DoS on the network.
\vspace{-2ex} \end{myBullets} &
\begin{myBullets} 
\item Spoofing: Identity theft through signature compromise
\item Tampering: Block alterations exploiting quantum vulnerabilities
\item Repudiation: Forged transactions leading to disputes
\item \gls{DoS}: Mining disruption due to compromised algorithms
\item Elevation of Privilege: Unauthorized control of mining resources
\vspace{-2ex} \end{myBullets} &
\high & \high & \high &
\begin{myBullets} 
\item Research on memory-hard hashing functions secure against classical and quantum attacks
\item Adopt quantum-resistant signature schemes to replace ECDSA.
\item Explore quantum resistant alternative for \gls{PoW} mechanisms such as proof-of-space or memory-hard PoW models to enhance resistance.
\item Evaluate Scrypt’s quantum resilience and improve hashing where necessary.

\vspace{-2ex} \end{myBullets} \\ \hline
\multirow{7}{*}{\begin{minipage}[c]{\linewidth}
 Zcash (Privacy Coin)
\end{minipage}} &
\begin{myBullets} 
\item Standard Transactions:
\begin{itemize}
\item ECDSA/SHA-256: Used for signatures and hashing, similar to Bitcoin.
\end{itemize}
\vspace{-2ex} \end{myBullets} &
\begin{myBullets} 
\item Vulnerable to signature forgery and block tampering.
\vspace{-2ex} \end{myBullets} &
\begin{myBullets} 
\item Spoofing: Signature forgery enabling identity theft
\item Tampering: Modification of transaction data
\item Repudiation: Disputes from altered or forged transactions
\item \gls{DoS}: Disruption of standard transaction processing
\vspace{-2ex} \end{myBullets} &
\high & \high & \high &
\begin{myBullets} 
 \item Transition from ECDSA to post-quantum signature schemes (e.g., lattice-based cryptography)
 \item Replace SHA-256 with SHA-512 or alternatives such as SHA-3~\cite{nist2024transition}
\vspace{-2ex} \end{myBullets} \\ \cline{2-8}
& 
\begin{myBullets} 
\item Shielded Transactions:
\begin{itemize}
\item Groth16 zk-SNARKs$^{*}$ (Elliptic curve and pairing-based cryptography): Used for private transactions.
\item ECC-based key generation: Supports Groth16 and transaction privacy.
\end{itemize}
\vspace{-2ex} \end{myBullets} 
& 
\begin{myBullets} 
\item Loss of anonymity if Groth16 is broken by quantum attacks.
\item Potential manipulation of shielded transactions if proofs or keys are compromised.
\item Disruption of shielded transaction validation due to computational overload.\vspace{-2ex} \end{myBullets} 
& 
\begin{myBullets} 
\item Information Disclosure: Loss of privacy due to \gls{ZKP} compromise
\item Tampering: Unauthorized modification of shielded transactions
\item Repudiation: Disputes from forged proofs
\item \gls{DoS}: Disruption of shielded transaction processing
\item Elevation of Privilege: Exploiting \gls{ZKP} weaknesses to gain unauthorized control

\vspace{-2ex} \end{myBullets} 
& 
\high 
& 
\high 
& 
\high 
& 
\begin{myBullets} 
\item Research and implement post-quantum zk-SNARKs (e.g., Virgo \cite{PQZKP1}, Ligero \cite{PQZKP2} and Aurora \cite{PQZKP3}) and post-quantum signatures.
\item Develop fallback mechanisms to handle computational overload or proof failures
\item Enhance network-wide transition plans to quantum-resistant cryptography
\vspace{-2ex} \end{myBullets} \\ \hline
\end{tabular}
}
\vspace{3pt}
\newline \scriptsize{$^*$In May 2022, Zcash introduced the Orchard shielded payment protocol with Network Upgrade 5 (NU5), using the Halo 2 zero-knowledge proving system. Halo 2 removes the trusted setup and supports scalable private payments. While it improves efficiency over previous zk-SNARKs like Groth16, its security depends on assumptions not proven secure against quantum attacks.}
  \vspace{-0.3cm}
\end{table*}

It is important to emphasize that the likelihood ratings assigned in the Table represent a scenario in which large-scale quantum computers are available and capable of breaking or weakening current cryptographic primitives. Our earlier analysis (Figure~\ref{fig:Likelihood}) indicates that the expected likelihood of such powerful quantum threats remains relatively low in the short term (within the next 10 years), increases to a medium level by around 15 years, and becomes high beyond a 20-year horizon. Thus, the \textit{High} likelihood ratings in Table~\ref{tab:quantum_impact} should be interpreted as long-term projections rather than immediate certainties. In the near term, due to the limited maturity of \gls{QC}, these threats are far less likely to materialize.

\subsection{Security Evaluation of Major \gls{BC} Platforms: Lessons Learned and Strategic Roadmap}\label{sec:patform-lesson}

{Comparative evaluation shows that major \gls{BC} platforms share fundamental quantum vulnerabilities despite architectural differences. All analyzed platforms (Bitcoin, Ethereum, Ripple, Litecoin, Zcash) critically depend on \gls{ECDSA} digital signatures, which Shor's algorithm could break, enabling signature forgery and unauthorized transactions. Their underlying hash functions also face quantum-accelerated attacks via Grover's algorithm: Bitcoin and Ripple rely on SHA-256, Ethereum uses Keccak-256, and Litecoin employs Scrypt for mining, offering only partial resistance. }

{Platform-specific risk profiles require tailored mitigation. Bitcoin has unique exposure with about 30\% of total supply in public-key-revealed addresses, combined with mining centralization risks if SHA-256 is broken. Ethereum adds complexity through its smart contract ecosystem, gas-constrained environments affecting post-quantum signature verification efficiency, and \gls{DeFi} applications requiring quantum-resistant privacy mechanisms. Ripple’s semi-centralized validator model creates governance-based risks if validator keys are compromised. Litecoin benefits from Scrypt’s partial quantum resistance but remains vulnerable to signature-based attacks and potential quantum mining acceleration. Zcash combines standard transaction vulnerabilities with privacy-specific threats, as Groth16 zk-SNARK or Halo 2 breaks could compromise anonymity and shielded transaction integrity.}

{Strategic roadmap priorities by platform: {(i)} Bitcoin: develop migration protocols for exposed public-key addresses (\textasciitilde30\% of total supply), transition from SHA-256 to quantum-resistant alternatives (e.g., SHA-512 or lattice-based \gls{PoW}), and deploy backward-compatible post-quantum signature schemes preserving operational continuity; {(ii)} Ethereum: implement gas-efficient lattice-based signature verification~\cite{refc12}, develop quantum-secure smart contract infrastructures, and integrate post-quantum \glspl{ZKP} for privacy-preserving \gls{DeFi} applications operating within gas-constrained environments~\cite{refc27}; {(iii)} Ripple: establish quantum-resistant validator selection mechanisms and consensus protocols resilient to quantum adversaries while ensuring performance~\cite{refc5}; {(iv)} Litecoin: assess and enhance Scrypt’s quantum resistance while exploring ASIC-resistant, memory-hard \gls{PoW} alternatives; {(v)} Zcash: develop and integrate advanced post-quantum \glspl{ZKP} (Virgo, Ligero, Aurora), implement privacy-preserving lattice-based commitments~\cite{refc30}, and deploy quantum-resistant ring signatures~\cite{refc29} to uphold transactional anonymity. }

{Cross-platform strategies should include cryptographic agility frameworks for modular algorithm replacement, standardized migration protocols for ecosystem-wide adoption, quantum-secure interoperability enabling safe cross-chain communication, and emergency response mechanisms for rapid cryptographic updates in the face of quantum breakthroughs. These coordinated measures are essential to maintain \gls{BC} integrity, trust, and availability in the quantum era.}
\section{Post-Quantum \glspl{BC}: Current Availability and Their Roles}\label{sec:PQ-BC}

\begin{table*}[ht]
\caption{Post-Quantum \gls{BC} Platforms: Features, Applications, and STRIDE Threats with Corrected Risk Assessment}
\scriptsize
\resizebox{\textwidth}{!}{
\begin{tabular}{|m{0.09\linewidth}|m{0.06\linewidth}|m{0.08\linewidth}|m{0.14\linewidth}|m{0.08\linewidth}|m{0.09\linewidth}|m{0.1\linewidth}|m{0.4\linewidth}|c|c|c|}
\hline
\textbf{Platform} & \textbf{Structure} & \textbf{\gls{BC} Type} & \textbf{Consensus Mechanism} & \textbf{Crypto Type} & \textbf{Signature Alg.} & \textbf{Application} & \textbf{STRIDE Threats} & \textbf{L} & \textbf{I} & \textbf{R} \\
\hline

QRL \cite{qrl} & \gls{BC} & Permissionless Public & \gls{PoS} (QRL \gls{PoS})& Hash & XMSS & Digital asset safety & 
\begin{myBullets} 
 \item {Spoofing}: Strong resistance via XMSS but risks arise from weak supporting infrastructure.
 \item {\gls{DoS}}: Resource exhaustion from computational overhead of PQC algorithms.
\vspace{-2ex} \end{myBullets} & \med & \high & \high \\
\hline

Komodo \cite{komodo} & \gls{BC} & Permissionless Public & \gls{PoW} & Lattice & Dilithium & Multi-chain smart contracts & 
\begin{myBullets} 
 \item {Spoofing}: Challenges in integrating Dilithium signatures may introduce vulnerabilities.
 \item {Tampering}: Risks during cryptographic transitions.
 \item {\gls{DoS}}: Overhead of PQC algorithms.
\vspace{-2ex} \end{myBullets} & \med & \med & \med \\
\hline

Nexus \cite{nexus} & \gls{BC} & Permissionless Public & \gls{PoW}, \gls{PoS} & Lattice & FALCON & Decentralized routing & 
\begin{myBullets} 
 \item {Tampering}: Integrity risks due to decentralized routing.
 \item {\gls{DoS}}: Overhead from FALCON scheme may lead to overload.
\vspace{-2ex} \end{myBullets} & \med & \high & \high \\
\hline

Tidecoin \cite{tidecoin} & \gls{BC} & Permissionless Public & \gls{PoW} & Lattice & FALCON-512 & Cryptocurrency & 
\begin{myBullets} 
 \item {Spoofing}: FALCON-512 integration challenges may lead to vulnerabilities.
 \item {Tampering}: Cryptographic update processes may expose integrity risks.
 \item {\gls{DoS}}: Performance constraints from \gls{PQC} implementations.
\vspace{-2ex} \end{myBullets} & \med & \high & \high \\
\hline

Abelian \cite{abelian} & \gls{BC} & Permissionless Public & \gls{PoW} & Lattice & Dilithium & Privacy-focused cryptocurrency & 
\begin{myBullets} 
 \item {Information Disclosure}: Potential for privacy leaks during \gls{PQC} transitions.
 \item {\gls{DoS}}: High resource demand may be exploited.
\vspace{-2ex} \end{myBullets} & \med & \high & \high \\
\hline

LACChain \cite{lacnet} & \gls{BC} & Permissioned Public$^*$ & \gls{PoA} & Lattice & FALCON-512 & Latin American adoption & 
\begin{myBullets} 
 \item {Repudiation}: Insufficient cryptographic logging during key exchanges may arise.
 \item {\gls{DoS}}: Medium likelihood due to potential insider threats and governance vulnerabilities, despite a controlled environment.
\vspace{-2ex} \end{myBullets} & \med & \med & \med \\
\hline

QAN \cite{qanplatform} & \gls{BC} & Hybrid & \gls{PoR} & Lattice & Dilithium & Smart contracts, DApps & 
\begin{myBullets} 
 \item {Tampering}: Cryptographic transition risks during hybrid operations.
 \item {Elevation of Privilege}: Weak hybrid integration could enable unauthorized access.
\vspace{-2ex} \end{myBullets} & \med & \high & \high \\
\hline

HCASH \cite{hcash} & \gls{BC}, DAG & Public & \gls{PoW}, PoS & Lattice & BLISS & Cross-ecosystem data flow & 
\begin{myBullets} 
 \item {Information Disclosure}: Vulnerabilities in data flow management across ecosystems.
 \item {\gls{DoS}}: High likelihood from resource-intensive dual structures.
\vspace{-2ex} \end{myBullets} & \med & \high & \high \\
\hline

IOTA \cite{iota} & DAG & Permissionless Public & PoS - OTV (On-Tangle-Voting) & Hash & W-OTS$^{**}$ & IoT micro-transactions & 
\begin{myBullets} 
 \item {Spoofing}: Transition to EdDSA signatures may increase impersonation risks.
 \item {Tampering}: Vulnerabilities in the Tangle's data integrity mechanisms.
 \item {\gls{DoS}}: Scalability features may be targeted.
\vspace{-2ex} \end{myBullets} & \high & \high & \high \\
\hline
\end{tabular}}
\vspace{3pt} \newline
\scriptsize{$^*$ Permissioned Public: Publicly accessible for reading/interacting, but validator roles require authorization by a governing body.\\
$^{**}$ In April 2021, they transitioned to using EdDSA as their digital signature, replacing W-OTS.}  \vspace{-0.5cm}
\label{table:postquantumblockchains}
\end{table*}

As the threat of \gls{QC} intensifies, \gls{BC} projects are increasingly integrating \gls{PQC} to future-proof their systems. This evolution addresses critical challenges such as integrating quantum-safe cryptography, managing performance trade-offs, and ensuring interoperability with existing \gls{BC} platforms, all central to the themes discussed in this section. {The evolutionary milestones of post-quantum \gls{BC} platforms are depicted in Figure~\ref{fig:quantum_blockchain_timeline}.}
\begin{figure}[htbp]
 \centering
 \resizebox{\linewidth}{!}{
  \begin{tikzpicture}[every node/.style={scale=0.85}, node distance=0.6cm and 0.6cm]
   
    \tikzset{
    milenode/.style={
     align=center,
     text width=2.2cm,
     minimum height=1.8cm,
     font=\small,
     inner sep=6pt,
     fill=white,
     rounded corners=4pt,
     draw=black!60,
     line width=0.8pt,
     drop shadow={shadow xshift=0.5pt, shadow yshift=-0.5pt, opacity=0.3}
    },
    timeline/.style={
     line width=2pt,
     draw=blue!70,
     rounded corners=2pt
    },
    connector/.style={
     line width=1pt,
     opacity=0.8
    }
   }

   \draw[timeline, orange] (0,0) -- (14,7);

   \foreach \x/\y/\year in {
    0/0/2014, 2/1/2016, 4/2/2018, 6/3/2020,
    8/4/2022, 10/5/2023, 12/6/2024, 14/7/2025
   } {
    \node[above=2pt] at (\x,\y) {\bfseries \year};
    \draw[very thick, orange] (\x,\y-0.1) -- (\x,\y+0.1);
   }

   \node[milenode, fill=purple!15, above left=0.7cm and 0.4cm of {(0.75,0)}] (komodo2014) {\textbf{Komodo}\\Atomic Swap Dev Begins};
   \draw[thin, purple!50] (0,0) -- 
   ($(komodo2014.south east)+(-0.75,0)$);
   \fill[purple!50] (0,0) circle (3pt);

   \node[milenode, fill=blue!15, below right=0.5cm and 0.3cm of {(0.6,1)}] (qrl2016) {\textbf{QRL}\\Development Begins};
   \draw[thin, blue!50] (2,1) -- ($(qrl2016.north west)+(1.4,0)$);
   \fill[blue!50] (2,1) circle (3pt);

   \node[milenode, fill=purple!15, above left=0.5cm and 0.3cm of {(2.5,1.25)}] (komodo2016) {\textbf{Komodo}\\Genesis Block};
   \draw[thin, purple!50] (2.5,1.25) -- (komodo2016.south east);
   \fill[purple!50] (2.5,1.25) circle (3pt);

   \node[milenode, fill=purple!15, below right=0.6cm and 0.3cm of {(2.5,1.75)}] (komodo2017) {\textbf{Komodo}\\Mainnet, dPoW \& Atomic Swaps};
   \draw[thin, purple!50] (3.5,1.75) --  ($(komodo2017.north west)+(1,0)$);
   \fill[purple!50] (3.5,1.75) circle (3pt);

   \node[milenode, fill=blue!15, above left=0.6cm and 0.3cm of {(4.5,2.25)}] (qrl2018) {\textbf{QRL}\\PoW Mainnet};
   \draw[thin, blue!50] (4.5,2.25) -- (qrl2018.south east);
   \fill[blue!50] (4.5,2.25) circle (3pt);

   \node[milenode, fill=blue!15, below right=0.6cm and 0.3cm of {(4.2,2.75)}] (qrlpos) {\textbf{QRL}\\PoS Migration};
   \draw[thin, blue!50] (5.5,2.75) -- ($(qrlpos.north west)+(1.3,0)$);
   \fill[blue!50] (5.5,2.75) circle (3pt);

   \node[milenode, fill=green!15, above left=0.6cm and 0.3cm of {(6,3)}] (tidecoin) {\textbf{Tidecoin}\\Falcon-512 Integration};
   \draw[thin, green!50] (6,3) --($(tidecoin.south east)+(0,0)$);
   \fill[green!50] (6,3) circle (3pt);

   \node[milenode, fill=red!15, below right=0.6cm and 0.3cm of {(6,3.5)}] (hcash) {\textbf{HCASH}\\BLISS \& DAG /PoW/PoS};
   \draw[thin, red!50] (7,3.5) -- ($(hcash.north west)+(1,0)$);
   \fill[red!50] (7,3.5) circle (3pt);

   \node[milenode, fill=purple!15, above left=0.6cm and 0.3cm of {(8,3.75)}] (komodo2021) {\textbf{Komodo}\\dPoW to Litecoin};
   \draw[thin, purple!50] (7.5,3.75) -- ($(komodo2021.south east)+(-0.5,0)$);
   \fill[purple!50] (7.5,3.75) circle (3pt);

   \node[milenode, fill=gray!15, below right=0.6cm and 0.3cm of {(8,4.25)}] (abelian) {\textbf{Abelian}\\Privacy-focused PQC Coin};
   \draw[thin, gray!50] (8.5,4.25) -- ($(abelian.north west)+(0.5,0)$);
   \fill[gray!50] (8.5,4.25) circle (3pt);

   \node[milenode, fill=yellow!15, above left=0.6cm and 0.3cm of {(10,4.65)}] (qan) {\textbf{QAN}\\Hybrid PoR with Dilithium};
   \draw[thin, yellow!50] (9.3,4.65) --    ($(qan.south east)+(-0.7,0)$);
   \fill[yellow!50] (9.3,4.65) circle (3pt);

   \node[milenode, fill=orange!15, below right=0.6cm and 0.3cm of {(10,5)}] (lacchain) {\textbf{LACChain}\\Falcon-512 Smart Contracts};
   \draw[thin, orange!50] (10,5) -- (lacchain.north west);
   \fill[orange!50] (10,5) circle (3pt);

   \node[milenode, fill=teal!15, above left=0.6cm and 0.3cm of {(11.7,5.5)}] (iota) {\textbf{IOTA}\\EdDSA Transition};
   \draw[thin, teal!50] (11,5.5) --    ($(iota.south east)+(-0.7,0)$);
   \fill[teal!50] (11,5.5) circle (3pt);

   \node[milenode,fill=cyan!15, below right=0.6cm and 0.3cm of {(12.7,7)}] (nexus2025) {\textbf{Nexus}\\Layer 1 \& Testnet III};
   \draw[thin, cyan!50] (14,7) -- ($(nexus2025.north west)+(1.3,0)$);
   \fill[cyan!50] (14,7) circle (3pt);

  \end{tikzpicture}
 }
 \caption{{Timeline of Key Developments in Post-Quantum  \gls{BC} platforms}}
 \vspace{-0.5cm}
 \label{fig:quantum_blockchain_timeline}
\end{figure}
Post-quantum \gls{BC} platforms employ cryptographic algorithms designed to withstand quantum attacks, primarily leveraging lattice-based and hash-based schemes. Representative implementations include QRL~\cite{qrl}, Komodo~\cite{komodo}, Nexus~\cite{nexus}, Tidecoin~\cite{tidecoin}, Abelian~\cite{abelian}, LACChain~\cite{lacnet}, QAN~\cite{qanplatform}, HCASH~\cite{hcash}, and IOTA~\cite{iota}, with shared objectives of quantum-safe cryptographic integration, performance optimization, and interoperability assurance. Table~\ref{table:postquantumblockchains} comprehensively summarizes each platform's architecture, cryptographic primitives, applications, and risk assessments.

QRL adopts XMSS hash-based signatures for digital asset security, providing forward secrecy at the cost of larger signatures ($\sim 2$ KB versus 64 bytes for ECDSA). Komodo employs Dilithium lattice-based signatures for multi-chain smart contracts, balancing efficiency with post-quantum robustness. Nexus integrates FALCON signatures for decentralized routing, benefiting from compact signatures (666 bytes for FALCON-512) while demanding robust key management. Tidecoin combines FALCON-512 with CPU-efficient \gls{PoW} consensus, maintaining hardware compatibility while mitigating quantum threats. Abelian leverages Dilithium to enhance cryptocurrency transaction privacy.
LACChain incorporates post-quantum X.509 certificates and quantum-secure \gls{TLS} protocols within a permissioned-public architecture, enabling public read access while restricting validator roles to authorized participants. QAN, a hybrid platform, combines Dilithium signatures with \gls{PoR} consensus for smart contracts and DApps, offering operational flexibility but introducing cryptographic transition risks. HCASH merges \gls{BC} and DAG structures using BLISS lattice-based signatures for secure cross-ecosystem data exchange. IOTA initially implemented probabilistic DAG-based consensus with W-OTS hash-based signatures optimized for IoT micro-transactions, but transitioned to EdDSA in April 2021, trading quantum resistance for operational efficiency, a decision that reintroduces elliptic curve vulnerabilities against future quantum attacks.

Post-quantum \gls{BC} adoption faces significant implementation challenges. Performance overheads from larger cryptographic primitives (e.g., XMSS signatures exceeding 2 KB) increase computational demands, affecting transaction throughput and straining resource-constrained devices. Interoperability complexity emerges as hybrid systems combining classical and quantum-resistant methods introduce vulnerabilities during cryptographic transitions, requiring robust synchronization mechanisms as demonstrated by QAN's hybrid approach. Regulatory uncertainty compounds these challenges, as formal \gls{PQC} integration guidelines for \gls{BC} applications remain underdeveloped despite ongoing NIST standardization efforts for CRYSTALS-Dilithium and FALCON.

The likelihood and impact of security threats vary significantly across post-quantum \gls{BC} platforms. For the post-quantum platforms analyzed in Table \ref{table:postquantumblockchains},  QRL, Komodo, Nexus, Tidecoin, Abelian, LACChain, QAN, and HCASH exhibit \textit{Medium} attack likelihood due to partial vulnerabilities: resource-intensive algorithms (XMSS, FALCON-512) enable DoS attacks but benefit from rate-limiting and limited user bases; multi-layer consensus architectures provide partial resilience despite complexity; mature codebases and network segmentation contain transition risks; privacy implementations may expose metadata if mishandled; permissioned models reduce external threats while remaining vulnerable to insider attacks; hybrid approaches face tampering during transitions but benefit from niche adoption; integrated architectures complicate data flow yet benefit from ecosystem isolation. IOTA faces \textit{High} likelihood due to EdDSA's quantum vulnerability combined with extensive IoT deployment exposure.

Impact assessment shows Komodo and LACChain experience \textit{Medium} disruption levels due to multi-chain design and controlled environments enabling effective breach containment and recovery. Other platforms (QRL, Tidecoin, Nexus, Abelian, QAN, HCASH, IOTA) face \textit{High} impact from successful exploits, encompassing severe operational downtime, compromised cryptographic integrity, and significant reputational damage. These risk profiles underscore the critical importance of continuous security audits, robust governance frameworks, and proactive threat mitigation for quantum-safe \gls{BC} ecosystems.

\subsection{{Post-Quantum \glspl{BC}: Lessons Learned and Strategic Roadmap}}\label{sec:PQ-BC-lesson}

{Comparative evaluation demonstrates that algorithm choice, architectural design, and migration strategy jointly determine quantum resilience and operational viability. Hash-based signatures (XMSS~\cite{nist800-208} in QRL) provide strong forward security but incur bandwidth and storage penalties from large signatures ($>2$~KB). Lattice-based deployments using Dilithium and FALCON (Komodo, Abelian, QAN, Nexus, Tidecoin) achieve compact signatures (666~bytes for FALCON-512) with superior performance yet require robust side-channel protections. Platforms using non-NIST primitives (BLISS in HCASH, quantum-vulnerable EdDSA in IOTA) face regulatory uncertainty and heightened exposure, underscoring migration urgency to standardized schemes.}

{Architectural choices strongly shape attack surfaces. Permissioned-public models (LACChain) reduce external threats but remain vulnerable to insider compromise, necessitating stronger governance. Multi-chain systems (Komodo) and DAG-based frameworks~\cite{refc1} (HCASH, IOTA) improve throughput but risk cross-chain spoofing, synchronization faults, and data tampering, requiring robust verification safeguards. Hybrid migration designs (QAN) offer flexibility but must harden defenses against rollback and privilege escalation during transitions. Early adopters (QRL, Komodo) gain operational maturity but risk cryptographic lock-in without crypto-agility mechanisms~\cite{alnahawi2023state,petrenko2019assessing}.}

{Strategic roadmap priorities by platform: {(i)} HCASH and IOTA: migrate from BLISS and EdDSA to NIST-approved schemes for compliance and quantum resistance; {(ii)} QRL: reduce XMSS overhead via signature compression and efficient state handling; {(iii)} Komodo, Abelian, QAN, Nexus, and Tidecoin: integrate hardware- and software-level side-channel protections without performance degradation~\cite{singh2024end,paiva2025tu}; {(iv)} QRL and Komodo: implement cryptographic agility mechanisms~\cite{alnahawi2023state,petrenko2019assessing} to prevent lock-in and facilitate future migrations; {(v)} Komodo, HCASH, IOTA, and QAN: strengthen cross-chain verification and synchronization safeguards~\cite{refml17,refc1} to mitigate inter-network attack surfaces. These measures confirm that robust post-quantum \gls{BC} security requires coordinated alignment of cryptographic primitives, architectural design, and migration strategy within a cohesive operational framework.}

\section{{Challenges and Future Research Directions}}\label{sec:future}

{The convergence of \gls{BC} with emerging technologies such as Web3~\cite{RAY2023213,refml28}, quantum-enhanced \gls{AI}~\cite{11018396,10988887}, and advanced privacy-preserving cryptographic protocols~\cite{9524814} represents a fundamental transformation in decentralized computing. As \gls{QC} capabilities mature, this convergence introduces novel security vulnerabilities requiring interdisciplinary research spanning cryptography, \gls{AI}, and distributed systems design. Building upon the risk modeling framework in Section~\ref{sec:risk}, hybrid architectural designs in Section~\ref{sec:hybrid}, and platform-specific evaluations in Sections~\ref{sec:platform} and \ref{sec:PQ-BC}, this section identifies critical challenges and priority research directions for quantum-resilient \gls{BC} ecosystems.}

\subsection{{Quantum-Enhanced \gls{AI} and \gls{BC} Integration}}\label{quantum-ai-integration}

{The convergence of \gls{QC} and \gls{AI} presents both unprecedented security threats and transformative defensive opportunities for post-quantum \gls{BC} systems~\cite{11018396}. \gls{QML} enhances cryptanalysis by combining classical \gls{ML} pattern recognition with quantum computational advantages, enabling adversaries to exploit statistical biases, solve large algebraic systems, and explore key spaces with greater efficiency. Quantum Deep Neural Networks (DNNs) uncover complex non-linear relationships between plaintext bits and side-channel leakages, amplifying differential and linear cryptanalysis. These capabilities are further reinforced through quantum algorithms such as the HHL algorithm~\cite{harrow2009quantum,chen2022quantum} for exponential speedups in sparse linear systems and the Quantum Approximate Optimization Algorithm (QAOA)~\cite{BLEKOS20241} for combinatorial key recovery.}

{Quantum-enhanced cryptanalytic techniques pose distinct, scheme-specific challenges requiring tailored defenses across \gls{PQC} families. \textit{Lattice-based schemes} face accelerated basis reduction via quantum random walk-aided sieving, necessitating increased lattice dimensions and quantum-augmented proofs~\cite{refml29}. 
\textit{Code-based schemes} are threatened by Grover-enhanced error pattern searches combined with deep learning side-channel analysis, requiring longer code lengths and obfuscation techniques~\cite{refml32,refml31}. \textit{Hash-based cryptography} is vulnerable to quantum-accelerated collision finding via amplitude amplification, mandating extended hash outputs and strict domain separation~\cite{refml33}.}

{Beyond cryptography, advanced \gls{AI} introduces additional attack vectors against \gls{BC} infrastructures~\cite{refml13}. These include automated smart contract vulnerability discovery, analysis of transaction patterns, identification of optimal conditions for consensus manipulation {(e.g., 51\% attacks in} \gls{PoW} {or one-third threshold violations in} \gls{BFT}/\gls{PoS}), and orchestration of coordinated Sybil or \gls{DDoS} attacks with adaptive real-time strategies~\cite{refml4,refml7,refml22}. Conversely, \gls{AI} provides significant defensive capabilities~\cite{refml2,refml3,refml9,refml10,refml11}, including cryptographic parameter optimization, real-time anomaly detection, network intrusion prevention, and automated quantum-safe key lifecycle management~\cite{refml21}.}


{Several critical research areas require immediate attention to address the evolving quantum-AI threat landscape. These priorities include developing quantum-resistant adversarial training frameworks against adaptive quantum-classical attacks like SALSA on LWE schemes~\cite{refml25}, creating \gls{BC}-specific \gls{AI} auditing tools for decentralized validation, and designing \gls{AI}-resistant cryptographic primitives. Furthermore, it is essential to implement zero-knowledge \gls{ML} proof systems (i.e., \gls{ZKP} {protocols attesting to ML inference correctness and model integrity}~\cite{10613422,10.1145/3627703.3650088}) built on quantum-resistant lattice-based cryptography for zero-trust validation environments~\cite{refc14,refc31} and establish ethical frameworks addressing data privacy and algorithmic bias as \gls{AI} integrates with Web3 infrastructure.}

\subsection{{Web3, Quantum Security and Interoperability}}\label{web3-quantum-security}
{Web3 technologies\cite{RAY2023213,refml28}, including \gls{DeFi}~\cite{al2024defi}, \gls{NFT}~\cite{refml20}, and metaverse platforms, face increasing threats from \gls{QC} advancements that compromise cryptographic primitives. The "Harvest Now, Decrypt Later" attacks endanger long-term encrypted data, while adoption is hindered by high implementation costs, limited asset visibility, and deployment challenges. Decentralized governance complicates consensus on quantum-safe upgrades, and \gls{NIST} standards require multi-stakeholder coordination amid a shortage of post-quantum expertise.}

{Vulnerabilities persist across core components: \textit{DID:Web standards}~\cite{did1.1} depend on quantum-vulnerable elliptic curves and SHA-256; oracles like Chainlink present centralized signing points of failure; and cross-chain bridges jeopardize Web3 composability.
Mitigation strategies include hybrid cryptographic stacks, cryptographic agility for rapid PQC transitions, quantum-resistant signatures, and secure key exchange via quantum key distribution. Platforms such as Ethereum 3.0 and Hyperledger Ursa are already exploring PQC integration.}
{Future research should prioritize the standardization of \gls{PQC}-based decentralized identity protocols with cross-chain interoperability~\cite{LI2025100286,Jia2023Cross-Chain}. Additionally, developing threshold \gls{PQC} oracles utilizing \gls{MPC} techniques can eliminate single points of failure~\cite{refc34}. Equally important is the formal verification of bridge protocols' security under quantum threat models~\cite{ethereum_formal_verification,lewis2023formal,tolmach2023securing} to ensure the stability of \gls{DeFi} ecosystems and facilitate robust Web3 interoperability.}

\subsection{{Advanced Privacy-Preserving Cryptographic Systems}}\label{privacy-preserving-systems}

{\gls{QC} fundamentally undermines cryptographic privacy mechanisms crucial for \gls{BC} applications. The majority of the current \glspl{ZKP}, such as Groth16 zk-SNARK, depend on elliptic curve cryptography, which is vulnerable to quantum algorithms. Moreover, privacy-preserving computation requires homomorphic encryption schemes that are secure against quantum adversaries~\cite{10.1007/978-3-031-91124-8_1}, particularly for sensitive applications such as medical \gls{BC} platforms.
Future research in this domain should prioritize two key areas. Firstly, a  comprehensive evaluation of quantum-secure \glspl{ZKP}~\cite{refc14,refc27}, particularly lattice-based constructions such as the LaZer Library implementations~\cite{refc31}, is essential.
This evaluation should encompass an analysis of proof size scalability, computational complexity, and the trade-offs associated with their \gls{BC} integration. Secondly, developing quantum-secure Homomorphic Encryption (HE) solutions~\cite{10.1007/978-3-031-91124-8_1} optimized for smart contract runtimes is critical~\cite{9408585}. {HE enables privacy-preserving smart contracts through computations on encrypted data without exposure, though current schemes remain inefficient for production-scale deployment, necessitating research into practical quantum-resistant implementations.}}

\subsection{{Federated Learning, Edge Computing, and IoT Integration}}\label{federated-edge-iot}

{\gls{FL} represents a decentralized paradigm that keeps sensitive data localized, reducing privacy risks~\cite{refml8}. When combined with \gls{BC} and quantum cryptographic protocols, \gls{FL} can significantly enhance data protection, especially in \gls{IoT} ecosystems~\cite{radanliev2024cyber,9631953,10756206}.  \gls{QC} can optimize \gls{FL} training through quantum-enhanced algorithms, strengthen encryption via quantum cryptography, enable efficient distributed data aggregation, and protect against inference-based attacks by disrupting adversarial pattern recognition~\cite{10988887}. \gls{BC} provides decentralized, auditable coordination for \gls{FL} networks~\cite{refml5,refml8}, addressing malicious participant detection, data integrity verification, and single point of failure elimination.}
{On the other hand, edge computing proliferation creates distributed attack surfaces requiring quantum-resistant cryptographic operations at edge nodes, including secure key management, authenticated firmware updates, and protected communication channels~\cite{karakaya2024survey}. \gls{IoT} deployments create vast attack surfaces with resource-constrained devices traditionally using lightweight cryptographic protocols~\cite{chawla2023roadmap}. The quantum threat amplifies \gls{IoT} security challenges due to cryptographic update difficulties across large device populations, requiring novel approaches balancing security with practical deployment constraints~\cite{mansoor2025securing}.}

{Research priorities include quantum-resistant protocols for secure \gls{FL} aggregation, \gls{BC}-based incentive mechanisms deterring adversarial behavior, privacy-preserving techniques protecting participant data against quantum adversaries, lightweight \gls{PQC} implementations for resource-constrained edge devices~\cite{10756206}, distributed key management systems for edge networks, quantum-resistant \gls{IoT} communication protocols, \gls{BC}-based \gls{IoT} identity management, and secure update mechanisms enabling \gls{PQC} migration across \gls{IoT} populations\cite{chawla2023roadmap}.}




\subsection{{Interdisciplinary  Technical and Standardization Challenges}}\label{cross-cutting-challenges}

{Integrating \gls{PQC}, \gls{ZKP}, and \gls{AI} into \gls{BC} systems introduces considerable performance overhead, affecting throughput and transaction finality. These issues are magnified in quantum-\gls{AI} and Web3 contexts, where latency-sensitive operations, such as \gls{ZKP} generation, are especially impacted~\cite{refml28,refml2,refml5,refml6}. Although \gls{NIST} has standardized several quantum-resistant algorithms, notable gaps remain in \gls{BC}-specific implementations, particularly those involving Web3 and \gls{AI}. A critical barrier to adoption is the lack of interdisciplinary expertise across cryptography, distributed systems, and \gls{AI}.
Key research directions include hardware acceleration for \gls{PQC}-\gls{ZKP} stacks using GPUs and TPUs, quantum-aware \gls{BC} sharding, and standardized performance benchmarks for \gls{DApp} environments. Coordinating  across various standards including \gls{NIST}, IEEE, ISO, IETF, and ETSI is essential to ensure interoperability. To address the skills gap, targeted educational programs, certification tracks, and collaborative academic-industry initiatives are needed. These directions are pivotal to preserving \gls{BC} security as \gls{QC} transitions from theory to deployment.}

\section{Conclusion}\label{sec:conclude}

The convergence of \gls{BC} technology and \gls{QC} presents both formidable challenges and transformative opportunities. While quantum algorithms such as Shor’s and Grover’s pose systemic threats to cryptographic foundations, proactive defense and structured risk assessment can ensure the continued resilience of \gls{BC} systems in the quantum era. Our analysis highlights that vulnerabilities extend across all components of the \gls{BC} ecosystem, including networks, mining pools, transaction verification mechanisms, smart contracts, and user wallets. Leading platforms such as Bitcoin, Ethereum, Ripple, Litecoin, and Zcash must adopt multifaceted mitigation strategies. These include transitioning to NIST-standardized \gls{PQC} algorithms, reinforcing consensus protocols to withstand quantum-enabled disruptions, and strengthening smart contract security through advanced analysis and verification techniques. Safeguarding \gls{BC} infrastructures demands collective effort from developers, researchers, wallet providers, and governance bodies. Continuous monitoring of quantum advancements, combined with proactive migration to quantum-resistant cryptography, is essential for preserving trust and long-term viability. Hybrid migration strategies, balancing classical and post-quantum primitives, further enable secure adaptation while minimizing operational risks during transition. In summary, although \gls{QC} poses a profound challenge, it also catalyzes innovation, collaboration, and standardization. Continued research in post-quantum cryptography, hybrid blockchain design, and quantum-aware security frameworks is critical for navigating this transition. By combining structured risk assessment, proactive defense, and cross-stakeholder collaboration, the \gls{BC} ecosystem can achieve quantum resilience and sustain its role as a cornerstone of secure, decentralized digital infrastructure in the quantum era.
\bibliographystyle{IEEEtran}
\bibliography{bibliography}

\end{document}